\documentclass[superscriptaddress,nofootinbib,aps,prd,10pt]{revtex4-1}
\usepackage{amsmath,amssymb,marvosym}
\usepackage{graphicx}
\usepackage{braket}
\newcommand{\omu}{\ensuremath{\overline \mu}}

\renewcommand{\d}{\ensuremath{\mathrm{d}}}

\newcommand{\MSbar}{\overline{\mbox{MS}}}
\renewcommand{\d}{\ensuremath{\mathrm{d}}}

\newcommand{\p}{\partial}

\begin{document}

\title{{\Large {\bf The No-Pole Condition in Landau gauge: \\[2mm]
                    Properties of the Gribov Ghost Form-Factor \\[2mm]
                    and a Constraint on the $2d$ Gluon Propagator \\[1mm]
      }}}

\author{Attilio~Cucchieri}
\email{attilio@ifsc.usp.br}
\affiliation{Ghent University, Department of Physics and Astronomy, Krijgslaan 281-S9, \\ 9000 Gent, Belgium}
\affiliation{Instituto de F\'\i sica de S\~ao Carlos, Universidade de S\~ao Paulo, Caixa Postal 369, \\
             13560-970 S\~ao Carlos, SP, Brazil}
\author{David~Dudal}
\email{david.dudal@ugent.be}
\affiliation{Ghent University, Department of Physics and Astronomy, Krijgslaan 281-S9, \\ 9000 Gent, Belgium}
\author{Nele~Vandersickel$\,$}
\email{nele.vandersickel@ugent.be}
\affiliation{Ghent University, Department of Physics and Astronomy, Krijgslaan 281-S9, \\ 9000 Gent, Belgium}

\begin{abstract}
We study general properties of the Landau-gauge Gribov ghost form-factor $\sigma(p^2)$ for SU($N_c$)
Yang-Mills theories in the $d$-dimensional case. We find a qualitatively different behavior for
$d=3,4$ with respect to the $d=2$ case. In particular, considering any (sufficiently regular) gluon
propagator $\mathcal{D}(p^2)$ and the one-loop-corrected ghost propagator, we prove in the $2d$ case
that the function $\sigma(p^2)$ blows up in the infrared limit $p \to 0$ as
$- \mathcal{D}(0) \ln(p^2)$. Thus, for $d=2$, the no-pole condition $\sigma(p^2) < 1$ (for $p^2 > 0$)
can be satisfied only if the gluon propagator vanishes at zero momentum, that is, $\mathcal{D}(0) = 0$.
On the contrary, in $d=3$ and 4, $\sigma(p^2)$ is finite also if $\mathcal{D}(0) > 0$. The same results are
obtained by evaluating the ghost propagator $\mathcal{G}(p^2)$ explicitly at one loop, using fitting forms
for $\mathcal{D}(p^2)$ that describe well the numerical data of the gluon propagator in two, three
and four space-time dimensions in the SU(2) case. These evaluations also show that, if one considers the
coupling constant $g^2$ as a free parameter, the ghost propagator admits a one-parameter family of behaviors
(labelled by $g^2$), in agreement with previous works by Boucaud et al. In this case the condition
$\sigma(0) \leq 1$ implies $g^2 \leq g^2_c$, where $g^2_c$ is a ``critical'' value. Moreover, a free-like
ghost propagator in the infrared limit is obtained for any value of $g^2$ smaller than $g^2_c$, while
for $g^2 = g^2_c$ one finds an infrared-enhanced ghost propagator. Finally, we analyze the
Dyson-Schwinger equation for $\sigma(p^2)$ and show that, for infrared-finite ghost-gluon vertices,
one can bound the ghost form-factor $\sigma(p^2)$. Using these bounds we find again that only in the $d=2$
case does one need to impose $\mathcal{D}(0) = 0$ in order to satisfy the no-pole condition. The
$d=2$ result is also supported by an analysis of the Dyson-Schwinger equation using a spectral
representation for the ghost propagator.
Thus, if the no-pole condition is imposed, solving the $d=2$ Dyson-Schwinger equations cannot lead to a
massive behavior for the gluon propagator. These results apply to any Gribov copy inside the so-called
first Gribov horizon, i.e.\ the $2d$ result $\mathcal{D}(0) = 0$ is not affected by Gribov noise. These
findings are also in agreement with lattice data.
\end{abstract}

\maketitle

%%%%%%%%%%%%%%%%%%%%%%%%%%%%%%%%%%%%%%%%%%%%%%%%%%%%%%%%%%%%%%%%%%%%%%%%%%%%%%%%%%%%%%%%%%%%%%%%%%%%%%%%%%%%%%%

\section{Introduction}

Green functions of Yang-Mills theories are gauge-dependent quantities. They can, however, be used as a
starting point for the evaluation of hadronic observables (see for example \cite{Alkofer:2000wg,Maris:2003vk,
Natale:2006nv,Dudal:2010cd}). Thus, the study of the infrared (IR) behavior of propagators and vertices is an
important step in our understanding of QCD. In particular, the confinement mechanism for color charges
\cite{Greensite} could reveal itself in the IR behavior of (some of) these Green functions. This IR
behavior should also be relevant for the description of the deconfinement transition and of the deconfined
phase of QCD. Indeed, at high temperature color charges are expected to be Debye-screened and the (electric
and magnetic) screening masses should be related to the IR behavior of the gluon propagator (see for example
\cite{PRINT-80-0538 (PRINCETON),Kraemmer:2003gd,Schafer:2005ff}).

\vskip 3mm
Among the gauge-fixing conditions employed in studies of Yang-Mills Green functions, a very popular choice
is the Landau gauge, which in momentum space reads $p_\mu A_{\mu}^b(p)=0$. From the continuum perspective
this gauge has various important properties, including its renormalizability, various associated
nonrenormalization theorems \cite{Piguet:1995er} and a ghost-antighost symmetry \cite{Alkofer:2000wg}. In
the past few years many analytic studies of Green functions in Landau gauge have focused on the solution of the
Yang-Mills Dyson-Schwinger equations (DSEs), which are the exact quantum equations of motion of the theory
(see for example \cite{Alkofer:2000wg,Maris:2003vk,ADP-94-8-T-150,Swanson:2010pw}). Since the DSEs are an
infinite set of coupled equations, any attempt of solving them requires a truncation scheme. Then, some Green
functions (usually the gluon and the ghost propagators) are obtained self-consistently from the considered
equations, while all the other Green functions entering the equations are given as an input. For the
coupled DSEs of gluon and ghost propagators two solutions have been extensively analyzed (see for example
Chapter 10 in Ref.\ \cite{Greensite} and Ref.\ \cite{Boucaud:2011ug} for recent short reviews). The scaling
solution \cite{Alkofer:2000wg,von Smekal:1997vx,Zwanziger:2001kw,Lerche:2002ep,Huber:2007kc} finds in $d=2, 3$
and 4 an IR-enhanced ghost propagator $\mathcal{G}(p^2)$ and a vanishing gluon propagator $\mathcal{D}(p^2)$
at zero momentum.\footnote{For the explanation of color confinement based on the scaling solution see for example
Section 3.4 in \cite{Alkofer:2006fu} and references therein.} In particular, the IR behavior of the two propagators
should be given respectively by $\mathcal{G}(p^2) \sim (p^2)^{-\kappa_G -1}$ and by $\mathcal{D}(p^2) \sim (p^2)^{2\kappa_D
+ (2-d)/2}$ with $\kappa_G = \kappa_D \approx 0.2 (d-1)$ \cite{Zwanziger:2001kw,Huber:2007kc}. On the other
hand, the massive solution \cite{Aguilar:2004sw,Boucaud:2006if,Aguilar:2008xm,Boucaud:2008ji,Boucaud:2008ky,
Binosi:2009qm,Aguilar:2010zx,Pennington:2011xs,Aguilar:2011xe} gives (for $d=3$ and 4) a free-like ghost propagator
in the IR limit, i.e.\ $\kappa_G = 0$, and a massive behavior for the gluon propagator,\footnote{The possible
existence of a dynamical mass for the gluons, as well as its relation to quark confinement through vortex
condensation, has been discussed a long time ago in Ref.\ \cite{Cornwall:1981zr}.} that is, $\mathcal{D}(0) > 0$
and $\kappa_D = (d-2)/4$.

The existence of two different types of solution for the coupled gluon and ghost DSEs is now understood
as due to the use of different auxiliary {\em boundary conditions}.\footnote{The possibility of
having different non-perturbative solutions for DSEs in relation with different boundary conditions
has been discussed for example in Refs.\ \cite{PRINT-74-1282 (SALERNO),270387} (see also Section 1.4 in
\cite{rivers} and Sections 3 and 3.1 in Ref.\ \cite{Alkofer:2000wg}).} These conditions can be given in terms
of the value of the ghost dressing function $\mathcal{F}(p^2) = p^2 \mathcal{G}(p^2)$ at a given momentum
scale $p$ \cite{Boucaud:2008ky,Fischer:2008uz}. In particular, if one considers $p = 0$, it is clear that
$1 / \mathcal{F}(0) = 0$ gives an IR-enhanced ghost propagator $\mathcal{G}(p^2)$ while
$1 / \mathcal{F}(0) > 0$ yields a free-like behavior for $\mathcal{G}(p^2)$ at small momenta. As
stressed in Ref.\ \cite{Boucaud:2008ky} (see also the discussion in Section 4.2.2 of Ref.\ \cite{Boucaud:2011ug}),
the scaling condition $1 / \mathcal{F}(0) = 0$ relies on a particular cancellation in the ghost DSE which,
in turn, implies a specific ``critical'' value $g^2_c$ for the coupling constant $g^2$ \cite{Boucaud:2011ug,
Bloch:2003yu}. Thus, at least from the mathematical point of view, there is a one-parameter family of
solutions for the gluon and ghost DSEs, labelled by $g^2$ or, equivalently, by $1 / \mathcal{F}(0)$:
in the case $g^2 = g^2_c$ one recovers the scaling solution while, for all cases $g^2 < g^2_c$, the
solution is a massive one.\footnote{If $g^2 > g^2_c$ one gets a negative ghost propagator
at small momenta.} In $4d$, the SU(3) physical value of the coupling seems to select\footnote{Recently,
in Ref.\ \cite{Weber:2011nw}, it has also been shown --- using a renormalization-group approach --- that
in three and four space-time dimensions only the decoupling solution is expected to be physically realized.}
the massive solution \cite{Boucaud:2008ji,arXiv:1103.0904}.

\vskip 3mm
At this point we should recall that, when considering gauge-dependent quantities in non-Abelian gauge theories,
one has to deal with the existence of Gribov copies \cite{Gribov:1977wm} (see also Ref.\ \cite{Vandersickel:2012tz}
for a recent review). Indeed, for compact non-Abelian Lie groups defined on the 4-sphere \cite{Singer:1978dk}
or on the 4-torus \cite{Killingback:1984en}, it is impossible
to find a continuous choice of one (and only one) connection $A_{\mu}(x)$ on each gauge orbit. The effect of
Gribov copies is not seen in perturbation theory \cite{Gribov:1977wm}, i.e.\ the usual Faddeev-Popov-quantization
procedure is correct at the perturbative level. However, these copies could be relevant at the non-perturbative
level, i.e.\ in studies of the IR properties of Yang-Mills theories.

Different approaches have been proposed in order to quantize a Yang-Mills theory while taking into account the
existence of Gribov copies (see for example \cite{Gribov:1977wm,Hirschfeld:1978yq,Zwanziger:1981kg,ROME-750-1990,
Dell'Antonio:1991xt,Zwanziger:1993dh,Friedberg:1995ty,Baulieu:2000fr}). The one usually considered, both in the
continuum and on the lattice, is based on restricting the functional integration to a subspace of the hyperplane
of transverse configurations. The original proposal, made by Gribov \cite{Gribov:1977wm}, was based on the
observation that the Landau gauge condition $\partial_{\mu} A_{\mu}^b(x)=0$ allows for (infinitesimally) gauge-equivalent
configurations if the Landau-gauge Faddeev-Popov operator $\mathcal{M}^{bc}(x,y) = - \delta(x-y) \partial_{\mu}
D_\mu^{bc}$ has zero modes. (Here $D_{\mu}^{bc}$ is the usual covariant derivative.) Indeed, since an infinitesimal
gauge transformation $\delta \omega(x)$ gives $A_{\mu}^b(x) \to A_{\mu}^b(x) + D_{\mu}^{bc} \omega^c(x)$, it is clear
that the exclusion (in the path-integral measure) of the zero modes of $\mathcal{M}^{bc}(x,y)$ implies that gauge
copies connected by such infinitesimal gauge transformations are ignored in the computation of expectation values.
In order to exclude these zero modes, Gribov considered a stronger condition by requiring that the functional
integration be restricted to the region $\Omega$ of gauge configurations $A^b_{\mu}(x)$ defined as
\begin{equation}\label{defgribovregion}
\Omega \, \equiv \, \left\{ \, A^b_{\mu}(x) : \; \partial_{\mu} A^b_{\mu}(x)=0 \; , \;
                                       \mathcal{M}^{bc}(x,y) > 0 \right\} \; .
\end{equation}
This set, known as the (first) Gribov region, clearly includes the vacuum configuration $A^b_{\mu}(x)=0$, for which
the Faddeev-Popov operator is given by $- \delta(x-y) \delta^{bc} \partial_{\mu}^2$. The region $\Omega$ can also be
defined (see for example \cite{Zwanziger:1982na,Dell'Antonio:1989jn}) as the whole set of local minima\footnote{For
this reason this gauge condition is often indicated in the literature as {\em minimal Landau gauge}.} of the
functional $\mathcal{E}[A] \, = \, \int \, \d^d x \, A_{\mu}^b(x) \, A_{\mu}^b(x)$. Since usually each orbit allows
for more than one local minimum of $\mathcal{E}[A]$, it is clear that the region $\Omega$ is not free of Gribov
copies. On the contrary, in the interior of the so-called fundamental modular region $\Lambda$, given by the set
of the absolute minima of the functional $\mathcal{E}[A]$, no Gribov copies occur \cite{Dell'Antonio:1991xt,
vanBaal:1991zw}.

The characterization of the fundamental modular region $\Lambda$, i.e.\ finding the absolute minima of the {\em energy}
functional $\mathcal{E}[A]$, is a problem similar to the determination of the ground state of a spin glass system
\cite{Marinari:1991zv,Parisi:1994fj}. Thus, a local formulation of a Yang-Mills theory, with the functional
measure delimited to $\Lambda$, is not available, whereas a practical way of restricting the physical configuration
space to the region $\Omega$ was introduced by Gribov \cite{Gribov:1977wm}. To this end, he required that the ghost
dressing function $\mathcal{F}(p^2)$ cannot have a pole at finite nonzero momenta. After setting
\begin{equation}
\mathcal{G}(p^2) \, = \, \frac{\mathcal{F}(p^2)}{p^2} \, = \, \frac{1}{p^2} \frac{1}{1- \sigma(p^2)} \; ,
\label{eq:Gsigma}
\end{equation}
this condition can be written as
\begin{equation}
\sigma(p^2) < 1 \qquad \mbox{for} \qquad p^2 > 0 \; ,
\label{eq:nopole}
\end{equation}
where $\sigma(p^2)$ is the so-called Gribov ghost form-factor \cite{Gribov:1977wm}. Indeed, since the ghost
propagator is given by
\begin{equation}
\mathcal{G}(p^2) \, = \, \frac{\delta^{bc}}{N^2_c - 1} \, \left< \, p \,
            \left| \, \left(\mathcal{M}^{-1}\right)^{bc} \, \right| \, p \, \right> \; ,
\end{equation}
i.e.\ it is related to the inverse of the Faddeev-Popov matrix $\mathcal{M}^{bc}(x,y)$, the above
inequality --- known as the no-pole condition --- should be equivalent to the restriction of the
functional integration to the Gribov region $\Omega$, i.e.\ to the condition $\mathcal{M}^{bc}(x,y) > 0$.

From the discussion above, it is clear that both scaling and massive solutions of DSEs satisfy the
no-pole condition, i.e.\ $1 / \mathcal{F}(p^2) = 1- \sigma(p^2) > 0$ for $p^2 > 0$. Indeed, in the scaling
case \cite{Alkofer:2000wg,von Smekal:1997vx,Zwanziger:2001kw,Lerche:2002ep,Huber:2007kc}, this condition
[together with the condition $1 / \mathcal{F}(0) = 0$] is imposed from the beginning to the solution
of the DSEs. On the contrary, for the massive solution, the no-pole condition is either verified
a posteriori, as in Ref.\ \cite{Aguilar:2008xm}, or used [together with the condition
$1 / \mathcal{F}(0) > 0$] as an input for the solution of the DSEs, as in Refs.\ \cite{Boucaud:2006if,
Pennington:2011xs,Aguilar:2011xe}. In particular, in Ref.\ \cite{Aguilar:2011xe}, the value of
$1/\mathcal{F}(0)$ is fixed using lattice data.

\vskip 3mm
The restriction to the first Gribov region $\Omega$ is also always implemented in lattice numerical
simulations of Green functions in Landau gauge by (numerically) finding local minima of the functional
$\mathcal{E}[A]$. Results obtained using very large lattice volumes \cite{Bogolubsky:2007ud,Cucchieri:2007md,
Sternbeck:2007ug,Bogolubsky:2009dc} (see also Chapter 10 in Ref.\ \cite{Greensite}, Section 3 in Ref.\
\cite{Boucaud:2011ug} and Ref.\ \cite{Cucchieri:2010xr} for recent short reviews) have shown that in
$d=3$ and 4 the gluon propagator $\mathcal{D}(p^2)$ is finite and nonzero in the limit $p \to 0$
while the ghost propagator $\mathcal{G}(p^2)$ behaves as $1/p^2$. On the contrary, for $d=2$ the lattice
data \cite{Maas:2007uv,Cucchieri:2007rg,arXiv:1101.4779,Cucchieri:2011ig} are in quantitative agreement
with the scaling solution and one finds $\kappa_D = \kappa_G \approx 0.2$.

Since the region $\Omega$ is not free of Gribov copies, their (possible) influence on the numerical
evaluation of gluon and ghost propagators has been studied by various groups \cite{Cucchieri:1997dx,
Silva:2004bv,Bogolubsky:2005wf,Bornyakov:2008yx,Maas:2008ri,arXiv:1112.4975}. It has been found that these
effects are usually observable only the IR limit and that any attempt to restrict the functional integration
to the fundamental modular region $\Lambda$ gives a stronger suppression at small momenta for both
propagators, i.e.\ reducing the value of $\mathcal{D}(0)$ and increasing that of $1 / \mathcal{F}(0)$.
More recently, it has been suggested \cite{Maas:2009se,Maas:2009ph,Maas:2010wb,Maas:2011ba} that the
one-parameter family of solutions obtained for the gluon and ghost DSEs should be related\footnote{This
identification is, however, based on several (unproven) hypotheses, as already stressed in Ref.\
\cite{arXiv:1101.4779}.} to Gribov-copy effects and that the value of $1 / \mathcal{F}(0)$ could be used
as a gauge-fixing parameter. This analysis finds indeed IR-enhanced ghost propagators (and sometimes a
disconcerting over-scaling\footnote{Let us recall that the scaling solution is supposed to be unique
\cite{Fischer:2008uz}.}). On the other hand, the gluon propagator still shows a finite nonzero value at
zero momentum, that is, $\mathcal{D}(0) > 0$. Moreover, this approach does not explain why the
numerical results found in $d=2$ are different from those obtained in $d=3$ and 4, even though
Gribov copies inside the first Gribov region $\Omega$ are clearly present in any space-time dimension
$d > 1$.

\vskip 3mm
From the analytical point of view, following Gribov's approach, Zwanziger modified the usual Yang-Mills
action in order to restrict the path integral to the first Gribov region $\Omega$ \cite{Zwanziger:1989mf}
Although this restriction is obtained using a non-local term, the Gribov-Zwanziger (GZ) action\footnote{See
again Ref.\ \cite{Vandersickel:2012tz} for a comprehensive review of the GZ action.} can be
written as a local action and it is proven \cite{Zwanziger:1992qr,Maggiore:1993wq,Dudal:2010fq} to be
renormalizable. At tree level the GZ gluon propagator is given by $\mathcal{D}(p^2) = p^2 / (p^4 + \lambda^4) $,
where $\lambda$ is a parameter with mass-dimension 1. At the same time, the ghost propagator is given by
$ \mathcal G(p^2) \sim 1/p^4$. Thus, as in the scaling solution of the gluon and ghost DSEs, the gluon
propagator is null at zero momentum\footnote{One can, however, obtain a finite nonzero value for
$\mathcal D(0)$ within the GZ approach by considering a non-analytic behavior for the free energy of
the system \cite{arXiv:1012.2859,arXiv:1103.1137}.} and the ghost propagator is IR-enhanced
\cite{Zwanziger:1993dh}. These tree-level results, also in agreement with the original work by Gribov
\cite{Gribov:1977wm}, have been confirmed by one-loop calculations in the three- and four-dimensional
cases \cite{Gracey:2005cx,Gracey:2006dr,Gracey:2007vv,Ford:2009ar,Gracey:2010df}.

More recently, the GZ action has been modified by considering (for $d=3$ and 4) dimension-two condensates
\cite{Dudal:2008sp,Dudal:2008rm,Vandersickel:2011zc,Dudal:2011gd}. The corresponding action, called the
Refined Gribov-Zwanziger (RGZ) action, still imposes the restriction of the functional integration
to the region $\Omega$ and it is renormalizable. However, the RGZ action allows for a finite nonzero
value of $\mathcal{D}(0)$ and for a free-like ghost propagator $\mathcal G(p^2)$ in the IR limit. Thus,
nonzero values for these dimension-two condensates yield for the gluon and ghost propagators an IR
behavior in agreement with the massive solution of the gluon and ghost DSEs.\footnote{Let us note here
that a massive behavior for these two propagators has also been obtained in Refs.\ \cite{Frasca:2007uz,
Kondo:2009ug} using different analytic approaches.} Indeed, the RGZ tree-level gluon propagator
describes well the numerical data in the SU(2) case \cite{Cucchieri:2011ig,Cucchieri:2012gb}, for $d=3$ and
4, and in the SU(3) case \cite{{Dudal:2010tf}} with $d=4$. It is also interesting to note that the fitting
values for the dimension-two condensates are very similar for the SU(2) and SU(3) gauge groups in the
four-dimensional case.

\vskip 3mm
As stressed above, the restriction of the functional integration to the first Gribov region $\Omega$ and
the no-pole condition (\ref{eq:nopole}) are key ingredients in the study of the IR sector of Yang-Mills
theories in Landau gauge. However, to our knowledge, a detailed investigation of the properties of
the Gribov form-factor $\sigma(p^2)$ as well as of the possible implications of the no-pole condition was
missing up to now, although some interesting one-loop results were already presented in Refs.\
\cite{Vandersickel:2011zc,Dudal:2008xd,Sobreiro:2005ec,Gomez:2009tj}. In particular, in Appendix B.2 of
\cite{Vandersickel:2011zc} it was shown that, if the gluon propagator $\mathcal{D}(p^2)$ is
positive, then in the $2d$ case the derivative $\partial \sigma(p^2) / \partial p^2$ is negative for
all values of $p^2$, i.e.\ $\sigma(p^2)$ is largest at $p^2 = 0$. Also, in Ref.\ \cite{Dudal:2008xd} it
was proven that in the RGZ framework the form-factor $\sigma(p^2)$ presents a logarithmic IR singularity
$- \ln(p^2)$ for $d=2$. This result precluded the use of the RGZ action in the two-dimensional
case, leading to a first interpretation of the different behavior found in lattice numerical
simulations for the $2d$ case, compared to the $d=3$ and 4 cases. Similar findings have
been (more recently) presented in Refs.\ \cite{Weber:2011nw,Tissier:2010ts,Tissier:2011ey}.

In this work we collect some general properties of the Gribov form-factor $\sigma(p^2)$ and we
study the consequences of imposing the no-pole condition.
In particular, in Section \ref{sec:ghost-prop}, using the expression for $\sigma(p^2)$ obtained from
the evaluation of the ghost propagator at one loop, we prove that $\sigma(p^2)$ attains its maximum
value at $p^2 = 0$ for any dimension $d \geq 2$. Since this expression for $\sigma(p^2)$ depends on the
gluon propagator $\mathcal{D}(p^2)$, in the same section we also investigate (for a general $d$-dimensional
space-time) which IR behavior of the gluon propagator is necessary in order to satisfy the no-pole condition
$\sigma(p^2) < 1$. By considering a generic (and sufficiently regular) gluon propagator $\mathcal{D}(p^2)$,
we find in the $d=2$ case that $\sigma(p^2)$ is unbounded unless the gluon propagator is null at zero
momentum. More exactly, we find $\sigma(p^2) \sim - \mathcal{D}(0) \ln(p^2)$ in the $p \to 0$ limit, in
agreement with \cite{Dudal:2008xd}.  This result does not apply to the $d=3$ and 4 cases. Indeed, in
these cases one can introduce, for all values of $p^2$, simple finite upper bounds for the Gribov
form-factor. In Section \ref{sec:sigma234} we present explicit one-loop calculations for $\sigma(p^2)$
using for the gluon propagator $\mathcal{D}(p^2)$ linear combinations of Yukawa-like propagators
(with real and/or complex-conjugate poles), which have been recently used to model lattice data
of the gluon propagator in the SU(2) case \cite{Cucchieri:2011ig,Cucchieri:2012gb}. Besides confirming the
results obtained in Section \ref{sec:ghost-prop}, we also find that the ghost propagator admits a
one-parameter family of behaviors \cite{Boucaud:2008ji} labelled by the coupling constant $g^2$,
considered as a free parameter. Moreover, the massive solution $\mathcal{G}(p^2) \sim 1/p^2$,
corresponding to $\sigma(0) < 1$, is obtained for all values of $g^2$ smaller than a ``critical''
value $g^2_c$. At the ``critical'' value $g^2_c$, implying $\sigma(0) = 1$, one finds an
IR-enhanced ghost propagator. (As already stressed above, the case $g^2 > g^2_c$ corresponds to
$\sigma(0) > 1$ and one obtains a negative ghost propagator at small momenta.)
Finally, in Section \ref{sec:DSE}, we analyze the DSE for $\sigma(p^2)$. We stress that in this
case we do not try to solve the DSE but we focus only on general properties of this equation.
As we will see, considering IR-finite ghost-gluon vertices, we confirm and extend the one-loop
analysis of the no-pole condition presented in Section \ref{sec:ghost-prop}. In particular, after
introducing bounds for the Gribov form-factor, we show again for $d=2$ that the gluon propagator
$\mathcal{D}(p^2)$ must vanish at zero momentum in order to keep $\sigma(p^2)$ finite. On the
contrary, such a constraint does not apply in the three- and four-dimensional cases. We also present
alternative evidence for the $d=2$ result using a spectral representation for the ghost propagator
in the DSE.

It is important to note that all our results in Sections \ref{sec:ghost-prop} and \ref{sec:DSE}
apply irrespective of which set of Gribov copies (inside the region $\Omega$) is considered, i.e.\
they are not affected by the so-called Gribov noise. We end with our Conclusion in Section \ref{sec:concl}.
Some technical details have been collected in four Appendices. In particular, in Appendix \ref{sec:hyper}
we present properties of the Gauss hypergeometric function $_2F_1\left(a,b;c;z\right)$ that are
relevant to prove some of our results.

%%%%%%%%%%%%%%%%%%%%%%%%%%%%%%%%%%%%%%%%%%%%%%%%%%%%%%%%%%%%%%%%%%%%%%%%%%%%%%%%%%%%%%%%%%%%%%%%%%%%%%%%%%%%%%%

\begin{figure}
   \centering
       \includegraphics[width=8cm]{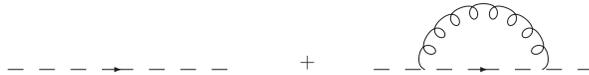}
   \caption{Feynman diagrams for the one-loop-corrected Landau-gauge ghost propagator. Dashed
            lines represent ghosts, the curly line represents gluons. \label{figghost}}
\end{figure}

\section{The One-Loop-Corrected Ghost Propagator and the Gribov Form-Factor}
\label{sec:ghost-prop}

In this Section, as well as in Section \ref{sec:sigma234} below, we consider the one-loop-corrected
Landau-gauge ghost propagator, diagrammatically represented in Figure \ref{figghost}. This
propagator can be written [for the SU($N_c$) gauge group in the $d$-dimensional case] as
\begin{equation}
\mathcal{G}(p^2) \, = \, \frac{1}{p^2} \, - \, \frac{\delta^{ab}}{N_c^2 - 1} \,
                         \frac{1}{p^4} \, g^2 f^{adc} f^{cdb} \int \frac{\d^d q}{(2\pi)^d}
                         \, (p - q)_{\mu} \, p_{\nu} \, \mathcal{D}(q^2) \, P_{\mu \nu}(q)
                         \, \frac{1}{(p - q)^2} \; ,
\label{eq:Gini}
\end{equation}
where $ \delta^{ab} \, \mathcal{D}(q^2) \, \mathcal{P}_{\mu\nu}(q) $ is the tree-level gluon
propagator [not necessarily given by $\mathcal{D}(q)=1/q^2$] and $\mathcal{P}_{\mu\nu}(q) =
\left(\delta_{\mu\nu}-q_\mu q_\nu/q^2\right)$ is the usual projector onto the transverse
sub-space, i.e.\ $q_{\mu} \mathcal{P}_{\mu\nu}(q) = 0$. We have also considered the tree-level
ghost-gluon vertex $\, i g f^{adc} p_{\nu} $, where $p$ is the outgoing ghost momentum. The
indices $a, d, c$ refer, respectively, to the incoming ghost, to the gluon and to the outcoming
ghost. After using $f^{adc} \, f^{cdb} = - N_c \delta^{ab}$, valid for the adjoint representation,
we obtain
\begin{equation}
\mathcal{G}(p^2) \, = \, \frac{1}{p^2} \, \left[ \, 1 \, + \, \sigma(p^2) \, \right] \; ,
\end{equation}
where $\sigma(p^2)$ is the momentum-dependent function
\begin{equation}\label{sigma1}
\sigma(p^2) \, = \, g^2 N_c \frac{p_{\mu} p_{\nu}}{p^2} \int \frac{\d^d q}{(2\pi)^d}
                   \frac{1}{(p-q)^2} \mathcal{D}(q^2) \, \mathcal{P}_{\mu\nu}(q) \; .
\end{equation}
Finally, we can write [as in Eq.\ (\ref{eq:Gsigma})]
\begin{equation}
\mathcal{G}(p^2) \, = \, \frac{1}{p^2} \frac{1}{1- \sigma(p^2)} \; ,
\label{eq:finalG}
\end{equation}
which corresponds to the usual resummation of an infinite set of diagrams into the self-energy.
Note that this resummation makes sense only when $\sigma(k^2) < 1$, i.e.\ when the no-pole
condition (\ref{eq:nopole}) is satisfied.

Clearly the function $\sigma(p^2)$ is dimensionless and it should go to zero for $p \to \infty$,
modulo possible logarithmic corrections. Also, this function coincides with the so-called Gribov
ghost form-factor \cite{Gribov:1977wm,Sobreiro:2005ec,Gomez:2009tj}, even though the latter is
obtained in a slightly different way.\footnote{Note that in the Gribov ghost form-factor there is
usually an extra factor $1/(N_c^2-1)$ \cite{Sobreiro:2005ec,Gomez:2009tj} compared to our Eq.\
(\ref{sigma1}). However, this is due to the fact that in Eq.\ (\ref{eq:Gini}) above we considered
for the Landau-gauge gluon propagator the usual expression $D^{ab}_{\mu\nu}(q^2) = \delta^{ab} \,
\mathcal{P}_{\mu\nu}(q) \, \mathcal{D}(q^2)$ while, in the derivation of the Gribov ghost form-factor,
one usually writes $A^{a}_{\mu}(q) A^{a}_{\nu}(-q) = \omega(q^2) \mathcal{P}_{\mu\nu}(q)$,
as in Eq.\ (255) of Ref.\ \cite{Sobreiro:2005ec}.} As discussed in the Introduction, the no-pole
condition $\sigma(p^2) < 1$ for $p^2 > 0$ should be equivalent to the restriction of the path
integral to the first Gribov region $\Omega$ [defined in Eq.\ (\ref{defgribovregion})]. In this section
we will derive general properties of $\sigma(p^2)$ in $d \geq 2$ space-time dimensions. In
particular, as we will see below, the weaker condition $\sigma(p^2) < + \infty$ is already sufficient
to obtain a strong constraint on the IR behavior of the gluon propagator $\mathcal{D}(p^2)$ in the
$d=2$ case.

%%%%%%%%%%%%%%%%%%%%%%%%%%%%%%%%%%%%%%%%%%%%%%%%%%%%%%%%%%%%%%%%%%%%%%%%%%%%%%%%%%%%%%%%%%%%%%%%%%%%%%%%%%%%%%%

\subsection{First Derivative of $\sigma(p^2)$ for $d=2$}
\label{sec:2Dderivative}

Following Appendix B.2 of Ref.\ \cite{Vandersickel:2011zc} one can show that, if the gluon
propagator $\mathcal{D}(q^2)$ is positive, then in the $2d$ case and for all values of $p^2$
we have
\begin{equation}
  \frac{\partial \sigma(p^2)}{\partial p^2} \, < \, 0 \; ,
\end{equation}
with $\sigma(p^2)$ defined in Eq.\ (\ref{sigma1}). This implies that $\sigma(p^2)$ attains
its maximum value at $p^2 = 0$. To this end, we choose the positive $x$ direction parallel
to the external momentum $p$ and write (using polar coordinates)
\begin{eqnarray}
\frac{\sigma(p^2)}{g^2 N_c} & = & \frac{p_{\mu} p_{\nu}}{p^2} \int \frac{\d^2 q}{(2\pi)^2}
                                \frac{1}{(p-q)^2} \mathcal{D}(q^2) \, \mathcal{P}_{\mu\nu}(q)
                          \, = \, \int_0^{\infty} \frac{q \, \d q}{4 \pi^2} \, \mathcal{D}(q^2)
                          \, \int_0^{2 \pi} \d\theta
                          \, \frac{1 - \cos^2(\theta)}{p^2 + q^2 - 2 \, q \, p \, \cos(\theta)} \; .
\end{eqnarray}
The integral in $\d\theta$ can be  evaluated using contour integration on the unit circle and the
residue theorem. This yields
\begin{eqnarray}\label{2}
\int_0^{2 \pi} \d\theta \, \frac{1 - \cos^2(\theta)}{p^2 + q^2 - 2 \, q \, p \, \cos(\theta)}
\, = \, \oint \, \frac{i \, \d z}{4 p q} \, \frac{2 - z^2 - \overline{z}^2}{z^2 - z
        \left(p/q \, + \, q/p\right) + 1}
\, = \, \frac{\pi}{p^2} \, \theta(p^2-q^2)  \, + \, \frac{\pi}{q^2} \, \theta(q^2-p^2) \; ,
\end{eqnarray}
where $\oint \d z$ represents the integral on the unit circle $|z|=1$, we indicated with $\overline{z}$
the complex-conjugate of $z = e^{i\theta}$ and $\theta(x)$ is the step function. This integral is
also evaluated in Eqs.\ (\ref{eq:2F1-uno}) and (\ref{eq:2F1-due}) in the Appendix \ref{app:hsc} (for
the general $d$-dimensional case). Considering also Eq.\ (\ref{eq:2F1-D2}) and Eq.\ (\ref{eq:omegaD}),
with $d=2$, we find
\begin{eqnarray}
\frac{\sigma(p^2)}{g^2 N_c} & = & \frac{1}{4 \pi} \left[ \, \int_0^p \frac{\d q}{p^2} \, q \,
                                \mathcal{D}(q^2) \, + \, \int_p^{\infty} \frac{\d q}{q} \,
                                \mathcal{D}(q^2) \, \right]  \label{eq:sigma2D} \\[2mm]
                          & = & \frac{1}{8 \pi} \left[ \, \int_0^{p^2} \frac{\d q^2}{p^2}
                               \, \mathcal{D}(q^2) + \, \int_{p^2}^{\infty} \frac{\d q^2}{q^2}
                               \, \mathcal{D}(q^2) \, \right] \\[2mm]
            & = & \frac{1}{8 \pi} \int_0^{\infty} \, \d q^2 \, \mathcal{D}(q^2) \, \left[ \,
                  \frac{\theta(p^2 - q^2)}{p^2} \, + \, \frac{\theta(q^2 - p^2)}{q^2}
                  \, \right] \; .
\label{eq:D2res}
\end{eqnarray}
Then, by using $\partial_{x} \theta(x) = \delta(x)$, where $\delta(x)$ is the Dirac delta
function, the derivative of $\sigma(p^2)$ with respect to $p^2$ yields
\begin{equation}
 \frac{\partial \sigma(p^2)}{\partial p^2} \, = \, - \, \frac{g^2 N_c}{8 \pi} \int_0^{\infty}
                                 \, \d q^2 \, \mathcal{D}(q^2) \, \frac{\theta(p^2 - q^2)}{p^4}
       \, = \, - \, \frac{g^2 N_c}{8 \pi p^4} \int_0^{p^2} \, \d q^2 \, \mathcal{D}(q^2) \; ,
\label{eq:sigmader2}
\end{equation}
which is clearly negative, for any value of $p^2$, if $\mathcal{D}(q^2)$ is positive. We can evaluate
the limit $p^2 \to 0$ of this derivative using, for example, the trapezoidal rule\footnote{The
trapezoidal rule gives the numerical approximation
\begin{equation}
\int_a^b \,\d x \, f(x) \, = \, \frac{b-a}{2} \, \left[ \, f(b) + f(a) \, \right]
                           \, + \, {\cal O}(b-a)^3 \; ,
\end{equation}
which can be obtained by integrating $f(x) \approx f(a) \, + \, (x-a) \left[ \, f(b) - f(a)
\, \right] / (b-a)$. Thus, the trapezoidal rule is equivalent to using a linear Taylor expansion
$f(a) + (x - a) f'(a)$ for $f(x)$ with the first derivative $f'(a)$ approximated by a (first forward)
finite difference $\left[ \, f(b) - f(a)\, \right] / (b-a)$.}
\begin{equation}
\lim_{p^2 \to 0} \, \frac{\partial \sigma(p^2)}{\partial p^2}
   \, = \, - \, \lim_{p^2 \to 0} \, \frac{g^2 N_c}{8 \pi p^4} \, \frac{p^2}{2} \,
                \left[ \, \mathcal{D}(p^2) \, + \, \mathcal{D}(0) \, \right]
\, = \, - \, \lim_{p^2 \to 0} \, \frac{g^2 N_c \mathcal{D}(0)}{8 \pi p^2} \; .
\label{eq:lim-deriv}
\end{equation}

One arrives at the same result after writing Eq.\ (\ref{eq:sigmader2}) as
\begin{equation}
\frac{\partial \sigma(p^2)}{\partial p^2}
   \, = \, - \, \frac{g^2 N_c}{8 \pi p^2} \int_0^{1} \, \d x \; \mathcal{D}(x p^2)
   \, = \, - \, \frac{g^2 N_c}{8 \pi p^2} \, \frac{\hat{D}(p^2) \, - \, \hat{D}(0)}{p^2} \; ,
\end{equation}
where $\hat{D}(p^2)$ is a primitive of $\mathcal{D}(p^2)$, that is, $\hat{D}'(p^2) = \mathcal{D}(p^2)$
where we indicate with $'$ the first derivative with respect to the variable $p^2$. The limit
$p^2 \to 0$ then yields again Eq.\ (\ref{eq:lim-deriv}).

Clearly, one finds an IR singularity at $p^2=0$, unless $\mathcal{D}(0)=0$. If this condition is satisfied,
using again the trapezoidal rule, we have from Eq.\ (\ref{eq:lim-deriv}) that
\begin{eqnarray}
\lim_{p^2 \to 0} \, \frac{\partial \sigma(p^2)}{\partial p^2}
  & = & - \, \lim_{p^2 \to 0} \, \frac{g^2 N_c}{8 \pi p^4} \, \frac{p^2}{2} \, \mathcal{D}(p^2)
 \, = \, - \, \lim_{p^2 \to 0} \, \frac{g^2 N_c}{16 \pi} \, \frac{\mathcal{D}(p^2) - \mathcal{D}(0)}{p^2}
             = - \, \frac{g^2 N_c}{16 \pi} \, \lim_{p^2 \to 0} \, \mathcal{D}'(p^2) \; .
\end{eqnarray}
For a gluon propagator $\mathcal{D}(p^2)$ that is regular at small momenta, i.e.\ that can be
expanded as $\mathcal{D}(p^2) \approx \mathcal{D}'(0) \, p^2 + \mathcal{D}''(0) p^4 / 2$ at small
$p^2$, the above limit is finite. On the other hand, if the leading IR behavior of $\mathcal{D}(p^2)$
is proportional to $p^{2 \eta}$ with $1 > \eta > 0$, as found for example in the $2d$ case in Refs.\
\cite{Zwanziger:2001kw,Huber:2007kc,Maas:2007uv,Cucchieri:2011ig}, then the above limit gives a singular
value, due to the non-integer-power (and non-analytic) behavior of $\mathcal{D}(p^2)$.

%%%%%%%%%%%%%%%%%%%%%%%%%%%%%%%%%%%%%%%%%%%%%%%%%%%%%%%%%%%%%%%%%%%%%%%%%%%%%%%%%%%%%%%%%%%%%%%%%%%%%%%%%%%%%%%

\subsection{Infrared Singularity of $\sigma(p^2)$ for $d=2$}
\label{sec:d=2sing}

Here we prove that --- for $d=2$ and for any gluon propagator $\mathcal{D}(p^2)$ that goes to zero
sufficiently fast at large momenta, e.g.\ as $1/p^2$, and that is reasonably regular at small
momenta, e.g.\ that can be expanded at $p=0$ as $\mathcal{D}(p^2) \approx \mathcal{D}(0) +
B \, p^{2 \eta} + C p^{2 \xi}$ (with $\xi > \eta > 0$ and $\mathcal{D}(0), B$ and $C$
finite)\footnote{For example, in Eq.\ (\ref{eq:D2D}) below, the Taylor expansion of
$\mathcal{D}(p^2)$ at $p^2=0$ is of the type considered here with $\xi = 1$.} --- the ghost
form-factor (\ref{sigma1}) displays a logarithmic divergence for $p \to 0$ proportional to
$\mathcal{D}(0)$. Indeed, by considering Eq.\ (\ref{eq:sigma2D}), one obtains
\begin{eqnarray}
\frac{\sigma(p^2)}{g^2 N_c}
     & = & \frac{1}{4 \pi} \left[ \, \int_0^p \frac{\d q}{p^2} \, q \, \mathcal{D}(q^2)
            \, + \, \int_p^{\infty} \frac{\d q}{q} \, \mathcal{D}(q^2) \, \right] \\[2mm]
     & = & \frac{1}{8 \pi} \lim_{\Lambda \to \infty} \left\{ \, \int_0^{p^2}
            \frac{\d x}{p^2} \, \mathcal{D}(x) \, + \, 2 \, \int_p^{\Lambda} \frac{\d q}{q}
             \, \mathcal{D}(0) \, + \, 2 \, \int_p^{\Lambda} \frac{\d q}{q} \,
              \left[ \mathcal{D}(q^2) \, - \, \mathcal{D}(0) \right] \, \right\} \\[2mm]
     & = & \frac{1}{8 \pi} \lim_{\Lambda \to \infty} \left\{ \, \int_0^{p^2}
            \frac{\d x}{p^2} \, \mathcal{D}(x) \, + \, \mathcal{D}(0) \,
             \ln\left( \frac{\Lambda^2}{p^2} \right) + \, 2 \, \int_p^{\Lambda} \frac{\d q}{q}
              \, \frac{- q^{2 \eta}}{q^{2 \eta} + M} \, \left[ \frac{\mathcal{D}(q^2) \, - \,
              \mathcal{D}(0)}{- q^{2 \eta}} \, \left(q^{2 \eta} + M\right) \right] \, \right\} \\[2mm]
     & = & \frac{1}{8 \pi} \lim_{\Lambda \to \infty} \left\{ \, \frac{\hat{D}(p^2)
            - \hat{D}(0)}{p^2} \, + \, \mathcal{D}(0) \, \ln\left( \frac{\Lambda^2}{p^2} \right)
            \, - \, \int_{p^2}^{\Lambda^2} \d x \, \frac{x^{\eta-1}}{x^{\eta} + M} \, \left[
            \frac{\mathcal{D}(x) \, - \, \mathcal{D}(0)}{-x^{\eta}} \, \left(x^{\eta} + M\right)
                  \right] \, \right\} \; ,
\end{eqnarray}
where $x = q^2$, $\hat{D}(x)$ is again a primitive of $\mathcal{D}(x)$ and $M > 0$ is a (finite)
constant. If we indicate with $H(x)$ the quantity in square brackets in the last line, then we have
\begin{eqnarray}
\frac{\sigma(p^2)}{g^2 N_c}
        & = & \frac{1}{8 \pi} \lim_{\Lambda \to \infty} \Biggl\{ \, \frac{\hat{D}(p^2) -
              \hat{D}(0)}{p^2} \, + \, \mathcal{D}(0) \, \ln\left( \frac{\Lambda^2}{p^2} \right)
            - \, \frac{1}{\eta} \, H(x) \, \ln\left( x^{\eta} + M \right) \Bigr|_{p^2}^{\Lambda^2}
                         \nonumber \\[2mm]
        &   & \qquad \qquad \qquad \qquad \;
                  + \, \frac{1}{\eta} \, \int_{p^2}^{\Lambda^2} \, \d x \,
                        \ln\left( x^{\eta} + M \right) \, H'(x) \, \Biggr\} \; .
\label{eq:sigmaint}
\end{eqnarray}
Note that for $\mathcal{D}(x) = 1 / (x^{\eta} + M)$ we find $H(x) = 1/M = \mathcal{D}(0)$ and
the last term in Eq.\ (\ref{eq:sigmaint}) is zero. Since $\lim_{x \to \infty} \mathcal{D}(x) = 0$
we also have that $\lim_{x \to \infty} H(x) = \mathcal{D}(0)$ and the two logarithmic singularities
for infinite $\Lambda$ cancel each other. Thus, we get
\begin{eqnarray}
\frac{\sigma(p^2)}{g^2 N_c}
        & = & \frac{1}{8 \pi} \Biggl\{ \, \frac{\hat{D}(p^2) - \hat{D}(0)}{p^2}
              \, - \, \mathcal{D}(0) \, \ln\left( p^2 \right) \, + \, \frac{1}{\eta}
                    \, H(p^2) \, \ln\left( p^{2 \eta} + M \right) \nonumber \\[2mm]
        & & \qquad \qquad \; + \, \frac{1}{\eta} \, \int_{p^2}^{\infty} \, \d x \,
                 \ln\left( x^{\eta} + M \right) \,
                     \frac{\eta M \left[ \mathcal{D}(x) \, - \, \mathcal{D}(0) \right] \, - \, x \,
                      \left( x^{\eta} + M \right) \, \mathcal{D}'(x)}{x^{1+\eta}} \, \Biggr\} \; .
\label{eq:sigmafin}
\end{eqnarray}
If $\mathcal{D}(x) \sim 1/x$ at large $x$, it is easy to check\footnote{See details in Appendix
\ref{app:relax}.} that $\sigma(p^2)$ is null for $p^2 \to \infty$, as expected. At the same
time, in the limit $p^2 \to 0$ we obtain
\begin{eqnarray}
\frac{\sigma(0)}{g^2 N_c}
    & = & \, \frac{1}{8 \pi} \Biggl\{ \, \mathcal{D}(0) \, - \, \lim_{p^2 \to 0} \,
                \mathcal{D}(0) \, \ln\left( p^2 \right) \, + \, \frac{1}{\eta} \,
                      H(0) \, \ln\left( M \right) \nonumber \\[2mm]
    &   & \quad + \, \int_{0}^{\infty} \, \d x \, \ln\left( x^{\eta} + M \right) \,
                    \frac{\eta M \left[ \mathcal{D}(x) \, - \, \mathcal{D}(0) \right] \, - \,
                    x \, \left( x^{\eta} + M \right) \, \mathcal{D}'(x)}{x^{1+\eta}} \, \Biggr\} \; ,
\label{eq:sigmafin2}
\end{eqnarray}
where we used
\begin{equation}
\lim_{p^2 \to 0} \,\frac{\hat{D}(p^2) - \hat{D}(0)}{p^2} \, = \, \hat{D}'(0) \, = \, \mathcal{D}(0)
\end{equation}
and $H(0) = - M B$ is a finite constant.\footnote{Note that here we used the IR expansion
$\mathcal{D}(p^2) \approx \mathcal{D}(0) + B \, p^{2 \eta} + C p^{2 \xi}$ for the gluon
propagator.} Finally, in Appendix \ref{app:relax} we show that, under the
assumptions made for the gluon propagator,\footnote{In the same Appendix we will also show that the
hypotheses considered above for the gluon propagator $\mathcal{D}(x)$ can be relaxed.} the last term
on the r.h.s.\ of Eq.\ (\ref{eq:sigmafin2}) is finite. Thus, the only IR singularity in the ghost
form-factor $\sigma(p^2)$ is proportional to $- \mathcal{D}(0) \,\ln(p^2)$. This result is in
qualitative agreement with \cite{Dudal:2008xd}. An IR singularity plaguing the $2d$ calculation has
also been recently obtained in Ref.\ \cite{Tissier:2010ts}.

\vskip 3mm
An alternative (equivalent) proof\footnote{Note that both proofs are singling out the singularity
$-\mathcal{D}(0) \, \ln(p^2)$ by essentially doing a Taylor expansion of
$\mathcal{D}(p^2)$ at $p^2=0$.} can be done by performing an integration by parts. Then,
Eq.\ (\ref{eq:sigma2D}) becomes
\begin{eqnarray}
\frac{\sigma(p^2)}{g^2 N_c}
       & = & \frac{1}{8 \pi} \left[ \, \int_0^{p^2} \frac{\d x}{p^2} \, \mathcal{D}(x) \, + \,
              \int_{p^2}^{\infty} \frac{\d x}{x} \, \mathcal{D}(x) \, \right] \label{eq:sigma-1L-ini} \\[2mm]
       & = & \frac{1}{8 \pi} \left[ \, \frac{\hat{D}(p^2) - \hat{D}(0)}{p^2} \, + \,
              \ln\left( x \right) \, \mathcal{D}(x) \, \Bigr|_{p^2}^{\infty}  \, - \,
               \int_{p^2}^{\infty} \d x \, \ln\left( x \right) \, \mathcal{D}'(x) \, \right] \\[2mm]
       & = & \frac{1}{8 \pi} \left[ \, \frac{\hat{D}(p^2) - \hat{D}(0)}{p^2} \, - \,
              \ln\left( p^2 \right) \, \mathcal{D}(p^2) \, - \, \int_{p^2}^{\infty} \d x \,
               \ln\left( x \right) \, \mathcal{D}'(x) \, \right] \; ,
\label{form2}
\end{eqnarray}
where we used the assumption $\mathcal{D}(x) \sim 1/x$ at large $x$. Note that the above result
coincides with Eq.\ (\ref{eq:sigmafin}) when $M = 0$, which implies $H(x) = \mathcal{D}(0) -
\mathcal{D}(x)$. A second integration by parts yields
\begin{eqnarray}
\!\!\!\!\!\!\!\!\!\frac{\sigma(p^2)}{g^2 N_c} \!
      & = & \! \frac{1}{8 \pi} \Biggl\{ \, \frac{\hat{D}(p^2) - \hat{D}(0)}{p^2} \, - \,
               \ln\left( p^2 \right) \, \mathcal{D}(p^2)  \, - \, \left[ \, x \,
               \ln\left( x \right) \,-\,  x\, \right] \, \mathcal{D}'(x) \, \Bigr|_{p^2}^{\infty}
               \, + \,  \int_{p^2}^{\infty} \d x \, \left[ \, x \,
               \ln\left( x \right) \,-\,  x\, \right] \, \mathcal{D}''(x) \, \Biggr\} \\[2mm]
      & = & \! \frac{1}{8 \pi} \Biggl\{ \, \frac{\hat{D}(p^2) - \hat{D}(0)}{p^2} \, - \,
               \ln\left( p^2 \right) \, \mathcal{D}(p^2)  \, + \, \left[ \, p^2 \,
               \ln\left( p^2 \right) \,-\,  p^2\, \right] \, \mathcal{D}'(p^2) \, + \,
               \int_{p^2}^{\infty} \d x \, \left[ \, x \, \ln\left( x \right) \,-\,  x\, \right]
                          \, \mathcal{D}''(x) \, \Biggr\} \; .
\end{eqnarray}
Here we used the hypothesis that $\mathcal{D}'(x)$ goes to zero sufficiently fast at large momenta,
e.g.\ as $1/x^2$. As before, one easily sees that $\sigma(p^2)$ is null for $p^2 \to \infty$ (see
Appendix \ref{app:relax}). At the same time, under the assumptions made for the gluon propagator
$\mathcal{D}(p^2)$, in the limit $p^2 \to 0$ we obtain
\begin{equation}
\frac{\sigma(0)}{g^2 N_c} \, = \, \frac{1}{8 \pi} \,
   \Biggl\{ \, \mathcal{D}(0) \, - \, \lim_{p^2 \to 0} \, \ln\left( p^2 \right) \, \mathcal{D}(0)
             \, + \, \int_{0}^{\infty} \d x \, \left[ \, x \, \ln\left( x \right) \,-\,  x\, \right]
                  \, \mathcal{D}''(x) \, \Biggr\}
\label{eq:sigma0due}
\end{equation}
and we again find\footnote{In Appendix \ref{app:relax} we will prove that the integral on the r.h.s.\
of Eq.\ (\ref{eq:sigma0due}) is finite under the hypotheses made for the gluon propagator $\mathcal{D}(p^2)$.
In the same Appendix we will also show how these hypotheses can be relaxed in this case.} an IR singularity
proportional to $- \mathcal{D}(0) \ln(p^2)$, unless one has $\mathcal{D}(0)=0$.

\vskip 3mm
Thus, in the $2d$ case and using a generic (sufficiently regular) gluon propagator, a null value for
$\mathcal{D}(0)$ is a necessary condition to obtain a finite value for $\sigma(0)$ at one loop. As a
consequence, the condition $\mathcal{D}(0)=0$ must be imposed if one wants to satisfy the no-pole condition
(\ref{eq:nopole}) and keep the functional integration inside the first Gribov region $\Omega$. It is
important to stress again that our proofs apply to any Gribov copy inside the first Gribov horizon, i.e.\
the result $\mathcal{D}(0)=0$ is not affected by the Gribov noise.

%%%%%%%%%%%%%%%%%%%%%%%%%%%%%%%%%%%%%%%%%%%%%%%%%%%%%%%%%%%%%%%%%%%%%%%%%%%%%%%%%%%%%%%%%%%%%%%%%%%%%%%%%%%%%%%

\subsection{Properties of $\sigma(p^2)$ in $d$ Dimensions: Approximate Calculation}
\label{sec:dapprox}

We can easily extend the result
\begin{equation}
  \frac{\partial \sigma(p^2)}{\partial p^2} \, < \, 0
\label{eq:ineq}
\end{equation}
to the $d$-dimensional case by using for the integral in $\d^dq$ the so-called $y$-max approximation or
angular approximation (see for example \cite{von Smekal:1997vx,Atkinson:1997tu,Aguilar:2004sw}). The same
approach allows us to show that the IR singularity $- \mathcal{D}(0) \ln(p^2)$ is present only
in the two-dimensional case. Indeed, by using hyperspherical coordinates (see Appendix \ref{app:hsc}) and
by considering the positive $x_1$ direction parallel to the external momentum $p$, we can write the
$d$-dimensional ghost form-factor (\ref{sigma1}) as
\begin{equation}
\frac{\sigma(p^2)}{g^2 N_c} \, = \, \frac{p_{\mu} p_{\nu}}{p^2} \int \frac{\d^d q}{(2\pi)^d} \frac{1}{(p-q)^2}
                                 \mathcal{D}(q^2) \, \mathcal{P}_{\mu\nu}(q)
                          \, = \, \int_0^{\infty} \, \d q \, \frac{q^{d-1}}{(2\pi)^d} \, \mathcal{D}(q^2)
                                 \, \int \, \d\Omega_d \, \frac{1 - \cos^2(\phi_1)}{(p-q)^2} \; .
\label{eq:sigmaDangular}
\end{equation}
In the $y$-max approximation one substitutes $1/(p-q)^2$ with $1/p^2$, for $q^2 < p^2$, and with $1/q^2$, for
$p^2 < q^2$. Then, we obtain
\begin{eqnarray}
\frac{\sigma(p^2)}{g^2 N_c} & = & \frac{1}{(2\pi)^d} \, \left[ \,
                                \int_0^p \, \d q \, \frac{q^{d-1}}{p^2} \, \mathcal{D}(q^2) \, + \,
                                \int_p^{\infty} \, \d q \, \frac{q^{d-1}}{q^2} \, \mathcal{D}(q^2) \, \right] \,
                                \int \, \left[ 1 - \cos^2(\phi_1) \right] \, \d\Omega_d \\[2mm]
                          & = & \frac{\Omega_d}{(2\pi)^d} \, \frac{d-1}{2 \, d} \, \left[ \,
                                \int_0^{p^2} \, \d q^2 \, \frac{q^{d-2}}{p^2} \, \mathcal{D}(q^2) \, + \,
                                \int_{p^2}^{\infty} \, \d q^2 \, \frac{q^{d-2}}{q^2} \, \mathcal{D}(q^2) \,
                                \right] \label{eq:sigmaDD} \\[2mm]
                          & = & \frac{\Omega_d}{(2\pi)^d} \, \frac{d-1}{2 \, d} \, \int_0^{\infty} \, \d q^2 \,
                                q^{d-2} \, \mathcal{D}(q^2) \, \left[ \, \frac{\theta(p^2 - q^2)}{p^2}
                                \, + \, \frac{\theta(q^2 - p^2)}{q^2}  \, \right] \; ,
\label{eq:sigmaderiv}
\end{eqnarray}
where we used Eq.\ (\ref{eq:firstint}). Note that, for $d=2$ and using Eq.\ (\ref{eq:omegaD}) one gets the
exact result (\ref{eq:D2res}). By repeating the argument shown in the Section \ref{sec:2Dderivative}, the
proof of the inequality (\ref{eq:ineq}) follows directly from Eq.\ (\ref{eq:sigmaderiv}).

At the same time, we can write Eq.\ (\ref{eq:sigmaDD}) as
\begin{equation}
\frac{\sigma(p^2)}{g^2 N_c} \, = \, \frac{\Omega_d}{(2\pi)^d} \, \frac{d-1}{d} \, \left[ \,
                                  \int_0^{p} \, \d q \, q^{d-1} \, \frac{\mathcal{D}(q^2)}{p^2} \, + \,
                                  \int_{p}^{\infty} \, \d q \, q^{d-3} \, \mathcal{D}(q^2) \, \right]
                            \, = \, I_d(p^2,\infty) \; ,
\label{eq:sigmaDDbis}
\end{equation}
where the integral $I_d(p^2,\ell)$ is defined in Eq.\ (\ref{eq:Id-ell}). In Appendix \ref{sec:hyper} we have also
shown that, for $d > 2$, this integral is finite when the gluon propagator $\mathcal{D}(p^2)$ is finite and
nonzero at $p^2=0$. Thus, using the $y$-max approximation, we find that only in the $2d$ case the condition
$\mathcal{D}(0)=0$ is necessary in order to obtain a finite value for the Gribov form-factor $\sigma(p^2)$
for all values of $p^2$.

Of course, in case of ultraviolet (UV) divergences we should regularize the integral defining $\sigma(p^2)$,
as done for example in Section \ref{sec:4d} below for the $4d$ case using the modified minimal subtraction
($\MSbar$) scheme and dimensional regularization. One can also consider a fixed momentum $\mu$ and
subtract\footnote{This is equivalent to a momentum-subtraction (MOM) renormalization scheme defined by the
condition $\mathcal{G}(\mu^2) = 1/\mu^2$.} the value $\sigma(\mu^2)$ from the Gribov form-factor $\sigma(p^2)$.
Due to the use of the $y$-max approximation the result of the subtraction is very simple. Indeed, instead of
Eq.\ (\ref{eq:sigmaDDbis}) we have the relation
\begin{equation}
\frac{\sigma(p^2) \,-\, \sigma(\mu^2)}{g^2 N_c} \, = \, \frac{\Omega_d}{(2\pi)^d} \, \frac{d-1}{2 \, d}
  \, \left[ \, \int_0^{p^2} \, \d x \, \frac{x^{d/2-1}}{p^2} \, \mathcal{D}(x) \, + \,
               \int_{p^2}^{\mu^2} \, \d x \, x^{d/2-2} \, \mathcal{D}(x) \, - \,
               \int_0^{\mu^2} \, \d x \, \frac{x^{d/2-1}}{\mu^2} \, \mathcal{D}(x) \, \right] \; ,
\label{eq:sigmamom}
\end{equation}
which is valid for $p^2 \leq \mu^2$ as well as for $\mu^2 < p^2$. Then, we find again
\begin{equation}
  \frac{\partial \sigma(p^2)}{\partial p^2} \, = \, - \, g^2 N_c \, \frac{\Omega_d}{(2\pi)^d}
                            \, \frac{d-1}{2 \, d}
        \, \int_0^{p^2} \, \d x \, \frac{x^{d/2-1}}{p^4} \, \mathcal{D}(x) \, < \, 0
\end{equation}
if $\mathcal{D}(x)$ is positive. We can also easily check that, for $\mathcal{D}(0) > 0$ and $d > 2$, the
Gribov form-factor $\sigma(p^2)$ in Eq.\ (\ref{eq:sigmamom}) does not display an IR singularity.

%%%%%%%%%%%%%%%%%%%%%%%%%%%%%%%%%%%%%%%%%%%%%%%%%%%%%%%%%%%%%%%%%%%%%%%%%%%%%%%%%%%%%%%%%%%%%%%%%%%%%%%%%%%%%%%

\subsection{Properties of $\sigma(p^2)$ in $d$ Dimensions: Exact Calculation}
\label{sec:d-exact}

One can improve the results obtained in the previous Section by considering the formulae reported in
Appendices \ref{app:hsc} and \ref{sec:hyper}, which allow us to perform the angular integration in Eq.\
(\ref{eq:sigmaDangular}) without approximations. Indeed, we have\footnote{Again we make the
hypothesis that a regularization is introduced in the case of UV divergences.  In Appendix \ref{sec:mom}
we explicitly show how to extend the proof to a MOM scheme, i.e.\ by subtracting the value $\sigma(\mu^2)$
from $\sigma(p^2)$ where $\mu$ is a fixed momentum.}
\begin{equation}
\frac{\sigma(p^2)}{g^2 N_c} \, = \, \int_0^{\infty} \, \d q \, \frac{q^{d-1}}{(2\pi)^d} \, \mathcal{D}(q^2)
                                \, \int \, \d\Omega_d \, \frac{1 - \cos^2(\phi_1)}{p^2 \,+\, q^2 \, - \,
                                   2 \, p \, q \, \cos(\phi_1) }
                            \, = \, I(p^2, 1, d, \infty) \; ,
\label{eq:sigmaDini}
\end{equation}
with $I(p^2, \nu, d, \ell)$ defined in Eq.\ (\ref{eq:Igeneral-dell}). Since $\nu=1$ in this case, for
$d \geq 2$ we can also make use of the inequalities (\ref{eq:I-upper-nu1}) and write
\begin{equation}
\frac{d}{2 \, (d - 1)} \, I_d(p^2,\infty) \, \leq \, I(p^2,1,d,\infty) \, \leq \, I_d(p^2,\infty) \; .
\label{eq:ineqId}
\end{equation}
Note that $I_d(p^2,\infty)$ is the same integral obtained on the right-hand side of Eq.\ (\ref{eq:sigmaDDbis}).
Thus, the $y$-max approximation of the previous Section provides, for $d=3$ and 4, an upper bound for the
Gribov ghost factor. On the contrary, for $d=2$, the above inequalities become equalities. At the same
time, as one can see in Appendix \ref{sec:hyper}, the integral $I_d(p^2,\infty)$ is finite (for $d>2$) also
if $\mathcal{D}(0)$ is nonzero, i.e.\ we do not need to impose the condition $\mathcal{D}(0)=0$ in order
to attain a finite value for $\sigma(p^2)$ in the IR limit.

By evaluating the derivative with respect to $p^2$ of the result (\ref{eq:Igeneral-dell2}) we also obtain
\begin{eqnarray}
\frac{1}{g^2 N_c} \, \frac{\partial \sigma(p^2)}{\partial p^2} & = & \frac{\Omega_d}{(2\pi)^d} \,
    \frac{d-1}{d} \, \int_0^{\infty} \, \d q \, q^{d-1} \, \mathcal{D}(q^2) \, \left[ \,
    -\frac{\theta(p^2 - q^2)}{p^4} \, _2F_1\left(1, 1-d/2; 1+d/2; q^2/p^2\right) \right. \nonumber \\[2mm]
  & & \qquad \qquad \qquad \qquad \qquad \; - \, \frac{q^2 \, \theta(p^2 - q^2)}{p^6} \,
                                             _2F'_1\left(1, 1-d/2; 1+d/2; q^2/p^2\right) \nonumber \\[2mm]
  & & \left. \qquad \qquad \qquad \qquad \qquad \; + \, \frac{\theta(q^2 - p^2)}{q^4} \,
                                          _2F'_1\left(1, 1-d/2; 1+d/2; p^2/q^2\right) \, \right] \; ,
\label{eq:sigmaderd}
\end{eqnarray}
where $_2F'_1\left(a,b;c;z\right)$ indicates the derivative with respect to the variable $z$ of the Gauss
hypergeometric function $_2F_1\left(a,b;c;z\right)$ (see Appendix \ref{sec:hyper}). Here we used again the
properties of the theta and of the Dirac delta functions and Eq.\ (\ref{eq:gauss}). For $d=2$, the last
two terms in Eq.\ (\ref{eq:sigmaderd}) are null [see Eq.\ (\ref{eq:derivativeF})] and, using the result
(\ref{eq:2F1-D2}), we find again Eq.\ (\ref{eq:sigmader2}). In the $4d$ case one can use the expression
(\ref{eq:2F1-D4}) for the Gauss hypergeometric function $_2F_1\left(1, 1-d/2; 1+d/2; z\right)$. Then,
from Eq.\ (\ref{eq:sigmaderd}) --- or, equivalently, by evaluating the derivative with respect to $p^2$
of Eq.\ (\ref{eq:4dmom}) in Appendix \ref{sec:mom} --- we find that
\begin{eqnarray}
\frac{\partial \sigma(p^2)}{\partial p^2} & = & \frac{g^2 N_c}{32 \, \pi^2} \, \left[ \,
          \int_0^{p} \, \d q \, \mathcal{D}(q^2) \, \frac{2 q^5 \,-\, 3 p^2 q^3}{p^6}
          \, - \, \int_{p}^{\infty} \, \d q \, \frac{\mathcal{D}(q^2)}{q} \, \right] \\[2mm]
                                          & = & \frac{g^2 N_c}{32 \, \pi^2} \, \left[ \,
          \int_0^1 \, \d y \; \mathcal{D}(y^2 p^2) \, \left( 2 y^5 \,-\, 3 y^3 \right)
          \, - \, \int_{p}^{\infty} \, \d q \, \frac{\mathcal{D}(q^2)}{q} \, \right] \; ,
\end{eqnarray}
where $y = q/p$ and we have used Eq.\ (\ref{eq:omegaD}). For $\mathcal{D}(p^2) > 0$ both terms in square
brackets are negative, i.e.\ the derivative $\partial \sigma(p^2) / \partial {p^2}$ is negative for all values
of the momentum $p$. Let us note that in the original work by Gribov \cite{Gribov:1977wm} the same result
was proven [see comment after Eq.\ (37) in the same reference] under the much stronger hypothesis of a
gluon propagator $\mathcal{D}(q^2)$ decreasing monotonically with $q^2$ over the main range of integration.

\begin{figure}[t]
   \centering
       \includegraphics[width=8cm]{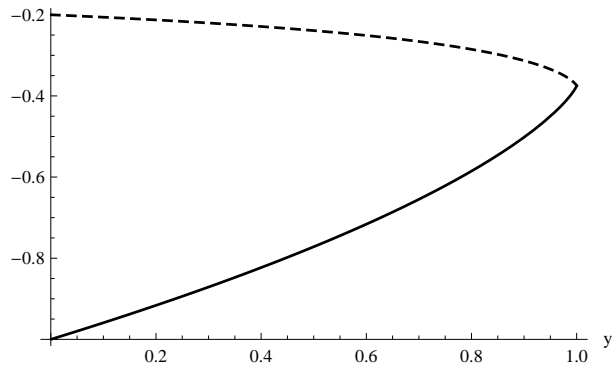}
   \caption{The functions $- \left[ \, \Psi(y) \, + \, y \, \Psi'(y) \, \right]$ (full line) and $\Psi'(y)$
            (dashed line) for $y \in [0,1]$.  \label{figdiff}}
\end{figure}

A similar analysis can be done in the $3d$ case using Eq.\ (\ref{eq:2F1-D3}).  In order to simplify the
notation we define
\begin{equation}\label{psi1}
\Psi(z) \, = \, _2F_1\left(1,-1/2;5/2;z\right) \, = \, \frac{3}{4} \, + \, \frac{3 \, \left(1 - z\right)}{8 z} \,
      \left[ \, 1 \,-\, \frac{1 - z}{\sqrt{z}} \, \text{arcsinh}\left(\sqrt{\frac{z}{1 - z}} \, \right) \,
                      \right] \; .
\end{equation}
This gives
\begin{equation}\label{psi2}
\Psi'(z) \, = \, \frac{3}{16 z^3} \, \left[ \, z \, \left(z-3\right)
                \, + \, \sqrt{z} \, \left(3 - 2 z - z^2\right) \,
                       \text{arcsinh}\left(\sqrt{\frac{z}{1-z}}\right) \, \right] \; .
\end{equation}
Then, after setting $d=3$ in Eq.\ \eqref{eq:sigmaderd} and using Eq.\ \eqref{eq:omegaD}, we obtain
\begin{eqnarray}
\label{psi3}
\frac{\partial \sigma(p^2)}{\partial p^2}
    & = & \frac{g^2 N_c}{3\pi^2} \,\Biggl\{ \, \int_0^{\infty} \, \d q \, q^{2} \, \mathcal{D}(q^2)
           \, \Biggl[ \, -\frac{\theta(p^2-q^2)}{p^4} \, \Psi\left(\frac{q^2}{p^2}\right) \,
                - \, \frac{q^2\theta(p^2-q^2)}{p^6} \, \Psi'\left(\frac{q^2}{p^2}\right) \nonumber \\[2mm]
    &   & \qquad \qquad \qquad \qquad \qquad \qquad \qquad \qquad
               + \, \frac{\theta(q^2-p^2)}{q^4} \, \Psi'\left(\frac{p^2}{q^2}\right) \, \Biggr] \, \Biggr\} \\[2mm]
    & = & \frac{g^2 N_c}{6\pi^2} \, \Biggl\{ \, \int_0^{p^2} \, \d x \, \sqrt{x} \,\mathcal{D}(x)
           \, \Biggl[ \, -\frac{1}{p^4} \, \Psi\left(\frac{x}{p^2}\right) \,
                - \, \frac{x}{p^6}\, \Psi'\left(\frac{x}{p^2}\right) \, \Biggr] \,
               + \, \int_{p^2}^{\infty} \, \frac{\d x}{x^{3/2}} \, \mathcal{D}(x) \, \Psi'\left(\frac{p^2}{x}\right)
                        \, \Biggr\} \; ,
\end{eqnarray}
where we also made the substitution $x = q^2$. Next, the change of variable $x = y p^2$ in the first integral and
$x = p^2/y$ in the second integral yield
\begin{eqnarray} \label{psi4}
\frac{\partial \sigma(p^2)}{\partial p^2}
    & = & \frac{g^2 N_c}{6\pi^2} \, \Biggl\{ \, - \, \int_0^1 \, \d y \, \frac{\sqrt{y}}{p} \, \mathcal{D}(y p^2)
           \, \Biggl[ \, \Psi(y) \, + \, y \, \Psi'(y) \, \Biggr] \, + \,
                    \int_{0}^{1} \, \frac{\d y}{p} \, \frac{1}{\sqrt{y}} \, \mathcal{D}(p^2/y)
                      \, \Psi'(y) \, \Biggr\} \; .
\end{eqnarray}
As one can see in Figure \ref{figdiff}, the factor $- \left[ \, \Psi(y) \, + \, y \, \Psi'(y) \, \right]$ is negative
for $y \in [0,1]$. At the same time, from Eq.\ (\ref{eq:d3derivative2F1}) we know that $ \Psi'(y) $ is negative for
$y \geq 0$ (see also the corresponding plot in Figure \ref{figdiff}). Thus, for a positive gluon propagator
$\mathcal{D}(p^2)$, the $3d$ derivative $\p \sigma(p^2) / \p p^2$ is negative for $p^2 > 0$.

Finally, we can consider a general $d > 2$ and, after suitable changes of variables (for $p^2 > 0$), we write
\begin{eqnarray}
\frac{1}{g^2 N_c} \, \frac{\partial \sigma(p^2)}{\partial p^2}
    & = & \frac{\Omega_d}{(2\pi)^d} \, \frac{d-1}{d} \, \frac{p^{d-4}}{2} \, \left\{
             \int_0^{1} \, \d y \, y^{2-d/2} \; \mathcal{D}(p^2/y) \;\; _2F'_1\left(1, 1-d/2; 1+d/2; y\right)
                \right. \nonumber \\[2mm]
    & &  \left. \;\;\; - \, \int_0^{1} \,
        \d y \, y^{d/2-1} \; \mathcal{D}(y p^2) \, \Bigl[ \;
             _2F_1\left(1, 1-d/2; 1+d/2; y\right) \, + \, y \; _2F'_1\left(1, 1-d/2; 1+d/2; y\right) \, \Bigr]
                   \right\} \; .
\label{eq:sigmaderd2}
\end{eqnarray}
Note that the dependence on $p^2$ is only in the global factor $p^{d-4}$ and in the argument of the gluon
propagator. From Appendix \ref{sec:hyper} we know that the derivative $_2F'_1\left(1, 1-d/2; 1+d/2; x\right)$
is negative for $x \in [0, 1]$ and $d > 2$ and that, under the same hypotheses, the expression in square
brackets is positive. Thus, for a positive gluon propagator, both term in the r.h.s.\ of the above expression
are negative and we have proven that, for any dimension $d \geq 2$, the Gribov form-factor $\sigma(p^2)$ (at one
loop) is monotonically decreasing with $p^2$, i.e.\ it gets its maximum value at $p^2 = 0$.

%%%%%%%%%%%%%%%%%%%%%%%%%%%%%%%%%%%%%%%%%%%%%%%%%%%%%%%%%%%%%%%%%%%%%%%%%%%%%%%%%%%%%%%%%%%%%%%%%%%%%%%%%%%%%%%

\section{Evaluation of the One-Loop Corrected Ghost Propagator Using (Linear Combinations of)
         Yukawa-Like Gluon Propagators}
\label{sec:sigma234}

In the previous Section we have proven that, at one-loop level and for a sufficiently regular gluon
propagator $\mathcal{D}(p^2)$, the Gribov ghost form-factor $\sigma(p^2)$ is always finite in three and
four space-time dimensions while, in $d=2$, one needs to impose $\mathcal{D}(0) = 0$ in order to avoid an
IR singularity of the type $- \mathcal{D}(0) \,\ln(p^2)$. In this Section we present an explicit
calculation of $\sigma(p^2)$ at one loop for $d=2, 3$ and 4 using, for the gluon propagator, results recently
presented in Ref.\ \cite{Cucchieri:2011ig,Cucchieri:2012gb} from fits to lattice data of $\mathcal{D}(p^2)$ in
the SU(2) case.  The expressions obtained below for the ghost propagator $\mathcal{G}(p^2)$ will be used in
a subsequent work \cite{ghost} to model lattice data of SU(2) ghost propagators.

In this Section, besides recovering the same results reported in Section \ref{sec:ghost-prop}, we also find
that the ghost propagator $\mathcal{G}(p^2)$ admits a one-parameter family of behaviors \cite{Boucaud:2008ji,
arXiv:1103.0904} labelled by the coupling constant $g^2$, considered as a free parameter. The no-pole
condition $\sigma(0) \leq 1$ implies $g^2 \leq g^2_c$, where $g^2_c$ is a ``critical'' value. Moreover,
for $g^2$ smaller than $g^2_c$ one has $\sigma(0) < 1$ and the ghost propagator shows a free-like behavior
in the IR limit, in agreement with the so-called massive solution of gluon and ghost DSEs
\cite{Aguilar:2004sw,Boucaud:2006if,Aguilar:2008xm,Boucaud:2008ji,Boucaud:2008ky,Binosi:2009qm,
Aguilar:2010zx,Pennington:2011xs}. On the contrary, for $g^2 = g^2_c$ one finds $\sigma(0) = 1$ and the
ghost propagator is IR enhanced \cite{Alkofer:2000wg,von Smekal:1997vx,Zwanziger:2001kw,Lerche:2002ep,
Huber:2007kc}.

%%%%%%%%%%%%%%%%%%%%%%%%%%%%%%%%%%%%%%%%%%%%%%%%%%%%%%%%%%%%%%%%%%%%%%%%%%%%%%%%%%%%%%%%%%%%%%%%%%%%%%%%%%%%%%%

\subsection{Yukawa-Like Gluon Propagators and Set Up}
\label{sec:yukawa}

In Ref.\ \cite{Cucchieri:2011ig,Cucchieri:2012gb} the SU(2) gluon propagator was fitted in 2, 3 and 4 space-time
dimensions using, respectively, the functions\footnote{Note that, for consistency with the notation used
in Ref.\ \cite{Cucchieri:2011ig}, in this Section the non-integer power of the momentum $p$ is $\eta$
and not $2 \eta$ as in the rest of the manuscript.}
\begin{equation}
\mathcal{D}(p^2) \, = \, C \, \frac{p^2 + l\,p^{\eta} + s}{p^4 + u^2 \, p^2 + t^2} \; ,
\label{eq:f2dgluon}
\end{equation}
\begin{equation}
\mathcal{D}(p^2) \, = \, C \, \frac{p^4 + (s+1) p^2 + s}{p^6 + (k + u^2) p^4 + (k u^2 + t^2) p^2 + k t^2}
\label{eq:fDvrgzsimple}
\end{equation}
and
\begin{equation}
\mathcal{D}(p^2) \, = \, C \, \frac{p^2 + s}{p^4 + u^2 \, p^2 + t^2} \; .
\label{eq:f4dgluon}
\end{equation}
The last two propagators are tree-level gluon propagators that arise in the study of the RGZ action
\cite{Dudal:2008sp,Dudal:2008rm,Vandersickel:2011zc,Dudal:2011gd}. The first one is a simple
generalization of the form (\ref{eq:f4dgluon}) that fits well the $2d$ data. Note that these
three functions can be written as a linear combination of propagators of the type $ 1 / (p^2+\omega^2)$,
where $\omega^2$ is in general a complex number. [In the $2d$ case we need to consider the more general
form $ p^{\eta} / (p^2+\omega^2)$ with $\eta \geq 0$.] Thus, in order to evaluate $\sigma(p^2)$ in
Eq.\ (\ref{sigma1}) using the above gluon propagators $\mathcal{D}(p^2)$, we first consider the integral
\begin{equation}
f(p, \omega^2) \, = \, \frac{p_{\mu} p_{\nu}}{p^2} \int \frac{\d^d q}{(2\pi)^d}
       \frac{1}{(p-q)^2} \frac{1}{q^2 + \omega^2} \left(\delta_{\mu\nu}-\frac{q_\mu q_\nu}{q^2}\right) \; .
\label{eq:deffofk}
\end{equation}
The evaluation of $f(p, \omega^2)$ can be done in three and four space-time dimensions by introducing
Feynman parameters and applying the usual shift in the momentum $q$. The integration then yields
\begin{eqnarray}
f(p^2,\omega^2) & = & \frac{1}{(4\pi)^{d/2}} \int_0^1 \d x \left[ \Delta^{d/2-2} \, \Gamma(2-d/2) \right]
                        \nonumber \\[2mm]
                &   & \; - \frac{1}{(4\pi)^{d/2}} \int_0^1 \d x \int_0^{1-x} \d y
                                   \left[ \frac{1}{2} \Theta^{d/2-2} \, \Gamma(2-d/2) \, + \,
                                           x^2p^2 \, \Theta^{d/2-3} \, \Gamma(3-d/2) \right]
\label{integr}
\end{eqnarray}
with
\begin{align}
    \Delta & = -x^2 p^2 + x p^2 + (1 - x) \omega^2 \; , &
    \Theta & = -x^2 p^2 + x p^2 + y \omega^2 \; .
\label{eq:DeltaOmega}
\end{align}
Since the Gamma function has the behavior $\Gamma(x) \approx 1/x$ for small $x$, it is clear that the first
two integrals are UV finite for $d<4$ while the third one is UV finite for $d<6$. Below we will calculate
the integral \eqref{integr} for $d=$ 3 and 4. We start from the case $d=3$, where all terms are finite, and
then we evaluate the integral for the case $d=$ 4, using the $\MSbar$ scheme. On the contrary, as stressed
above, in the $2d$ case one needs to evaluate the more general function
\begin{equation}
f(p, \omega^2, \eta) \, = \, \frac{p_{\mu} p_{\nu}}{p^2} \int \frac{\d^2 q}{(2\pi)^d} \frac{1}{(p-q)^2}
         \frac{q^{\eta}}{q^2 + \omega^2} \left(\delta_{\mu\nu}-\frac{q_\mu q_\nu}{q^2}\right)
\label{eq:deffofkwithe}
\end{equation}
with $\eta \geq 0$. This case will be treated (in a slightly different way) in Section \ref{sec:2Dghost}.

Most of the analytic results reported in this Section have been checked using {\tt Mathematica} and/or {\tt Maple}.

%%%%%%%%%%%%%%%%%%%%%%%%%%%%%%%%%%%%%%%%%%%%%%%%%%%%%%%%%%%%%%%%%%%%%%%%%%%%%%%%%%%%%%%%%%%%%%%%%%%%%%%%%%%%%%%

\subsection{Ghost Propagator in the 3$d$ Case}
\label{sec:3doneloop}

In the $3d$ case the residual $x$- and $y$-integrations in Eq.\ \eqref{integr} are straightforward and give
\begin{eqnarray}
f(p^2,\omega^2) &=& \left[ \, \frac{1}{4 \pi p} \arctan\left( \frac{p}{\sqrt{\omega^2}} \right) \, \right]
            \, + \, \left[ \, - \frac{(p^2 - \omega^2) \sqrt{\omega^2}}{32 \pi p^2 \omega^2}
                           + \frac{\pi p}{64 \pi \omega^2} - \frac{(p^2+\omega^2)^2}{32 \pi p^3 \omega^2}
                           \arctan\left( \frac{p}{\sqrt{\omega^2}} \right) \, \right] \nonumber \\[2mm]
                & & \quad + \, \left[ \, \frac{3 \, (p^2 - \omega^2) \sqrt{\omega^2}}{32 \pi p^2 \omega^2}
                      + \frac{3 p^4 - 2 p^2 \omega^2 + 3 (\omega^2)^2}{32 \pi p^3 \omega^2}
                            \arctan\left( \frac{p}{\sqrt{\omega^2}} \right)
                                - \frac{3 \pi p}{64 \pi \omega^2} \, \right] \; ,
\end{eqnarray}
where the three square brackets highlight the contribution from the three terms in Eq.\ \eqref{integr}.
Here we have only made the assumption $p^2 > 0$. By simplifying the above result, we find
\begin{eqnarray}
f(p^2,\omega^2) &=& \frac{1}{4 \pi p} \arctan\left( \frac{p}{\sqrt{\omega^2}} \right) \,
                   + \, \frac{(p^2 - \omega^2) \sqrt{\omega^2}}{16 \pi p^2 \omega^2} \,
                   - \, \frac{\pi p}{32\pi\omega^2} \,
                   + \, \frac{(p^2 - \omega^2)^2}{16 \pi p^3 \omega^2}
                         \arctan\left(  \frac{p}{\sqrt{\omega^2}} \right) \nonumber \\[2mm]
                &=& \frac{1}{32 \pi p^3 \omega^2} \, g(p^2,\omega^2)
\label{integr2}
\end{eqnarray}
where
\begin{equation}
g(p^2,\omega^2) \, = \, -\pi \, p^4 \,+ \, 2 \, p^3 \, \sqrt{\omega^2} \,- \,
                        2 \, p \, (\omega^2)^{3/2} \, + \, 2 \,
                        (p^2 \, + \, \omega^2)^2 \, \arctan\left(\frac{p}{\sqrt{\omega^2}}\right) \; .
            \label{eq:f3d}
\end{equation}

In order to use the result \eqref{integr2} we need to write the gluon propagator \eqref{eq:fDvrgzsimple}
as
\begin{equation}\label{gluonpropsimp}
\mathcal{D}(p^2) \, = \, \frac{\alpha}{p^2+\omega_1^2} + \frac{\beta}{p^2+\omega_2^2}
                             + \frac{\gamma}{p^2+\omega_3^2} \; .
\end{equation}
Here $\omega_1^2$, $\omega_2 ^2$ and $\omega_3^2$ are the three roots of the cubic equation, with respect
to the variable $p^2$, obtained by setting equal to zero the denominator of Eq.\ \eqref{eq:fDvrgzsimple}.
Thus, by combining Eqs.\ \eqref{sigma1}, \eqref{gluonpropsimp} and \eqref{eq:deffofk} we can write for
the function $\sigma(p^2)$ in the $3d$ case the expression
\begin{equation}\label{sigma2}
\sigma(p^2) \, = \, g^2 N_c \, \left[ \alpha f(p^2,\omega_1^2) \, + \, \beta f(p^2,\omega_2^2)
                                                \, + \, \gamma f(p^2,\omega_3^2) \right]
\end{equation}
or
\begin{equation}
\sigma(p^2) \, = \, \frac{g^2 N_c}{32\,\pi\,\omega^2\,p^3} \,
                      \left[ \alpha \, g(p^2,\omega_1^2) \, + \, \beta \, g(p^2,\omega_2^2)
                       \, + \, \gamma \, g(p^2,\omega_3^2) \right]
                   \label{eq:sigma3d}
\end{equation}
with $g(p^2,\omega^2)$ given in Eq.\ \eqref{eq:f3d}. In general, the roots $\omega_1^2$, $\omega_2^2$
and $\omega_3^2$ are all real or there is one real root, for example $\omega_1^2$, and two
complex-conjugate roots, i.e.\ $(\omega_2^{2})^* = \omega_3^2$, implying also $\beta = \gamma^*$.
Since the fits in Ref.\ \cite{Cucchieri:2011ig,Cucchieri:2012gb} support the latter case we write
\begin{equation}
    \beta  \, = \, a + i b \; , \qquad
    \gamma \, = \, a - i b
\end{equation}
and
\begin{equation}
\omega_2^2 \, = \, v + i w \; , \qquad
\omega_3^2 \, = \, v - i w \; .
\end{equation}
Then, following for example \cite{web1}, we find for $\omega_2^2$ the relations
\begin{align}
\sqrt{\omega_2^2}               & \, = \, \sqrt{v + i w}
   \, = \, \frac{1}{\sqrt{2}} \sqrt{\sqrt{v^2 + w^2} + v}
         + \frac{i}{\sqrt{2}} \sqrt{\sqrt{v^2 + w^2} - v} \; , \\
\left( \omega_2^2 \right)^{3/2} & \, = \, (v+iw)^{3/2}
   \, = \, (v + iw) \sqrt{v + i w} \; , \\
\frac{p}{\sqrt{\omega_2^2}}     & \, = \, \frac{p}{\sqrt{v + i w}}
   \, = \, \frac{p}{\sqrt{v^2 + w^2}} \, \sqrt{v - i w}
\end{align}
and similar results for $\omega_3^2$. We also use the expression (see for example \cite{GR})
\begin{equation}
\arctan\left( z \right) \, = \, \frac{1}{2} \arg \left( \frac{i - z}{i + z}\right)
         - \frac{i}{2} \ln\left| \frac{i-z}{i+z} \right|   \qquad   \forall z \not= \{i, -i\} \; .
\end{equation}
This allows us to write the function $\sigma(p^2)$ only in term of real quantities, i.e.\
\begin{equation}
\sigma(p^2) \, = \, \frac{g^2 N_c}{8}
                    \left[ \frac{\alpha g(p^2,\omega_1^2)}{4\,\pi\,\omega_1^2\,p^3} \, + \, f_R(p^2) \right]
\label{eq:sigma3d2}
\end{equation}
where $g(p^2,\omega^2)$ is given in Eq.\ (\ref{eq:f3d}) above. Also, we have
\begin{equation}
f_R(p^2) \, = \, f_1(p^2) \, + \, f_2(p^2) \, + \, f_3(p^2) \, + \, f_4(p^2) \, + \, f_5(p^2)
\end{equation}
with
\begin{eqnarray}
f_1(p^2) & = & - p \; \frac{a v \, + \, b w}{2 \, R^2}   \; ,\\[2mm]
f_2(p^2) & = & \frac{\left( a v \, + \, b w \right) \sqrt{R+v} \,-\,
                   \left( b v \, - \, a w \right) \sqrt{R-v}}{\sqrt{2} \, \pi \, R^2} \; , \\[2mm]
f_3(p^2) & = & - \frac{1}{p^2} \, \frac{a \sqrt{R+v} \,-\, b \sqrt{R-v}}{\sqrt{2} \, \pi} \; , \\[2mm]
f_4(p^2) & = & A(p^2) \; \frac{p^4 \left( a v \, + \, b w \right) \,+\, 2\,a\,p^2\,R^2
                   \,+\,R^2 \left( a v \, - \, b w \right)}{2 \, \pi \, R^2\, p^3}  \; ,\\[2mm]
f_5(p^2) & = &-L(p^2) \; \frac{p^4 \left( b v \, - \, a w \right) \,+\, 2\,b\,p^2\,R^2
                              \,+\,R^2 \left( b v \, + \, a w \right)}{2 \, \pi \, R^2\, p^3}
\end{eqnarray}
and
\begin{eqnarray}
A(p^2) & = & \left\{ \begin{array}{ll}
                \arctan\left( \frac{\sqrt{2} \, p \, \sqrt{R+v}}{R\,-\,p^2} \right) \qquad
                     & \mbox{if} \;\;\; R\,-\,p^2 > 0 \\[3mm]
            \pi \,+\, \arctan\left( \frac{\sqrt{2} \, p \, \sqrt{R+v}}{R\,-\,p^2} \right) \qquad
                     & \mbox{if} \;\;\; R\,-\,p^2 < 0
                 \end{array} \right.  \; ,\\[3mm]
L(p^2) & = & \ln\left[ \frac{\sqrt{p^4 \,+\, 2 \, p^2\,v \,+\, R^2}}{
                   R \,+\, p \, \left( p \,+\, \sqrt{2}\,\sqrt{R-v} \right)} \right] \; ,\\[3mm]
R      & = & \sqrt{v^2+w^2} \label{eq:r3d} \label{eq:f33d} \; .
\end{eqnarray}

One can check that $\sigma(p^2)$ is null in the limit $p \to \infty$.
Finally, by expanding $\sigma(p^2)$ around $p^2 = 0$ in Eqs.\ (\ref{eq:sigma3d2})--(\ref{eq:r3d}) we find
\begin{eqnarray}
\frac{\sigma(p^2)}{g^2 N_c} & = & \frac{\alpha}{6 \pi \sqrt{\omega_1^2}} \, + \,
                   \sqrt{R+v}\;\frac{9\,a\,R^2-\left(a\,v-b\,w\right)\,\left(2\,v-R\right)}{
                             24\,\sqrt{2}\,\pi\,R^3} \nonumber \\[3mm]
            &   & \quad +\sqrt{R-v}\;\frac{9\,b\,R^2-\left(b\,v+a\,w\right)\,\left(2\,v+R\right)}{
                             24\,\sqrt{2}\,\pi\,R^3} \, - \,
                   \frac{\alpha\,R^2 + 2 \, (a\,v+b\,w) \omega_1^2}{32 \, \omega_1^2 \, R^2} \, p \,
                  + \, O\left(p^2\right) \; ,
\label{eq:sigmair3d}
\end{eqnarray}
which implies $\mathcal{G}(p^2) \propto p^{-2}$ at very small momenta. However, if the constant term in
the above expression is equal to $1/(g^2 N_c)$, yielding $\sigma(0)=1$, then one gets in the IR limit
$\mathcal{G}(p^2) \propto p^{-4}$ or $\mathcal{G}(p^2) \propto p^{-3}$, depending on whether the term
$\alpha\,R^2 + 2 \, (a\,v+b\,w) \omega_1^2$ vanishes or not. In particular, in the original GZ case,
i.e.\ when the terms containing $\omega_1^2$ and $\alpha$ are absent, we do recover the usual
$1/p^4$ behavior. Also note that for purely imaginary poles, i.e.\ when $v = b = 0$ (and $R = w$),
the condition $\sigma(0)=1$ simplifies to
\begin{equation}
\frac{\alpha}{\sqrt{\omega_1^2}} \, + \, \frac{\sqrt{2} \, a}{\sqrt{w}} \, = \, \frac{6 \pi}{g^2 N_c} \; .
\end{equation}

Clearly, for a given value of $N_c$ and with a suitable choice of $g^2$, one can always set $\sigma(0) = 1$
in Eq.\ (\ref{eq:sigmair3d}). For example, using the numerical data in the second column of Table XI of Ref.\
\cite{Cucchieri:2011ig} and $N_c=2$, we find from Eq.\ (\ref{eq:sigmair3d}) the result\footnote{The errors in
brackets have been evaluated using a Monte Carlo analysis with 10000 samples (see Ref.\
\cite{Cucchieri:2011ig} for details).}
\begin{equation}
\frac{\sigma(p^2)}{2\,g^2} \, \approx \, 0.039(0.001) - 0.017(0.003) \, p \, + \, O\left(p^2\right)
\label{eq:sigma3dnum}
\end{equation}
and we have $\sigma(0)=1$ if $g^2 \approx 12.82$. Thus, if one considers $g^2$ as a free parameter,
then Eq.\ (\ref{eq:sigmair3d}) gives a one-parameter family of behaviors, labeled by $g^2$. For a
specific value of $g^2 = g^2_{c}$ we have $\sigma(0)=1$ and one finds an IR-enhanced ghost propagator
at one loop. On the contrary, for $g^2 < g^2_{c}$ we obtain $\sigma(0) < 1$ and $\mathcal{G}(p^2)
\propto p^{-2}$ in the IR limit. Finally, for $g^2 > g^2_{c}$ the no-pole condition $\sigma(0) \leq 1$ is not
satisfied, i.e.\ the ghost propagator is negative in the IR limit. These findings are in qualitative agreement
with the DSE results obtained in Refs.\ \cite{Boucaud:2008ji,arXiv:1103.0904}. Finally, note that at small
momenta the function $\sigma(p^2)$ in the above formula (\ref{eq:sigma3dnum}) is decreasing as $p^2$ increases,
as expected from Section \ref{sec:d-exact}.

%%%%%%%%%%%%%%%%%%%%%%%%%%%%%%%%%%%%%%%%%%%%%%%%%%%%%%%%%%%%%%%%%%%%%%%%%%%%%%%%%%%%%%%%%%%%%%%%%%%%%%%%%%%%%%%

\subsection{Ghost Propagator in the 4$d$ Case}
\label{sec:4d}

We want now to evaluate $f(p^2,\omega^2)$ in Eq.\ \eqref{integr} for $d=4$. As stressed above, in this case
we have to deal with UV divergences. We do the calculation in the $\MSbar$ renormalization scheme using
dimensional regularization with $d=4-\varepsilon$. For the first term in Eq.\ \eqref{integr} we have
\begin{equation}
\frac{(4 \pi)^{2-d/2}}{16 \pi^2} \int_0^1 \d x \left[ \Delta^{d/2-2} \, \Gamma(2-d/2) \right] \, = \,
\frac{1}{16 \pi^2} \int_0^1 \d x \left[ \frac{2}{\varepsilon} \, - \, \gamma_E \, + \, \ln\left(4 \pi\right) \,
                                                - \, \ln\left(\Delta\right) \right] \; ,
\end{equation}
where $\gamma_E$ is the Euler constant. Then, using the usual $\MSbar$ prescription, we find
\begin{equation}
\frac{-1}{16 \pi^2} \int_0^1 \d x \, \ln\left[ \frac{-x^2 p^2 + x p^2 + (1 - x) \omega^2}{\omu^2} \right] \, = \,
\frac{-1}{16 \pi^2} \int_0^1 \d x  \left[ \ln\left(\frac{p^2}{\omu^2}\right) \, + \,
                                       \ln\left(1-x\right) \, + \,
                                       \ln\left(x + \frac{\omega^2}{p^2}\right) \right]
\end{equation}
and the $\d x$ integration yields
\begin{eqnarray}
& & \frac{-1}{16 \pi^2} \left[ \ln\left(\frac{p^2}{\omu^2}\right)
                      \, - \, 2 \, - \, \frac{\omega^2}{p^2} \ln\left(\frac{\omega^2}{p^2}\right)
                      \, + \, \left(1 + \frac{\omega^2}{p^2}\right) \ln\left(1 + \frac{\omega^2}{p^2}\right)
                    \right] \nonumber  \\[2mm]
&=& \frac{-1}{16 \, p^2 \pi^2} \left[ - 2 p^2 \, + \, p^2 \, \ln\left(\frac{p^2 + \omega^2}{\omu^2}\right)
                               \, + \, \omega^2 \, \ln\left(\frac{p^2 + \omega^2}{\omega^2}\right) \right] \; ,
\label{4D6}
\end{eqnarray}
where $\omu$ is the renormalization scale. For the second term in Eq.\ \eqref{integr}, which is also divergent,
we first perform the $y$ integration exactly, obtaining
\begin{equation}
- \frac{\Gamma(2-d/2)}{(4 \pi)^{d/2} \, \omega^2 \, \left(d-2\right)} \int_0^1 \d x \,
                                      \left( -x^2 p^2 + x p^2 \right)^{d/2-1} \,
               \left[ \left( 1 + \frac{\omega^2}{x p^2} \right)^{d/2-1} \, - \, 1 \right] \; .
\label{eq:4dsecondterm}
\end{equation}
The $\varepsilon$ expansion then gives
\begin{equation}
\frac{1}{32 \pi^2} \int_0^1 \d x \, \left(1-x\right) \,
   \left[ \ln\left( \frac{-x^2 p^2 + x p^2}{\omu^2} \right) \, + \, \left( 1 + \frac{x p^2}{\omega^2} \right)
                                   \ln\left(1 + \frac{\omega^2}{x p^2} \right) \, - 1 \, \right] \; ,
\end{equation}
where we have already applied the $\MSbar$ prescription, and after integrating in $\d x$ we find
\begin{equation}
\frac{1}{192 p^4 \omega^2 \pi^2} \, \left\{
      p^4 \left(p^2 +3 \omega^2\right) \ln\left( \frac{\omega^2}{p^2} \right)
  \,+\, \left(p^2 + \omega^2\right)^3 \ln\left( \frac{p^2+\omega^2}{\omega^2} \right)
  \,+\, p^2 \omega^2 \left[-7 p^2 - \omega^2 + 3 p^2 \ln\left(\frac{p^2}{\omu^2}\right)\right]
                                    \right\} \; .
\label{4D7}
\end{equation}
Finally, the third term, which is finite, yields
\begin{eqnarray}
& & - \frac{1}{16 \pi^2} \int_0^1 \d x \int_0^{1-x} \d y
                            \left[ x^2 p^2 \, \Theta^{-1} \right]
 \, = \, - \frac{1}{16 \pi^2 \omega^2} \int_0^1 \d x \, x^2 p^2 \,
            \left[ \ln\left( x + \frac{\omega^2}{p^2} \right) \, - \,
                                                 \ln\left(x\right) \right] \\[2mm]
&=& - \frac{1}{96 p^4 \omega^2 \pi ^2} \, \left[ \,
       p^2 \omega^2 (p^2 - 2 \omega^2) \, + \, 2 \, p^6 \ln\left( \frac{\omega^2}{p^2} \right) \,
        + \, 2 \left(p^6 + \omega^6\right) \ln\left( \frac{p^2+\omega^2}{\omega^2} \right) \, \right] \; .
\label{4D5}
\end{eqnarray}
By summing the three results above we ultimately find (in the $\MSbar$ scheme)
\begin{equation}
f(p^2,\omega^2) \, = \, \frac{1}{64 p^4 \omega^2 \pi ^2} \, \left[
                     \, f_1(p^2,\omega) \, + \, f_2(p^2,\omega^2) \, + \, f_3(p^2,\omega^2) \, \right]
\label{4D8}
\end{equation}
with
\begin{eqnarray}
f_1(p^2,\omega) & = & p^4 \left(\omega^2 - p^2\right) \ln\left(\frac{\omega^2}{p^2}\right) \; , \\[2mm]
f_2(p^2,\omega) & = & - \, \left(p^6 - p^4 \omega^2 + 3 \, p^2 \omega^4 + \omega^6 \right)
                               \ln\left(\frac{p^2 + \omega^2}{\omega^2}\right)  \; , \\[2mm]
f_3(p^2,\omega) & = & p^2 \omega^2 \left[ 5 \, p^2 + \omega^2 + p^2 \, \ln\left(\frac{p^2}{\omu^2}\right)
                         - 4 \, p^2 \, \ln\left(\frac{p^2 + \omega^2}{\omu^2}\right) \right] \; .
\label{eq:f3}
\end{eqnarray}
As shown in Ref.\ \cite{Cucchieri:2011ig,Cucchieri:2012gb}, in the $4d$ case the fit of the gluon-propagator data is
done using the expression (\ref{eq:f4dgluon}). Thus, in order to use the above result \eqref{4D8}--\eqref{eq:f3},
we need to write the gluon propagator as
\begin{equation}\label{4D3}
\mathcal{D}(p^2) \, = \, \frac{\alpha_+}{p^2+\omega_{+}^2} + \frac{\alpha_-}{p^2+\omega_{-}^2} \; ,
\end{equation}
where $\omega_{\pm}$ are the roots of the quadratic equation, with respect to the variable $p^2$, obtained
by setting equal to zero the denominator of Eq.\ (\ref{eq:f4dgluon}). Then, the ghost form-factor in the
$\MSbar$ scheme is given by
\begin{equation}
\sigma^{\overline{\mbox{\tiny{MS}}}}(p^2) \, = \, g^2 N_c \,
                \left[ \, \alpha_+ \, f(p^2,\omega_+^2) \, + \, \alpha_- \, f(p^2,\omega_-^2) \,\right]
               \label{eq:sigma4d}
\end{equation}
and we have
\begin{equation}\label{4D11}
\mathcal{G}^{\overline{\mbox{\tiny{MS}}}}(p^2) \, = \, \frac{1}{p^2}
                  \, \left[ \, 1 \, - \, \sigma^{\overline{\mbox{\tiny{MS}}}}(p^2) \, \right]^{-1} \; .
\end{equation}
Note that the function $\sigma^{\overline{\mbox{\tiny{MS}}}}(p^2)$ is real. From \cite{Cucchieri:2011ig,
Cucchieri:2012gb} we know that $\omega_{\pm}^2$ are complex-conjugate roots, i.e.\
$\omega_{-}^2 = (\omega_{+}^2)^*$ and $\alpha_- = \alpha_+^*$. By writing
$\alpha_{\pm} = a \pm i b $ and $\omega_{\pm}^2 = v \pm i w $ we find
\begin{equation}
\sigma^{\overline{\mbox{\tiny{MS}}}}(p^2) \, = \, \frac{g^2 N_c}{32 \pi^2 R^2}
            \left[ -p^2 t_1(p^2) \, + \, R^2 t_2(p^2) \, + \,
                                p^{-2} t_3(p^2) \, - \, p^{-4} t_4(p^2) \right] \label{eq:sigma4d2}
\end{equation}
with
\begin{eqnarray}
t_1(p^2) & = & (a v+b w) [\ell_2(p^2)+\ell_3(p^2)] \,-\, (b v-a w) [a_1(p^2)-a_2(p^2)] \; , \\[2mm]
t_2(p^2) & = & a [5+\ell_1(p^2)+\ell_2(p^2)+\ell_3(p^2)-4 \ell_4(p^2)]
                                             \, - \, b[a_1(p^2)-a_2(p^2) - 4 a_3(p^2)] \; , \\[2mm]
t_3(p^2) & = & [1-3 \ell_3(p^2)] (a v^3-b w v^2+v a w^2-b w^3)
                                  \,- \, 3 a_2(p^2) (b v^3+a w v^2+v b w^2+a w^3) \; , \\[2mm]
t_4(p^2) & = & \ell_3(p^2) (a v^4-2 w b v^3-2 v b w^3-a w^4)
                                  \, + \, a_2(p^2) (b v^4+2 a w v^3+2 v a w^3-b w^4)
\end{eqnarray}
and
\begin{eqnarray}
\ell_1(p^2) & = & \ln\left(\frac{p^2}{\omu^2}\right) \; , \\
\ell_2(p^2) & = & \ln\left(\frac{R}{p^2}\right) \; , \\
\ell_3(p^2) & = & \ln\left( \frac{\sqrt{R^2 p^4+R^4+2 v R^2 p^2}}{R^2} \right) \; , \label{eq:l2} \\
\ell_4(p^2) & = & \ln\left( \frac{\sqrt{p^4+2 v p^2+R^2}}{\omu^2}\right) \; , \\
a_1(p^2)    & = & \arctan\left( \frac{w}{v} \right) \; , \\
a_2(p^2) & = & \arctan\left( \frac{w p^2}{R^2+v p^2}\right) \; , \label{eq:a2} \\
a_3(p^2) & = & \arctan\left( \frac{w}{v+p^2}\right) \; , \label{eq:a3} \\
R       & = & \sqrt{v^2+w^2} \label{eq:r} \; .
\end{eqnarray}
Also note that, at large momenta, one gets
\begin{equation}
\sigma^{\overline{\mbox{\tiny{MS}}}}(p^2) \approx - \frac{3 \, a \, g^2 \, N_c}{32 \, \pi^2}
                          \, \ln\left(\frac{p^2}{\omu^2}\right) \; .
\end{equation}
Finally, by expanding $\sigma^{\overline{\mbox{\tiny{MS}}}}(p^2)$ around $p^2 = 0$
in Eqs.\ (\ref{eq:sigma4d2})--(\ref{eq:r}) we obtain
\begin{eqnarray}
\frac{\sigma^{\overline{\mbox{\tiny{MS}}}}(p^2)}{g^2 N_c} & = &
                          - \frac{6\,a\, \ln\left( \frac{R}{\omu^2} \right)
                         -6\,b\, \arctan\left( \frac{w}{v} \right) -5\,a}{64\, \pi^2} \nonumber \\[2mm]
           &    & \quad + \frac{\left[ -11 + 6 \ln\left( \frac{p^2}{R} \right) \right]
                         \,\left(a\,v+w\,b \right) + 6 (b\,v-a\,w) \, \arctan\left( {\frac{w}{v}} \right)
                         }{192\, \pi^2 \, R^2} \, p^2 \, + \, O \left( {p}^{4} \right) \; .
\label{eq:sigmair}
\end{eqnarray}
Thus, if $\sigma^{\overline{\mbox{\tiny{MS}}}}(0)=1$ we have that $\mathcal{G}^{\overline{\mbox{\tiny{MS}}}}(p^2)
\sim 1/p^4$ at small momenta (plus logarithmic corrections). Clearly, also in $4d$, we obtain a one-parameter
family of behaviors, labelled by the value of $g^2$, and the IR-enhanced ghost propagator corresponds to the
upper value of $g^2$ allowed by the no-pole condition (\ref{eq:nopole}). With the numerical values reported in
the second column of Table IV of Ref.\ \cite{Cucchieri:2011ig} and $N_c=2$ we find\footnote{Again, the error in
parentheses have been evaluated using a Monte Carlo analysis with 10000 samples.}
\begin{equation}
\frac{\sigma^{\overline{\mbox{\tiny{MS}}}}(p^2)}{2 \, g^2} \, = \, 0.0240(0.0007) \, +\,
                        \left[ -0.0082(0.0003) \, + \, 0.0060(0.0002) \ln\left(p^2\right) \right] \, p^2
\label{eq:sigmq4d}
\end{equation}
and the condition $\sigma^{\overline{\mbox{\tiny{MS}}}}(0)=1$ corresponds\footnote{Clearly, different
renormalization schemes will modify the constant term in Eq.\ (\ref{eq:sigmq4d}) and the value
of $g^2_c$.} to $g^2_c \approx 20.83$. Note again the negative sign of the leading order corrections at
small momenta (see Section \ref{sec:d-exact}).

%%%%%%%%%%%%%%%%%%%%%%%%%%%%%%%%%%%%%%%%%%%%%%%%%%%%%%%%%%%%%%%%%%%%%%%%%%%%%%%%%%%%%%%%%%%%%%%%%%%%%%%%%%%%%%%

\subsection{Ghost Propagator in the 2$d$ Case}
\label{sec:2Dghost}

As stressed in Section \ref{sec:yukawa} above, Ref.\ \cite{Cucchieri:2011ig} has shown that the fit of the
gluon-propagator data in the $2d$ case can be done using the expression
\begin{equation}
\mathcal{D}(p^2) \, = \,
\frac{\alpha_+ \, + \, i c p^{\eta}}{p^2 + \omega_+^2} \, + \,
\frac{\alpha_- \, - \, i c p^{\eta}}{p^2 + \omega_-^2} \; ,
\label{eq:D2D}
\end{equation}
where $c$ is real, $\alpha_- = \alpha_+^*$, $\omega_{-}^2 = (\omega_{+}^2)^*$ and $\omega_{\pm}$ are the roots
of the quadratic equation, with respect to the variable $p^2$, obtained by setting equal to zero the denominator
of Eq.\ (\ref{eq:f2dgluon}). Thus, in order to evaluate the ghost form-factor $\sigma(p^2)$ we need to consider
the function $ f(p, \omega^2, \eta) $, defined in Eq.\ (\ref{eq:deffofkwithe}) above. To this end, we can choose
again the positive $x$ direction parallel to the external momentum $p$ and consider polar coordinates. Then,
after evaluating the angular integral we find
\begin{equation}
f(p, \omega^2, \eta) \, = \, \frac{1}{4 \pi} \left[ \, \int_0^p \frac{\d q}{p^2}
                          \, \frac{q^{1+\eta}}{q^2 + \omega^2} \, + \,
                          \int_p^{\infty} \frac{\d q}{q^{1-\eta}} \, \frac{1}{q^2 + \omega^2} \, \right] \; ,
\label{eq:deffofkwithesimpl}
\end{equation}
valid both for $\eta=0$ and for $\eta > 0$.

In the case $\eta=0$ the momentum integration is straightforward giving
\begin{eqnarray}
f(p, \omega^2) &=& \lim_{\Lambda \to \infty} \, \frac{1}{4 \pi} \left[ \, \int_0^p \frac{\d q}{p^2} \,
  \frac{q}{q^2 + \omega^2} \, + \, \int_p^{\Lambda} \frac{\d q}{q} \, \frac{1}{q^2 + \omega^2} \, \right] \\[2mm]
               &=& \lim_{\Lambda \to \infty} \, \frac{1}{4 \pi} \left\{ \,
            \frac{1}{2 p^2} \, \ln\left(1 + \frac{p^2}{\omega^2}\right)
           \, + \, \frac{1}{\omega^2} \, \left[ \ln\left(\Lambda\right) \, - \, \ln\left(p\right) \, - \,
                \frac{1}{2} \ln\left(\Lambda^2 + \omega^2\right) \,+ \,
                     \frac{1}{2} \ln\left(p^2 + \omega^2\right) \right] \, \right\} \\[2mm]
               &=& \frac{1}{8 \pi} \left[ \, \frac{1}{p^2} \, \ln\left(1 + \frac{p^2}{\omega^2}\right) \, + \,
                   \frac{1}{\omega^2} \, \ln\left(1 + \frac{\omega^2}{p^2}\right) \, \right] \; .
\label{eq:fwithe0}
\end{eqnarray}
Note that the second term above blows up logarithmically in the IR limit $p \to 0$, in agreement with
the result obtained in Section \ref{sec:d=2sing}.

For $\eta > 0$ the second integral in Eq.\ (\ref{eq:deffofkwithesimpl}) can be written, after the change of variable
$t = \omega^2/(q^2 + \omega^2)$, as
\begin{equation}
 \int_p^{\infty} \frac{\d q}{q^{1-\eta}} \, \frac{1}{q^2 + \omega^2} \,=\,\frac{1}{2 (\omega^2)^{1-\eta/2}}\,
                             B\left(\frac{\omega^2}{p^2+\omega^2}; 1-\frac{\eta}{2} , \frac{\eta}{2} \right) \; ,
\end{equation}
where
\begin{equation}
B\left(x;a,b\right) \, = \, \int_0^x \, \d t \, t^{a-1} \, \left( 1 - t \right)^{b-1}
\label{eq:Betainc}
\end{equation}
is the incomplete Beta function, which is defined for $a, b > 0$ \cite{GR}, implying $2 > \eta > 0$ in our case.
For the first integral in Eq.\ (\ref{eq:deffofkwithesimpl}) we cannot use directly the changes of variable $v = 1/q$
and $t = 1/(1+\omega^2 v^2)$ because we get an incomplete Beta function (\ref{eq:Betainc}) with $b < 0$. In this case
it is convenient to introduce a Feynman parameter (using non-integer exponents) and write
\begin{eqnarray}
\int_0^p \frac{\d q}{p^2} \, \frac{q^3}{q^{2-\eta}} \, \frac{1}{q^2 + \omega^2} & = &
\frac{1}{p^2} \left(1 - \frac{\eta}{2}\right) \int_0^1 \d x \, x^{-\eta/2} \,
                      \int_0^p \frac{q^3 \, \d q}{\left[q^2 + (1-x) \omega^2\right]^{2-\eta/2}} \\[2mm]
& = &\frac{1}{2 p^2} \int_0^1 \d x \, x^{-\eta/2} \,
              \Biggl\{ \, - \frac{p^2}{\left[p^2 + (1-x) \omega^2\right]^{1-\eta/2}} \nonumber \\[2mm]
  & & \qquad \qquad + \, \frac{2}{\eta} \left[p^2 + (1-x) \omega^2\right]^{\eta/2} \, - \, \frac{2}{\eta}
          \left[(1-x) \omega^2\right]^{\eta/2} \, \Biggr\} \; ,
\end{eqnarray}
where we have also done the integration in $\d q$. After suitable changes of variables, the last formula can
be written as
\begin{eqnarray}
\!\!\!\!\!\!\!\!\!\int_0^p \frac{\d q}{p^2} \, \frac{q^3}{q^{2-\eta}} \, \frac{1}{q^2 + \omega^2} & = & - \,
  \frac{1}{2 (\omega^2)^{1-\eta/2}}\, B\left(\frac{\omega^2}{p^2+\omega^2}; 1-\frac{\eta}{2} , \frac{\eta}{2}
                        \right) \nonumber \\[2mm]
& & \quad \, + \, \frac{p^2+\omega^2}{\eta \, p^2 \, (\omega^2)^{1-\eta/2}}\,
                 B\left(\frac{\omega^2}{p^2+\omega^2}; 1-\frac{\eta}{2} , 1+\frac{\eta}{2} \right)
\, - \, \frac{1}{\eta \, p^2 \, (\omega^2)^{-\eta/2}}\,B\left(1-\frac{\eta}{2} , 1+\frac{\eta}{2} \right) \; ,
\end{eqnarray}
where $B\left(a,b\right) = B\left(1;a,b\right) = \Gamma(a) \, \Gamma(b) / \Gamma(a+b)$ is the Beta function.
Thus, by summing the two results above, we find
\begin{eqnarray}
f(p, \omega^2, \eta) & = & \frac{(\omega^2)^{\eta/2}}{4 \pi \, \eta \, p^2} \,
      \left[ \frac{p^2+\omega^2}{\omega^2} \,
             B\left(\frac{\omega^2}{p^2+\omega^2}; 1-\frac{\eta}{2} , 1+\frac{\eta}{2} \right)
                        \, - \, B\left(1-\frac{\eta}{2} , 1+\frac{\eta}{2} \right) \right] \\[2mm]
     & = & \frac{(\omega^2)^{\eta/2-1}}{4 \pi \, \eta} \, B\left(1-\frac{\eta}{2} , 1+\frac{\eta}{2} \right)
                                                                    \nonumber \\[2mm]
          & & \quad + \, \frac{(\omega^2)^{\eta/2}}{4 \pi \, \eta \, p^2} \, \frac{p^2+\omega^2}{\omega^2}
               \, \left[ B\left(\frac{\omega^2}{p^2+\omega^2}; 1-\frac{\eta}{2} , 1+\frac{\eta}{2} \right)
                   \, - \, B\left(1-\frac{\eta}{2} , 1+\frac{\eta}{2} \right) \right] \; .
\label{eq:fwithe}
\end{eqnarray}
Note that for $p=0$ the incomplete Beta function $B\left(\omega^2/(p^2+\omega^2); 1-\eta/2 , 1+\eta/2 \right)$
becomes the Beta function $B\left(1-\frac{\eta}{2} , 1+\frac{\eta}{2} \right)$. Then, by Taylor expanding
$f(p, \omega^2, \eta)$ for small momenta $p$, we obtain
\begin{equation}
f(p, \omega^2, \eta) \, = \, \frac{(\omega^2)^{\eta/2-1}}{4 \pi \, \eta} \,
                B\left(1-\frac{\eta}{2} , 1+\frac{\eta}{2} \right) \, - \,
     \frac{p^{\eta}}{4 \pi \, \eta \, (1 + \eta/2) \, \omega^2} \left[ 1 - O \left( {p}^{2} \right) \right] \; ,
\label{eq:fat0}
\end{equation}
yielding a constant contribution at $p=0$.

Using the expression (\ref{eq:D2D}), the ghost form-factor in the $2d$ case is given by
\begin{equation}
\sigma(p^2) \, = \, g^2 N_c \, \left[ \, \alpha_+ \, f(p^2,\omega_+^2) \, + \, \alpha_- \, f(p^2,\omega_-^2)
     \, + \, i c f(p^2,\omega_+^2,\eta) \, - \, i c f(p^2,\omega_-^2,\eta) \,\right] \label{eq:sigma2d} \; ,
\end{equation}
with $f(p^2,\omega^2)$ and $f(p^2,\omega^2,\eta)$ defined, respectively, in Eqs.\ (\ref{eq:fwithe0}) and
(\ref{eq:fwithe}). Of course, the function $\sigma(p^2)$ is real. By writing $\alpha_{\pm} = a \pm i b $
and $\omega_{\pm}^2 = v \pm i w $ we get for the first two terms above
\begin{eqnarray}
\alpha_+ \, f(p^2,\omega_+^2) \, + \, \alpha_- \, f(p^2,\omega_-^2) & = & \frac{1}{8 \pi} \, \left\{ \,
            \frac{1}{p^2} \, \left[ a \, \ell_3(p^2) \, + \, b \, a_2(p^2) \right] \right. \nonumber \\[2mm]
& & \left. \qquad \, + \, \frac{1}{R^2} \, \left[ \left(a v + b w\right) \ell_5(p^2)
          \, - \, \left(b v - a w\right) a_3(p^2) \right] \, \right\} \; ,
\label{eq:sigmareal2d}
\end{eqnarray}
where $\ell_3(p^2)$, $a_2(p^2)$, $a_3(p^2)$ and $R$ have already been defined in Eqs.\ (\ref{eq:l2}),
(\ref{eq:a2}), (\ref{eq:a3}) and (\ref{eq:r}) and
\begin{equation}
\ell_5(p^2) \, = \, \ln\left( \frac{\sqrt{p^4+2 v p^2+R^2}}{p^2}\right) \; .
\end{equation}
As shown in Section \ref{sec:d=2sing}, there is a logarithmic singularity $\ell_5(p^2) \sim -\ln(p^2)$ at
small momenta proportional to the gluon propagator at zero momentum, that is, $\mathcal{D}(0)=2(a v + b w)/R^2$.
We also have
\begin{equation}
i c f(p^2,\omega_+^2,\eta) \, - \, i c f(p^2,\omega_-^2,\eta)
             \, = \, - 2 \, c \, \Im \left[ \, f(p^2,\omega_+^2,\eta) \, \right]  \; ,
\label{eq:imaginaryw}
\end{equation}
where we have indicated with $\Im$ the imaginary part of the expression in square brackets.

One can easily check that $\sigma(p^2)$ is null at large momenta. Finally, the results
(\ref{eq:sigmareal2d}) and (\ref{eq:imaginaryw}), together with the expressions (\ref{eq:fat0})
and (\ref{eq:sigma2d}), allow us to evaluate the behavior of the ghost propagator at small
momenta. We obtain
\begin{eqnarray}
\frac{\sigma(p^2)}{g^2 N_c} & = & \frac{1}{8 \pi} \left\{ \, \frac{a p^2}{2 R^2}
        \, + \, \frac{a v + b w}{R^2} \left[ \, 1 \, + \, \ln\left(\frac{R}{p^2}\right) \, \right]
        \, - \, \frac{b v - a w}{R^2} \left[ \, \arctan\left( \frac{w}{v} \right) \, - \,
               \frac{w p^2}{R^2} \, \right] \, + \, O \left( {p}^{4} \right) \right\} \nonumber \\[2mm]
         & & \qquad - 2 \, c \, \Im \left[ \frac{(\omega_+^2)^{\eta/2-1}}{4 \pi \, \eta} \,
                          B\left(1-\frac{\eta}{2} , 1+\frac{\eta}{2} \right)
                \, - \, \frac{p^{\eta}}{4 \pi \, \eta \, (1 + \eta/2) \, \omega^2_+}
                               \, + \, O \left( {p}^{2+\eta} \right) \right] \\[2mm]
   & = & \frac{1}{8 \pi} \left\{ \, \frac{a p^2}{2 R^2} \, + \, \frac{a v + b w}{R^2} \left[ \, 1 \, + \,
           \ln\left(\frac{R}{p^2}\right) \, \right] \, - \, \frac{b v - a w}{R^2}
                 \left[ \, \arctan\left( \frac{w}{v} \right) \, - \, \frac{w p^2}{R^2} \, \right]
                                    \, \right\} \nonumber \\[2mm]
   & & \quad - 2 \, c \, \sin\left[ \left(\frac{\eta}{2}-1\right) \arctan\left(\frac{w}{v}\right) \right]
               \frac{R^{\eta/2-1}}{4 \pi \, \eta} \, B\left(1-\frac{\eta}{2} , 1+\frac{\eta}{2} \right)
        \, - \, \frac{2 \, c \, w\, p^{\eta}}{4 \pi \, \eta \, (1 + \eta/2) \, R^2} \, + \,
                        O \left( {p}^{2+\eta} \right) \; .
\label{eq:sigma2dexp}
\end{eqnarray}
Note that, if $\sigma(0) = 1$, one finds a ghost propagator with a behavior $1/p^{2+\eta}$ in the IR
limit. As in $3d$ and in $4d$ we have a one parameter family of solutions labelled by the value of $g^2$.

As explained in Ref.\ \cite{Cucchieri:2011ig}, the $2d$ data for the gluon propagator suggest the relations
$a = -b$ and $v = w$, implying $a v + b w = 0$ and $R^2 = 2 v^2$. Then, we find
\begin{equation}
\frac{\sigma(p^2)}{g^2 N_c} \, = \, \frac{a}{32 v} \,
       - 2 \, c \, \sin\left[ \left(\frac{\eta}{2}-1\right) \frac{\pi}{4} \right]
              \frac{(2 v^2)^{\eta/2-1}}{4 \pi \, \eta} \, B\left(1-\frac{\eta}{2} , 1+\frac{\eta}{2} \right)
       \, - \, \frac{c \, p^{\eta}}{4 \pi \, \eta \, (1 + \eta/2) \, v} \, - \,
                                \frac{a p^2}{32 \pi v^2} \, + \, O \left( {p}^{2+\eta} \right) \; .
\end{equation}
Using the approximate result $\eta \approx 1$ (see again Ref.\ \cite{Cucchieri:2011ig}) this formula simplifies to
\begin{equation}
\frac{\sigma(p^2)}{g^2 N_c} \, = \, \frac{a + 4 c \sqrt{1-1/\sqrt{2}}}{32 v}
             \, - \, \frac{c \, p}{6 \pi \, v} \, - \,
                            \frac{a p^2}{32 \pi v^2} \, + \, O \left( {p}^{2+\eta} \right) \; .
\end{equation}
On the contrary, for $N_c=2$ and with the numerical values reported in \cite{Cucchieri:2011ig} --- see the
second column of Table XIV and, for the exponent $\eta$, the last line of Table XIII --- we find for Eq.\
(\ref{eq:sigma2dexp}) the numerical results
\begin{equation}\label{152}
\frac{\sigma(p^2)}{2 \, g^2} \, \approx \, 0.029 (0.004)
                       \, - \, 0.029 (0.005) \, p^{0.909(0.049)} \, - \, 0.023 (0.004) \, p^2 \; .
\end{equation}
The coefficient $(av+bw)/R^2 \propto \mathcal{D}(0)$, multiplying the logarithmic IR singularity, is zero
within error and we have omitted the corresponding term. Note that $\sigma(p^2)$ decreases for increasing
momenta $p^2$, as proven in Section \ref{sec:2Dderivative} above.  Also note that we have $\sigma(0)=1$
for $g^2_c \approx 17.24$ and in this case the ghost propagator behaves as $ \sim 1/p^{2.9}$ in the IR limit.

%%%%%%%%%%%%%%%%%%%%%%%%%%%%%%%%%%%%%%%%%%%%%%%%%%%%%%%%%%%%%%%%%%%%%%%%%%%%%%%%%%%%%%%%%%%%%%%%%%%%%%%%%%%%%%%

\section{The Ghost Propagator Beyond Perturbation Theory}
\label{sec:DSE}

The one-loop analysis above has shown that, in the $2d$ case, an IR singularity $- \mathcal{D}(0) \,\ln(p^2)$
appears in the Gribov form-factor $\sigma(p^2)$ when $p^2 \to 0$. Thus, one needs a null gluon propagator at
zero momentum in order to satisfy the no-pole condition $\sigma(0) \leq 1$. On the contrary, for $d=3$ and 4,
we found that $\sigma(p^2)$ is finite also for $\mathcal{D}(p^2) > 0$.

In this section we improve our analysis by considering the DSE for the ghost propagator $\mathcal{G}(p^2)$
(see for example \cite{von Smekal:1997vx,Atkinson:1997tu,Aguilar:2004sw}). As stressed in the Introduction,
here we do not try to solve the ghost propagator DSE, but instead we concentrate on general properties of this
equation for different space-time dimensions. In particular, the results obtained in Section
\ref{sec:ghost-prop} are confirmed by considering a generic (sufficiently regular) gluon propagator
$\mathcal{D}(p^2)$ and an IR-finite ghost-gluon vertex $ i g f^{adc} p_{\lambda} \Gamma_{\lambda \nu}(p,q) $.

%%%%%%%%%%%%%%%%%%%%%%%%%%%%%%%%%%%%%%%%%%%%%%%%%%%%%%%%%%%%%%%%%%%%%%%%%%%%%%%%%%%%%%%%%%%%%%%%%%%%%%%%%%%%%%%

\subsection{The 2$d$ Case}

In the $2d$ Landau gauge the DSE for the ghost propagator is written as
\begin{equation}
\frac{1}{\mathcal{G}(p^2)} \, = \, p^2 \, - \, g^2 \, N_c \,
           \int \frac{\d^2 q}{(2\pi)^2} \, p_{\lambda} \, \Gamma_{\lambda \nu}(p,q) \,
              s_{\mu} \, \mathcal{D}(q^2) \, P_{\mu \nu}(q) \, \mathcal{G}(s^2) \; ,
\label{eq:GiniDSE}
\end{equation}
where $s=p-q$, the gluon and the ghost propagators --- respectively $\mathcal{D}(p^2)$ and $\mathcal{G}(p^2)$ ---
are full propagators and we indicated with $ i g f^{adc} p_{\lambda} \Gamma_{\lambda \nu}(p,q) $ the full
ghost-gluon vertex. The above result implies
\begin{equation}
\sigma(p^2) \, = \, \frac{g^2 \, N_c}{p^2} \,
           \int \frac{\d^2 q}{(2\pi)^2} \, p_{\lambda} \, \Gamma_{\lambda \nu}(p,q) \,
              s_{\nu} \, \mathcal{D}(q^2) \, P_{\mu \nu}(q) \, \frac{1}{s^2} \frac{1}{1 - \sigma(s^2)}
\label{eq:sigmafull}
\end{equation}
if one uses Eq.\ (\ref{eq:finalG}). For a tree-level ghost-gluon vertex $\Gamma_{\lambda \nu}(p,q) =
\delta_{\lambda \nu}$ and using the transversality of the gluon propagator we finally find
\begin{equation}
\sigma(p^2) \, = \, g^2 \, N_c \, \frac{p_{\mu} p_{\nu}}{p^2} \, \int \frac{\d^2 q}{(2\pi)^2} \,
              \mathcal{D}(q^2) \, P_{\mu \nu}(q) \, \frac{1}{s^2} \frac{1}{1 - \sigma(s^2)} \; ,
\label{eq:sigma2}
\end{equation}
which should be compared to the one-loop result (\ref{sigma1}). As in Section \ref{sec:2Dderivative} above,
we can choose the $x$ direction along the external momentum $p$ obtaining (using polar coordinates)
\begin{equation}
\frac{\sigma(p^2)}{g^2 N_c} \, = \, \int_0^{\infty} \frac{q \, \d q}{4 \pi^2} \, \mathcal{D}(q^2) \,
      \int_0^{2 \pi} \d\theta \, \frac{1 - \cos^2(\theta)}{s^2 \, \left[ 1 - \sigma(s^2) \right]} \; ,
\label{eq:sigmaDS}
\end{equation}
with $s^2 = p^2 + q^2 - 2\,p\,q\,\cos(\theta)$.

This equation will be analyzed below using two different approaches. A first result can, however, be easily
obtained using again the $y$-max approximation, as in Section \ref{sec:dapprox} above. This gives us
\begin{equation}
\frac{\sigma(p^2)}{g^2 N_c} \, = \, \frac{1}{8 \pi} \, \left\{ \,
        \int_0^{p^2} \, \d x \, \frac{\mathcal{D}(x)}{p^2 \, \left[ 1 - \sigma(p^2) \right]}
\, + \, \int_{p^2}^{\infty} \, \d x \, \frac{\mathcal{D}(x)}{x \, \left[ 1 - \sigma(x) \right]}
                                          \, \right\} \; ,
\end{equation}
where we have done the angular integration and set $x=q^2$. In the limit of small momenta $p^2$ we
then obtain
\begin{equation}
\frac{\sigma(p^2)}{g^2 N_c} \, = \, \frac{1}{8 \pi} \, \left\{ \,
        \lim_{p^2 \to 0} \, \frac{p^2}{2} \, \frac{\mathcal{D}(p^2)
                   + \mathcal{D}(0)}{p^2 \, \left[ 1 - \sigma(p^2) \right]}
\, + \, \int_{0}^{\infty} \, \d x \, \frac{\mathcal{D}(x)}{x \, \left[ 1 - \sigma(x) \right]}
                                          \, \right\} \; .
\end{equation}
In order to avoid IR singularities in the above equation we have to impose $\mathcal{D}(p^2) \approx
B p^{2 \eta}$, i.e.\ the gluon propagator should be null at zero momentum. In particular, if
$\sigma(0) < 1$, i.e.\ for a free-like ghost propagator at small momenta, it is sufficient to have
$\eta > 0$. On the contrary, if the ghost propagator is IR enhanced and $1 - \sigma(0) \propto x^{\kappa}$
for small $x$ with $\kappa > 0$, then the condition $\eta > \kappa$ should be satisfied. Note that the
predictions of the scaling solution \cite{Zwanziger:2001kw,Lerche:2002ep,Huber:2007kc}, i.e.\ $\eta = 0.4$
and $\kappa = 0.2$, are consistent with the above inequality.
The same results can also be obtained by setting $p^2 = 0$ directly in Eq.\ (\ref{eq:sigmaDS}). This makes
the $\theta$ integral trivial and gives
\begin{equation}
\frac{\sigma(0)}{g^2 N_c} \, = \, \int_0^{\infty} \frac{q \, \d q}{4 \pi} \,
               \frac{\mathcal{D}(q^2)}{q^2 \, \left[ 1 - \sigma(q^2) \right]} \; .
\label{eq:sigma0wrong}
\end{equation}
Note, however, that in both cases we essentially miss the logarithmic IR singularity $- \ln(p^2)$ which
is found below. In the first case this is probably related to the very crude $y$-max approximation. On the
contrary, in Eq.\ (\ref{eq:sigma0wrong}), this is due to the (improper) exchange of the $q$ integration with
the $p^2 \to 0$ limit \cite{Dudal:2008xd}.

%%%%%%%%%%%%%%%%%%%%%%%%%%%%%%%%%%%%%%%%%%%%%%%%%%%%%%%%%%%%%%%%%%%%%%%%%%%%%%%%%%%%%%%%%%%%%%%%%%%%%%%%%%%%%%%

\subsubsection{Bounds on the Gribov Form-Factor}
\label{sec:2dbounds}

Since the Gribov form-factor is non-negative, we can easily construct a lower bound for the
l.h.s.\ of Eq.\ (\ref{eq:sigmaDS}) by writing
\begin{eqnarray}
\frac{\sigma(p^2)}{g^2 N_c} & \geq & \int_0^{\infty} \frac{q \, \d q}{4 \pi^2} \, \mathcal{D}(q^2) \,
      \int_0^{2 \pi} \d\theta \, \frac{1 - \cos^2(\theta)}{s^2} \, = \, I(p^2, 1, 2, \infty)
                       \label{eq:2dbound1} \nonumber \\[2mm]
               & = & I_2(p^2, \infty) \, = \, \frac{1}{4 \pi} \left[ \,
                      \int_0^p \frac{\d q}{p^2} \, q \, \mathcal{D}(q^2)
                        \, + \, \int_p^{\infty} \frac{\d q}{q} \, \mathcal{D}(q^2) \, \right] \; ,
                       \label{eq:2dbound2}
\end{eqnarray}
where we use the definitions (\ref{eq:Igeneral-dell}), (\ref{eq:Id-ell}) and the relations
(\ref{eq:I-upper-nu1}). The last integral in the above equation has already been analyzed in Section
\ref{sec:d=2sing}, where it was shown that $I_2(p^2, \infty)$ develops an IR singularity proportional
to $-\ln(p^2)$ if $\mathcal{D}(0) \neq 0$. Thus, $\sigma(p^2)$ also is IR singular, unless
$\mathcal{D}(0)=0$.

One can also find an upper bound for $\sigma(p^2)$ and check that the IR singularity is indeed
only logarithmic. To this end we can notice that, if $\sigma(0) < 1$, one can write\footnote{Recall
that, in the $2d$ case and in the one-loop approximation, the function $\sigma(p^2)$ is decreasing as
$p^2$ increases, i.e.\ the maximum value of $\sigma(p^2)$ is obtained for $p^2=0$ (see Section
\ref{sec:2Dderivative}). However, the proof presented here can be easily modified for the case when
$\sigma(p^2) < 1$ for all momenta $p$ and the maximum value of $\sigma(p^2)$ is not attained at $p=0$.
Finally, one should note that in the DSE (\ref{eq:sigma2}) one uses explicitly Eq.\ (\ref{eq:finalG}).
Thus, when estimating the integral in Eq.\ (\ref{eq:sigmaDS}), we cannot simply impose $\sigma(p^2) <
+ \infty$ but we have to consider the stronger condition $\sigma(p^2)
\leq 1$.}
\begin{equation}
\frac{\sigma(p^2)}{g^2 N_c} \, \leq \, \int_0^{\infty} \frac{q \, \d q}{4 \pi^2} \, \mathcal{D}(q^2) \,
      \int_0^{2 \pi} \d\theta \, \frac{1 - \cos^2(\theta)}{s^2 \, \left[\,1 - \sigma(0)\,\right]}
            \, = \, \frac{I_2(p^2, \infty)}{1 - \sigma(0)} \; ,
\end{equation}
where we have also used Eqs.\ (\ref{eq:2dbound1}) and (\ref{eq:2dbound2}) above. Therefore, the upper bound
also blows up as $-\ln(p^2)$ in the IR limit. At the same time, if $\sigma(0)=1$, with $\sigma(p^2)
\approx 1 - c p^{2\kappa}$ at small momenta we find
\begin{eqnarray}
\frac{\sigma(p^2)}{g^2 N_c} & = & \int_0^{\infty} \frac{q \, \d q}{4 \pi^2} \, \mathcal{D}(q^2) \,
        \int_0^{2 \pi} \d\theta \, \frac{1 - \cos^2(\theta)}{c \, s^{2+2\kappa}} \nonumber \\[2mm]
  & & \qquad \quad + \, \int_0^{\infty} \frac{q \, \d q}{4 \pi^2} \, \mathcal{D}(q^2) \,
        \int_0^{2 \pi} \d\theta \, \frac{1 - \cos^2(\theta)}{s^2} \, \left[ \,
                \frac{1}{1 - \sigma(s^2)} \, - \, \frac{1}{c \, s^{2\kappa}} \, \right] \; .
\end{eqnarray}
Note that the quantity in square brackets in the last integral is finite at $s=0$ if the behavior of
$\sigma(p^2)$ is given by $1 - c p^{2\kappa} + {\cal O}(p^{\tau})$ with $\tau \geq 4\kappa$.
Moreover, this quantity goes to 1 at large momenta and its absolute value is clearly bounded
from above by some positive constant $M$ if $\sigma(p^2) \in [0,1]$. Hence, we have
\begin{equation}
\frac{\sigma(p^2)}{g^2 N_c} \, \leq \, \int_0^{\infty} \frac{q \, \d q}{4 \pi^2} \, \mathcal{D}(q^2) \,
       \int_0^{2 \pi} \d\theta \, \frac{1 - \cos^2(\theta)}{c s^{2+2\kappa}}
                  \, + \, M \, I_2(p^2, \infty) \, = \, \frac{1}{c} \, I(p^2, 1+\kappa, 2, \infty)
                  \, + \, M \, I_2(p^2, \infty) \; .
\label{eq:sigmaupper}
\end{equation}
For $ 1/2 > \kappa$ we can also use the upper bound in Eq.\ (\ref{eq:Igeneral-ell-upper}) and write
\begin{equation}
\frac{\sigma(p^2)}{g^2 N_c} \, \leq \, \left( \, \frac{ M'' }{c} \, + \, M \, \right)
                               \, I_2(p^2, \infty) \; ,
\label{eq:sigmaupper2}
\end{equation}
where $ M'' $ is a positive constant. Thus, we have again an IR singularity proportional to
$-\ln(p^2)$ if $\mathcal{D}(0)$ is not zero. We conclude that $\sigma(p^2)$ can be finite solely if
$\mathcal{D}(0)=0$.

Let us remark that the only hypothesis considered in this case is the IR expansion $\sigma(p^2) = 1 -
c p^{2\kappa} + {\cal O}(p^{\tau})$ with $1 > 2\kappa$ and $\tau \geq 4\kappa$. Also note that the $2d$
lattice data \cite{Maas:2007uv} show for the ghost propagator an IR behavior in good agreement with the
so-called scaling solution \cite{Zwanziger:2001kw,Lerche:2002ep,Huber:2007kc} that predicts $\kappa = 0.2$.
Thus, the condition $1 > 2\kappa$ is verified in both cases. One can also note that, by considering in Eq.\
(\ref{eq:sigmafull}) the full ghost-gluon vertex $\Gamma_{\lambda \nu}(p,q)$, instead of the tree-level one
$\delta_{\lambda \nu}$, the above results still applies for an IR-finite vertex. This hypothesis is usually
adopted in DSE studies of gluon and ghost propagators \cite{Boucaud:2011ug,Aguilar:2008xm,Pennington:2011xs,
Fischer:2008uz} and it is confirmed by lattice data \cite{Cucchieri:2004sq,Cucchieri:2006tf,Ilgenfritz:2006he,
Cucchieri:2008qm}.

%%%%%%%%%%%%%%%%%%%%%%%%%%%%%%%%%%%%%%%%%%%%%%%%%%%%%%%%%%%%%%%%%%%%%%%%%%%%%%%%%%%%%%%%%%%%%%%%%%%%%%%%%%%%%%%

\subsubsection{Analysis of the Gribov Form-Factor Using a Spectral Representation}
\label{sec:spectral}

In this section we analyze the DSE \eqref{eq:sigmaDS} in an alternative way, also avoiding the $y$-max
approximation. To this end, let us first consider the $\theta$-integral using contour integration. After
setting $z=e^{i\theta}$ we find
\begin{equation}\label{contextra1}
\int_0^{2 \pi} \d\theta \, \frac{1 - \cos^2(\theta)}{p^2 \, \left[ 1 - \sigma(p^2) \right]}
\, = \, \frac{i}{4} \oint \d z \, \frac{(z^2-1)^2}{z^2 \, (-q+kz) \, (k-qz)}
                \, \frac{1}{1 \,- \, \sigma\left[ \, (-q+kz) \, (k-qz)z^{-1} \, \right]} \; ,
\end{equation}
where the integral $\oint \d z$ is again taken on the unit circle $|z|=1$. Clearly, besides the poles at
$q=k/z$ and at $q=kz$ in the first denominator on the r.h.s.\ of the above equation, one has to consider
possible divergences in the function
\begin{equation}
f(z) \, \equiv \, \frac{1}{1 \,- \, \sigma\left[ \, (-q+kz) \, (k-qz)z^{-1} \, \right]} \; .
\end{equation}
In particular, if we assume ghost enhancement, i.e.\ $\sigma(0)=1$, then $f(z)$ is divergent at $z=q/k$ and
at $z=k/q$. Note that these divergences are not necessarily poles of the function $f(z)$. Indeed, $f(z)$ could
display a branch cut in the unit disc or one passing through it. For example, the usual $d=2$ DSE scaling solution
has $\mathcal{G}(k^2) \sim 1/(k^2)^{\nu}$ in the limit $k^2 \to 0$, where $\nu$ is a fractional number. This behavior
signals a non-analyticity for $\mathcal{G}(k^2)$ at the origin and implies a non-analyticity for the function
$f(z)$ at $z=k/q$ or at $z=q/k$. Also, since the ghost is ``massless'' we should expect that the ghost propagator
develops a branch cut along the real axis for $k^2 < 0$. Then, $z=q/k$ or $z=k/q$ would correspond to branch
points of the function $f(z)$, making quite difficult the evaluation of the contour integral in the above
expression.

In order to overcome this problem, we make the hypothesis that a spectral representation for the ghost propagator
can be introduced, i.e.\ we write\footnote{Since we are working in the $d=2$ case, the theory should be UV finite
and we do not need to consider renormalization factors here.}
\begin{equation}\label{contextra10}
\mathcal{G}(p^2) \, = \, \frac{1}{p^2} \frac{1}{1-\sigma(p^2)} \, = \, \int_0^{\infty} \, \d t \,
                           \frac{\rho(t)}{t+p^2} \; ,
\end{equation}
which reproduces the branch cut in $\mathcal{G}(k^2)$ for $k^2<0$ (see for example \cite{Dudal:2010wn}).
If we assume $\sigma(\infty)=0$ and write
\begin{equation}
\mathcal{G}(p^2) \, = \, \frac{1}{p^2} \, \int_0^{\infty} \, \d t \,
                           \frac{\rho(t)}{1 + t/p^2}
\label{eq:spectral}
\end{equation}
it is clear that the spectral density $\rho(t)$ must satisfy the normalization condition
\begin{equation}
1 = \int_0^{\infty} \d t \, \rho(t) \; .
\label{eq:normalization}
\end{equation}
Also note that the tree-level ghost propagator $\mathcal{G}(p^2) = 1/p^2$ corresponds to the spectral
density $\rho(t) = 2 \delta(t)$, where $\delta(t)$ is the Dirac delta function. This case will be used
below to recover results obtained in the one-loop analysis carried on in Sections \ref{sec:2Dderivative}
and \ref{sec:d=2sing}. In the general case, the spectral density $\rho(t)$ is proportional to the
discontinuity of the ghost propagator along the cut.\footnote{Note that, if $\mathcal{G}(k^2)$ has a
branch cut along a curve $\mathcal{C}$ in the complex plane and if it goes to zero sufficiently fast
at infinity, using Cauchy's theorem we could write down an integral relation similar to Eq.\ \eqref{contextra10}
with the variable $t$ running over the curve $-\mathcal{C}$, with $z\in-\mathcal{C} \Leftrightarrow -z\in\mathcal{C}$.
Also, possible poles can be included by adding $\delta$-functions to the spectral density $\rho(t)$ or,
equivalently, by pulling the pole terms out of the spectral integral.}

Considering Eqs.\ (\ref{eq:sigmaDS}) and (\ref{contextra10}) we can write
\begin{eqnarray}\
\int_0^{2 \pi} \d\theta \, \frac{1 - \cos^2(\theta)}{s^2 \, \left[ 1 - \sigma(s^2) \right]} & = &
  \frac{i}{4} \, \int_0^\infty \, \d t \, \rho(t) \, \oint \d z \,
     \frac{z^2 \, + \, \overline{z}^2 \, - \, 2}{-p q z^2 \, + \, \left(p^2 + q^2 + t\right) z - p q} \\[2mm]
    & = & - \frac{i}{4 p q} \, \int_0^\infty \, \d t \, \rho(t) \, \oint \frac{\d z}{z^2} \,
                      \frac{(z^2-1)^2}{z^2 \, - \, \left(p^2 + q^2 + t\right) z / \left(p q\right) + 1} \; ,
\label{contextra11}
\end{eqnarray}
where we indicated with $\overline{z}$ the complex-conjugate of $z = e^{i\theta}$. Thus, using the representation
\eqref{contextra11} we can avoid dealing directly with the integral of an unknown function along the branch cut.
In exchange, we have in our formulae an extra integration of the (also unknown) spectral density $\rho(t)$.
Nevertheless, as we will see below, the above equation will allow us to control the $p^2 \to 0$ limit [at least
in the case $\rho(t) > 0$]. To this end, let us first note that in the contour integral (\ref{contextra11})
there is a double pole at $z=0$ and there are single poles at
\begin{equation}
z_{\pm} \, = \, \frac{\left(p^2+q^2+t\right) \pm \sqrt{\left(p^2+q^2+t\right)^2
                     \, - \, 4 p^2 q^2}}{2 p q} \; .
\end{equation}
Since $p, q, t \geq 0$ we have that $p^2+q^2+t \geq 2 p q \geq 0  \,$ and one can check that the pole $z_{-}$ lies
within the unit disc while $z_+$ lies outside of it. Moreover, for $p^2+q^2+t = 2 p q$ (which implies $t=0$ and $p=q$)
the two poles coincide and we have $z_{\pm} = 1$. It is also easy to check that the residues, inside the
unit circle, for the $z$-integrand are
\begin{eqnarray}
 \mathcal{R}\text{es}_{z=0}     &=& -\frac{p^2+q^2+t}{p^2 q^2} \; , \\[2mm]
 \mathcal{R}\text{es}_{z=z_{-}} &=&  \frac{\sqrt{\left[(p+q)^2+t\right] \, \left[(p-q)^2+t\right]}}{
                           p^2 q^2} \; .
\end{eqnarray}
Then, using the residue theorem, we find
\begin{equation}\label{contextra12}
\int_0^{2 \pi} \d\theta \, \frac{1 - \cos^2(\theta)}{s^2 \, \left[ 1 - \sigma(s^2) \right]} \, = \,
  \frac{\pi}{2} \, \int_0^\infty \, \d t \, \rho(t) \,
    \frac{p^2+q^2+t \, - \, \sqrt{\left[(p+q)^2+t\right] \, \left[(p-q)^2+t\right]}}{p^2 q^2}
\end{equation}
and we can write the ghost DSE (\ref{eq:sigmaDS}) as
\begin{eqnarray}
\frac{\sigma(p^2)}{g^2 N_c} & = & \int_0^{\infty} \frac{q \, \d q}{8 \pi} \, \mathcal{D}(q^2) \,
     \int_0^\infty \, \d t \, \rho(t) \,
    \frac{p^2+q^2+t \, - \, \sqrt{\left[(p+q)^2+t\right] \, \left[(p-q)^2+t\right]}}{p^2 q^2} \\[2mm]
& = & \int_0^{\infty} \frac{\d x}{16 \pi} \, \mathcal{D}(x) \,
     \int_0^\infty \, \d t \, \rho(t) \,
    \frac{p^2+x+t \, - \, \sqrt{ t^2 \, + \, 2 t \left(p^2 + x\right) \, + \,
                         \left(p^2 - x\right)^2 }}{p^2 x}
     \label{contextra13} \; .
\end{eqnarray}
Note that, for $\rho(t) = 2 \delta(t)$ and using
\begin{equation}
\sqrt{\left[(p+q)^2\right] \, \left[(p-q)^2\right]} = \left\{ \begin{array}{ll}
                                               p^2 \, - \, q^2 \qquad & \mbox{if} \quad p^2 > q^2 \\[3mm]
                                               q^2 \, - \, p^2 \qquad & \mbox{if} \quad q^2 > p^2
                                                              \end{array} \right.
\end{equation}
we find from Eq.\ (\ref{contextra12}) the one-loop result (\ref{2}). Also note that, by Taylor expanding the
integrand at $p^2 = 0$, one finds
\begin{equation}\label{contextra14}
\frac{\sigma(0)}{g^2 N_c} \, = \, \frac{1}{8 \pi} \, \int_0^{\infty} \, \d x \, \mathcal{D}(x) \,
                    \int_0^{\infty} \, \d t \, \frac{\rho(t)}{t+x}
    \, = \, \frac{1}{8 \pi} \, \int_0^{\infty} \, \d x \, \frac{\mathcal{D}(x)}{x \, \left[ \,
                    1 - \sigma(x) \, \right]} \; ,
\end{equation}
where we used the definition (\ref{contextra10}). As shown above [see Eq.\ (\ref{eq:sigma0wrong})], this result
can also be obtained immediately by setting $p^2 = 0$ in Eq.\ \eqref{eq:sigmaDS}. However, as already pointed
out below Eq.\ (\ref{eq:sigma0wrong}) and in Ref.\ \cite{Dudal:2008xd}, one should not exchange the $q$
integration and the $p^2 \to 0$ limit. Therefore, in order to properly evaluate $\sigma(p^2)$ for small
momenta $p^2$, we write Eq.\ (\ref{contextra13}) as
\begin{eqnarray}
\frac{\sigma(0)}{g^2 N_c} & = & \lim_{p^2 \to 0} \, \int_0^{p^2} \frac{\d x}{16 \pi} \, \mathcal{D}(x) \,
     \int_0^\infty \, \d t \, \rho(t) \, \frac{p^2+x+t \, - \, \sqrt{ t^2 \, + \, 2 t \left(p^2 + x\right)
                       \, + \, \left(p^2 - x\right)^2 }}{p^2 x} \nonumber \\[2mm]
   & & \quad \, + \, \lim_{p^2 \to 0} \, \int_{p^2}^{\infty} \frac{\d x}{16 \pi} \, \mathcal{D}(x) \,
     \int_0^\infty \, \d t \, \rho(t) \, \frac{p^2+x+t \, - \, \sqrt{ t^2 \, + \, 2 t \left(p^2 + x\right)
                       \, + \, \left(p^2 - x\right)^2 }}{p^2 x} \; .
\label{nieuw1}
\end{eqnarray}
The first integral can be estimated using the the trapezoidal rule. We then obtain
\begin{eqnarray}
  &   & \lim_{p^2 \to 0} \, \int_0^{p^2} \frac{\d x}{16 \pi} \, \mathcal{D}(x) \,
     \int_0^\infty \, \d t \, \rho(t) \, \frac{p^2+x+t \, - \, \sqrt{ t^2 \, + \, 2 t \left(p^2 + x\right)
                       \, + \, \left(p^2 - x\right)^2 }}{p^2 x} \\[2mm]
  & = & \lim_{p^2 \to 0} \, \frac{p^2}{32 \pi} \, \int_0^\infty \, \d t \, \rho(t) \,
        \left[ \, \mathcal{D}(p^2) \, \frac{2 p^2+t \, - \, \sqrt{ t^2 \, + \, 4 t p^2 }}{p^4}
               \, + \, \frac{2 \, \mathcal{D}(0)}{p^2+t} \, \right] \\[2mm]
  & = & \lim_{p^2 \to 0} \, \frac{1}{32 \pi} \, \int_0^\infty \, \d t \, \rho(t) \,
        \left[ \, \mathcal{D}(p^2) \, \frac{2 p^2+t \, - \, \sqrt{ t^2 \, + \, 4 t p^2 }}{p^2}
        \, \right] \, + \, \lim_{p^2 \to 0} \frac{\mathcal{D}(0)}{16 \pi \left[ \,
                       1-\sigma(p^2) \right]} \; ,
\label{nieuw2}
\end{eqnarray}
where we used again Eq.\ (\ref{contextra10}). For the second integral we define
\begin{equation}\label{nieuw4}
\mathcal{G}(x, p^2) \, = \, \frac{\mathcal{D}(x)}{16 \pi} \, \int_0^\infty \d t \, \rho(t) \,
        \frac{p^2+x+t \, - \, \sqrt{ t^2 \, + \, 2 t \left(p^2 + x\right)
                       \, + \, \left(p^2 - x\right)^2 }}{p^2}
\end{equation}
and find
\begin{eqnarray}
\int_{p^2}^{\infty}\, \d x \, \frac{\mathcal{G}(x, p^2)}{x} & = &
    \left. \ln(x) \, \mathcal{G}(x,p^2) \, \right|_{p^2}^{\infty} \, - \,
    \int_{p^2}^{\infty} \, \d x \, \ln(x) \, \mathcal{G}'(x, p^2) \\[2mm]
    & = & - \ln(p^2) \, \frac{\mathcal{D}(p^2)}{16 \pi} \, \int_0^\infty \d t \, \rho(t) \,
        \frac{2 p^2+t \, - \, \sqrt{ t^2 \, + \, 4 t p^2 }}{p^2}
           \, - \, \int_{p^2}^{\infty} \, \d x \, \ln(x) \, \mathcal{G}'(x, p^2) \; ,
\label{nieuw3}
\end{eqnarray}
where $'$ refers to the derivative w.r.t.\ the $x$ variable and we used the fact that $\mathcal{D}(x)$
goes to zero at large $x$. Note that, in the one-loop case $\rho(t) = 2 \delta(t)$, we have
$\mathcal{G}(x, p^2) = \mathcal{D}(x) / (8 \pi)$ and Eq.\ (\ref{nieuw3}) becomes
\begin{equation}
\int_{p^2}^{\infty}\, \frac{\d x}{8 \pi} \, \frac{\mathcal{D}(x)}{x} \, = \,
      - \ln(p^2) \, \frac{\mathcal{D}(p^2)}{8 \pi}
           \, - \, \int_{p^2}^{\infty} \, \frac{\d x}{8 \pi} \, \ln(x) \, \mathcal{D}'(x) \; ,
\end{equation}
in agreement with Eqs.\ (\ref{eq:sigma-1L-ini}) and (\ref{form2}).

By collecting the above results we can therefore write
\begin{eqnarray}
\frac{\sigma(0)}{g^2 N_c} & = & \frac{1}{16 \pi} \, \lim_{p^2 \to 0} \, \Biggl\{ \,
  \frac{\mathcal{D}(0)}{1-\sigma(p^2)} \, + \,
  \mathcal{D}(p^2) \, \left[ \, \frac{1}{2} \, - \, \ln(p^2) \, \right]
  \, \int_0^\infty \, \d t \, \rho(t) \,
     \frac{2 p^2+t \, - \, \sqrt{ t^2 \, + \, 4 t p^2 }}{p^2} \nonumber \\[2mm]
  & & \qquad \qquad \qquad \qquad - \, \int_{p^2}^{\infty} \, \d x \, \ln(x)
                    \, \mathcal{G}'(x, p^2) \, \Biggr\} \; .
\label{nieuw5b}
\end{eqnarray}

We can now verify that the last integral in the above expression is finite. Indeed, we have
\begin{eqnarray}
\mathcal{G}'(x, p^2) & = &
     \frac{\mathcal{D}'(x)}{16 \pi} \, \int_0^\infty \d t \, \rho(t) \,
        \frac{p^2+x+t \, - \, \sqrt{ t^2 \, + \, 2 t \left(p^2 + x\right)
                       \, + \, \left(p^2 - x\right)^2 }}{p^2} \nonumber \\[2mm]
 & + & \frac{\mathcal{D}(x)}{16 \pi} \, \int_0^\infty \d t \, \rho(t) \, \left[ \,
       \frac{1}{p^2} \, - \, \frac{x + t - p^2}{
              p^2 \, \sqrt{ t^2 \, + \, 2 t \left(p^2 + x\right)
                            \, + \, \left(p^2 - x\right)^2 }} \, \right] \; .
\end{eqnarray}
Then, for large $x$ we find
\begin{equation}
\mathcal{G}'(x, p^2) \, \sim \,
   \frac{\mathcal{D}'(x)}{8 \pi} \, \int_0^\infty \d t \, \rho(t) \, + \,
   \frac{\mathcal{D}(x)}{8 \pi x^2} \, \int_0^\infty \d t \, \rho(t) \, t
   \, = \, \frac{1}{8 \pi} \, \left[ \, \mathcal{D}'(x) \, + \,
   \frac{\mathcal{D}(x)}{x^2} \, \int_0^\infty \d t \, \rho(t) \, t \, \right] \; ,
\end{equation}
where we used the normalization condition (\ref{eq:normalization}), while for small $x$ we have
\begin{equation}
\mathcal{G}'(x, p^2) \, \sim \,
   \frac{x \, \mathcal{D}'(x)}{8 \pi} \, \int_0^\infty \d t \, \frac{\rho(t)}{t + p^2} \, + \,
   \frac{\mathcal{D}(x)}{8 \pi} \, \int_0^\infty \d t \, \frac{\rho(t)}{t + p^2}
   \, = \, \frac{\mathcal{G}(p^2)}{8 \pi} \, \left[ \, x \, \mathcal{D}'(x) \, + \,
                 \mathcal{D}(x) \, \right] \; ,
\end{equation}
where we used the definition (\ref{eq:spectral}). Thus, the integral
$\int_{0}^{\infty} \, \d x \, \ln(x) \, \mathcal{G}'(x, p^2)$ has no IR and UV singularities if
$\mathcal{D}(x)$ and $\mathcal{D}'(x)$ goes to zero sufficiently fast when $x$ goes to zero and
to infinity. At the same time we need the integral
\begin{equation}
\int_0^\infty \d t \, \rho(t) \, t
\end{equation}
to be finite. We can also check that the integral
\begin{equation}
\int_0^\infty \, \d t \, \rho(t) \,
     \frac{2 p^2+t \, - \, \sqrt{ t^2 \, + \, 4 t p^2 }}{p^2}
\label{eq:integral}
\end{equation}
is finite and nonzero if $\rho(t)$ is non-negative.\footnote{Our results cannot be easily extended to
the general case of a spectral density $\rho(t)$ that is negative for some values of $t$. However, they
apply if one can explicitly verify that the integral in Eq.\ (\ref{eq:integral}) is indeed finite and
nonzero. Also note that we cannot simply consider the limit $p \to 0$ of the integral (\ref{eq:integral}),
since the factor multiplying $\rho(t)$  vanishes in this limit and we might erroneously conclude that the
above expression is equal to zero. Indeed, already in the tree-level case, i.e.\ for $\rho(t)=2\delta(t)$,
one finds that the integral (\ref{eq:integral}) is non-zero and equal to 2. Finally, one could also formally
expand the integrand in powers of $p^2$ leading to
\begin{equation}\label{nieuw8a}
  \frac{2 p^2+t \, - \, \sqrt{ t^2 \, + \, 4 t p^2 }}{p^2}
     \, = \, -4 \, \sum_{n=2}^{\infty} \, \left( \begin{array}{c}
                                                   1/2 \\
                                                     n \\
                                                 \end{array}
                                          \right) \, \left( \frac{4 p^2}{t} \right)^{n-1} \; ,
\end{equation}
where $\displaystyle \left( \begin{array}{c}
                                a \\
                                b \\
                           \end{array}
                     \right) = \frac{\Gamma(a+1)}{\Gamma(b+1)\Gamma(a-b+1)}$ is the usual
binomial coefficient. However, this series is not converging for all values of $t \in [0,
+ \infty)$. Moreover, for $n=2$ we have a term proportional to
\begin{equation}
\int_0^\infty \, \d t \; \frac{\rho(t)}{t}
\end{equation}
and this integral is divergent. Indeed, by (formally) setting $p^2=0$ in Eq.\
\eqref{contextra10} we find
\begin{equation}\label{nieuw8b}
\int_0^\infty \d t \; \frac{\rho(t)}{t} = \infty \; .
\end{equation}
}
To this end let us first note that
the numerator $2 p^2+t \, - \, \sqrt{ t^2 \, + \, 4 t p^2 }$ is non-negative since
$2 p^2+t \geq \sqrt{ t^2 \, + \, 4 t p^2 }$ when $t, p^2 \geq 0$. Moreover, if we define
\begin{equation}
\Phi(t,p^2) \, =  \, \frac{2 p^2 \, + \, t \, - \, \sqrt{ t^2 \, + \, 4 t p^2 }}{p^2}
\end{equation}
it is clear that $ 2 = \Phi(0,p^2) \geq \Phi(t,p^2) \geq 0$, since the quantity
$t \, - \, \sqrt{ t^2 \, + \, 4 t p^2 }$ is negative for $t > 0$. This implies
\begin{equation}
\int_0^\infty \, \d t \, \rho(t) \, \Phi(t,p^2) \, < \, 2 \int_0^\infty \, \d t \, \rho(t) \, = \, 2 \; ,
\end{equation}
where we used again the normalization condition (\ref{eq:normalization}). At the same time
we can write
\begin{equation}
\int_0^\infty \, \d t \, \rho(t) \, \Phi(t,p^2) \, = \,
\int_0^\infty \, \d t \, \frac{p^2 \, \rho(t)}{t + p^2} \, \frac{(t + p^2) \,  \Phi(t,p^2)}{p^2} \, > \,
\frac{3}{2} \, \int_0^\infty \, \d t \, \frac{p^2 \, \rho(t)}{t + p^2} \; ,
\end{equation}
where we use the fact that the function $(t + p^2) \Phi(t,p^2)/ p^2$ is positive and gets its
minimum value, equal to $3/2$, for $t/p^2 = 1/2$. Then, using the definition (\ref{eq:spectral})
and the condition $\sigma(p^2) \geq 0$, we can write
\begin{equation}
\int_0^\infty \, \d t \, \rho(t) \, \Phi(t,p^2) \, > \,
\frac{3}{2 [ 1-\sigma(p^2) ]} \, \geq \, \frac{3}{2} \; .
\end{equation}

From the above results we conclude that in Eq.\ (\ref{nieuw5b}) we have two possible IR singularities,
i.e.\ the term $\mathcal{D}(0) / [1-\sigma(p^2)]$, if $\sigma(0) = 1$, and the term proportional to
$- \mathcal{D}(p^2) \, \ln(p^2)$. In both cases we need to impose the condition $\mathcal{D}(0)=0$ in
order to avoid the singularity. Thus, we find again that a massive gluon propagator in the $d=2$ case is
not compatible with the restriction of the functional integration to the first Gribov region.

%%%%%%%%%%%%%%%%%%%%%%%%%%%%%%%%%%%%%%%%%%%%%%%%%%%%%%%%%%%%%%%%%%%%%%%%%%%%%%%%%%%%%%%%%%%%%%%%%%%%%%%%%%%%%%%

\subsection{The 3$d$ Case}

In the $3d$ case we expect no UV divergences when using dimensional regularization\footnote{As shown in
Section \ref{sec:3doneloop} above at one-loop level, the evaluation of the ghost propagator in $3d$
usually involves Gamma functions with half-integer arguments, which do not generate infinities. Indeed, for
nonnegative values of $n$ with $n$ integer, one has \cite{GR} $\Gamma\left(n+1/2\right) = \sqrt{\pi} \,
2^{-n} \, \left(2 n - 1\right)!! \,$ and $\Gamma\left(-n+1/2\right) = \left(-2\right)^n \, \sqrt{\pi} /
\left(2 n - 1\right)!! \,$, where $n!!$ denotes the double factorial.} and the DSE for the Gribov form-factor
is simply
\begin{eqnarray}
\frac{\sigma(p^2)}{g^2 N_c} & = & \frac{1}{p^2} \, \int \frac{\d^3 q}{(2\pi)^3} \, p_{\lambda} \,
             \Gamma_{\lambda \nu}(p,q) \, s_{\nu} \, \mathcal{D}(q^2) \, P_{\mu \nu}(q) \,
                \frac{1}{s^2} \frac{1}{1 - \sigma(s^2)} \label{eq:DSE3Dfull} \\[2mm]
                  & = & \int_0^{\infty} \, \d q \, \frac{q^2}{(2\pi)^3} \, \mathcal{D}(q^2)
                    \, \int \, \d\Omega_3 \, \frac{1 - \cos^2(\phi_1)}{s^2 \, \left[
                                            \, 1 \, - \, \sigma(s^2) \, \right]} \; ,
\label{eq:DSE3D}
\end{eqnarray}
where we used the tree-level ghost-gluon vertex $\Gamma_{\lambda \nu}(p,q) = \delta_{\lambda \nu}$
and $s^2 = p^2 + q^2 - 2\,p\,q\,\cos(\phi_1)$. We can now work as in Section \ref{sec:2dbounds} and
use the results of Appendix \ref{sec:hyper}. In this way we obtain the upper bounds
\begin{equation}
\frac{\sigma(p^2)}{g^2 N_c} \, \leq \, \frac{I_3(p^2, \infty)}{1 \, - \, \sigma(0)} \; ,
\label{eq:sigmaupper3d0}
\end{equation}
if $\sigma(p^2) \leq \sigma(0) < 1$, and
\begin{equation}
\frac{\sigma(p^2)}{g^2 N_c} \, \leq \,  \left( \, \frac{ M'' }{c} \, + \, M \, \right)
                               \, I_3(p^2, \infty) \; ,
\label{eq:sigmaupper3d}
\end{equation}
if $\sigma(p^2) \leq \sigma(0) = 1$ with $ \sigma(p^2) \approx 1 - c p^{2\kappa} + {\cal O}(p^{\tau})$
for small $p^2$. In the latter case we also need the conditions $1 > \kappa$ and $\tau \geq 4\kappa$.
As we saw in Eq.\ (\ref{eq:limitI}), under simple assumptions for the gluon propagator $\mathcal{D}(q^2)$,
the integral $I_3(p^2, \infty)$ is finite in the IR limit $p \to 0$. Thus, in both cases the upper bound
of $\sigma(p^2)$ is also finite\footnote{Using the fact that $\sigma(p^2)$ is nonnegative and Eqs.\
(\ref{eq:Igeneral-dell}) and (\ref{eq:I-upper-nu1}), the lower bound
\begin{equation}
\frac{3}{4} \, I_3(p^2, \infty) \, \leq \, I(p^2, 1, 3, \infty) \, \leq \, \frac{\sigma(p^2)}{g^2 N_c}
\label{eq:sigmaupper3d0ini}
\end{equation}
clearly applies. However, since $\sigma(p^2)$ is finite, this bound does not add any relevant information
to our analysis. Note that this observation can be made also for the $4d$ case described in Section
\ref{sec:4DDSE}.} and, in order to have a finite value for $\sigma(0)$ in the $3d$ case we do not need
to set $\mathcal{D}(0)=0$. This result also applies when an IR-finite ghost-gluon vertex is included in
the ghost DSE (\ref{eq:DSE3Dfull}). Let us also note that the scaling solution predicts in the 3d case
\cite{Zwanziger:2001kw,Lerche:2002ep,Huber:2007kc} a value $\kappa \approx 0.4$ for which the condition
$1 > \kappa > 0$ is satisfied.

%%%%%%%%%%%%%%%%%%%%%%%%%%%%%%%%%%%%%%%%%%%%%%%%%%%%%%%%%%%%%%%%%%%%%%%%%%%%%%%%%%%%%%%%%%%%%%%%%%%%%%%%%%%%%%%

\subsection{The 4$d$ Case}
\label{sec:4DDSE}

In $4d$ the DSE for $\sigma(p^2)$ is given by (see for example \cite{Atkinson:1997tu})
\begin{equation}\label{eq:DSE4D}
\sigma(p^2)  \, = \, 1 \, - \, \widetilde{\mathcal{Z}}_3 \, + \, \widetilde{\mathcal{Z}}_1 \, g^2 N_c
     \int_0^{\infty} \, \d q \, \frac{q^3}{(2\pi)^4} \, \mathcal{D}(q^2) \, \int \, \d\Omega_4 \,
                     \frac{1 - \cos^2(\phi_1)}{ s^2 \, \left[ \, 1 \, - \, \sigma(s^2) \, \right]} \; ,
\end{equation}
where $\widetilde{\mathcal{Z}}_3$ and $\widetilde{\mathcal{Z}}_1$ are the renormalization constants for the
ghost propagator and the ghost-gluon vertex respectively\footnote{Note that we are again considering a
tree-level ghost-gluon vertex $\Gamma_{\lambda \nu}(p,q) = \delta_{\lambda \nu}$.} and
$s^2 = p^2 + q^2 - 2\,p\,q\,\cos(\phi_1)$. In order to eliminate these constants from the expression
for $\sigma(p^2)$ we can subtract\footnote{Again, this corresponds to a MOM scheme with the condition
$\mathcal{G}(\mu^2) = 1/\mu^2$.} the same equation for some fixed value $p^2 = \mu^2$ and set
$\widetilde{\mathcal{Z}}_1=1$, using the non-renormalization of the ghost-gluon
vertex in Landau gauge \cite{Taylor:1971ff}. This gives
\begin{eqnarray}
\frac{\sigma(p^2)}{g^2 N_c} & = & \frac{\sigma(\mu^2)}{g^2 N_c} \, + \,
               \int_0^{\infty} \, \d q \, \frac{q^3}{(2\pi)^4} \, \mathcal{D}(q^2) \, \int \, \d\Omega_4 \,
             \frac{1 - \cos^2(\phi_1)}{ s^2 \, \left[ \, 1 \, - \, \sigma(s^2) \, \right]} \nonumber \\[2mm]
       & &  \qquad \qquad \; - \, \int_0^{\infty} \, \d q \, \frac{q^3}{(2\pi)^4} \, \mathcal{D}(q^2) \, \int \,
            \d\Omega_4 \, \frac{1 - \cos^2(\phi_1)}{ t^2 \, \left[ \, 1 \, - \, \sigma(t^2) \, \right]} \; ,
\label{eq:DSE4Dbis}
\end{eqnarray}
with $t^2 = \mu^2 + q^2 - 2\,\mu\,q\,\cos(\phi_1)$. Clearly, considering $\mathcal{D}(q^2) \sim 1/q^2$ at
large momenta, the UV (logarithmic) divergence of the two integrals cancels out. In order to derive upper
bounds for the above expression without spoiling the cancellation of UV divergences and since we are
interested in the IR limit $p \to 0$, we can consider a momentum scale $\ell$ sufficiently large and write
\begin{eqnarray}
\frac{\sigma(p^2)}{g^2 N_c} & = & \frac{\sigma(\mu^2)}{g^2 N_c} \, + \,
               \int_0^{\ell} \, \d q \, \frac{q^3}{(2\pi)^4} \, \mathcal{D}(q^2) \, \int \, \d\Omega_4 \,
             \frac{1 - \cos^2(\phi_1)}{ s^2 \, \left[ \, 1 \, - \, \sigma(s^2) \, \right]} \nonumber \\[2mm]
       & &  \;\; + \, \int_{\ell}^{\infty} \, \d q \, \frac{q^3}{(2\pi)^4} \, \mathcal{D}(q^2) \, \int \,
            \d\Omega_4 \, \frac{1 - \cos^2(\phi_1)}{ s^2 \, \left[ \, 1 \, - \, \sigma(s^2) \, \right]}
           \, - \, \int_0^{\infty} \, \d q \, \frac{q^3}{(2\pi)^4} \, \mathcal{D}(q^2) \, \int \,
            \d\Omega_4 \, \frac{1 - \cos^2(\phi_1)}{ t^2 \, \left[ \, 1 \, - \, \sigma(t^2) \, \right]} \; .
\label{eq:DSE4Dter}
\end{eqnarray}
For small momenta $p$ only the first integral on the r.h.s.\ of the above equation can produce an IR
singularity. Following the analysis presented in the 3d case above we can then write
\begin{equation}
\int_0^{\ell} \, \d q \, \frac{q^3}{(2\pi)^4} \, \mathcal{D}(q^2) \, \int \, \d\Omega_4 \,
             \frac{1 - \cos^2(\phi_1)}{ s^2 \, \left[ \, 1 \, - \, \sigma(s^2) \, \right]} \, \leq \,
     \frac{I_4(p^2, \ell)}{1 \, - \, \sigma(0)} \; ,
\end{equation}
if $\sigma(p^2) \leq \sigma(0) < 1$, and
\begin{equation}
\int_0^{\ell} \, \d q \, \frac{q^3}{(2\pi)^4} \, \mathcal{D}(q^2) \, \int \, \d\Omega_4 \,
             \frac{1 - \cos^2(\phi_1)}{ s^2 \, \left[ \, 1 \, - \, \sigma(s^2) \, \right]} \, \leq \,
  \left( \, \frac{ M'' }{c} \, + \, M \, \right) \, I_4(p^2, \ell) \; ,
\end{equation}
if $\sigma(p^2) = 1 - c p^{2\kappa} + {\cal O}(p^{\tau})$ with $\tau \geq 4\kappa$ and $3/2 > \kappa$.
Again, thanks to the result (\ref{eq:limitI}), both upper bounds are finite in the IR limit
$p \to 0$ also for $\mathcal{D}(0) > 0$.

An alternative proof can be given by working directly with Eq.\ (\ref{eq:DSE4D}) and using dimensional
regularization, i.e.\ considering a dimension $d=4-\varepsilon$. In this case we can write
\begin{equation}
\sigma(p^2)  \, = \, 1 \, - \, \widetilde{\mathcal{Z}}_3
                        \, + \, \widetilde{\mathcal{Z}}_1 \, \sigma_d(p^2)
\end{equation}
with\footnote{Of course, with $d=4-\varepsilon$ and $\varepsilon > 0$, the integral in Eq.\
(\ref{eq:sigmad}) is no longer dimensionless. To keep the dimensionality correct we should, as
always, scale out a dimensional factor $m^{4-d}$ where $m$ is a mass scale, which could then be
combined with the coupling constant $g^2$, making $\sigma(p^2)$ dimensionless also for
$\varepsilon > 0$. This is important when evaluating the $\varepsilon$ expansion in order to
single out UV divergencies. Since here we are mainly interested in the IR behavior of
$\sigma(p^2)$, we do not keep track explicitly of all the terms depending on $\varepsilon$ and
we simply consider the coupling $g^2$ dimensionful.}
\begin{equation}
\frac{\sigma_d(p^2)}{g^2 N_c} \, = \, \int_0^{\infty} \, \d q \, \frac{q^{d-1}}{(2\pi)^d} \,
                   \mathcal{D}(q^2) \, \int \, \d\Omega_d \, \frac{1 - \cos^2(\phi_1)}{
                          s^2 \, \left[ \, 1 \, - \, \sigma_d(s^2) \, \right]} \; .
\label{eq:sigmad}
\end{equation}
Then, if we can show that no IR singularities occur for $2 < d \leq 4$, the UV infinity that appears
for $d \to 4$ is taken care of by the renormalization factors. In order to show that $\sigma_d(p^2)$
is IR finite we can work as done above and write
\begin{equation}
\frac{\sigma_d(p^2)}{g^2 N_c} \, \leq \, \frac{I_4(p^2, \infty)}{1 \, - \, \sigma(0)} \; ,
\label{eq:sigmaupper4d0}
\end{equation}
or
\begin{equation}
\frac{\sigma_d(p^2)}{g^2 N_c} \, \leq \, \left( \, \frac{ M'' }{c} \, + \, M \, \right) \,
                         I_4(p^2, \ell) \; ,
\label{eq:sigmaupper4d}
\end{equation}
depending on the value of $\sigma_d(0)$. In the latter case we considered again the IR expansion
$ \sigma_d(p^2) \approx 1 - c p^{2\kappa} + {\cal O}(p^{\tau})$ and the conditions $\tau \geq 4\kappa$
and $3/2 > \kappa$ . We conclude that also in the $4d$ case, $\sigma(0)$ is finite if $\mathcal{D}(0)$
is also finite (but not necessarily null).

\vskip 3mm
Let us remark that the IR exponent usually obtained in the scaling solution \cite{Zwanziger:2001kw,
Lerche:2002ep,Huber:2007kc} is $\kappa \approx 0.6$ in the $4d$ case, i.e.\ the condition $\kappa <
3/2$ is satisfied. Also note that when $(d-1)/2 \leq \kappa$ the hypergeometric function
$_2F_1\left(1+\kappa, 1+\kappa-d/2; 1+d/2; z\right)$ is not convergent at $z=1$ and we cannot use
the above proofs in order to derive properties of the Gribov form-factor. However, these large values
of $\kappa$ imply for the ghost propagator $\mathcal{G}(p^2)$ a very strong IR enhancement with a
behavior at least as singular as $1/k^{5}$ in $4d$ and at least as singular as $1/k^{4}$ for $d=3$.

%%%%%%%%%%%%%%%%%%%%%%%%%%%%%%%%%%%%%%%%%%%%%%%%%%%%%%%%%%%%%%%%%%%%%%%%%%%%%%%%%%%%%%%%%%%%%%%%%%%%%%%%%%%%%%%

\section{Conclusion}
\label{sec:concl}

Summarizing, in this manuscript we have considered general properties of the Landau-gauge Gribov ghost
form-factor $\sigma(p^2)$ for SU($N_c$) Euclidean Yang-Mills theories in $d \geq 2$
space-time dimensions. This form-factor is in a one-to-one correspondence with the ghost propagator
$\mathcal{G}(p^2)$ via Eq.\ (\ref{eq:Gsigma}). Also, $\sigma(p^2)$ is bounded by 1 if the no-pole
condition (\ref{eq:nopole}) is imposed, i.e.\ if one restricts the functional integration to the
first Gribov region $\Omega$. The main result of this work is an exact proof of qualitatively different
behavior of $\sigma(p^2)$ for $d=3,4$ with respect to $d=2$. In particular, for $d=2$, the gluon propagator
$\mathcal{D}(p^2)$ needs to vanish at zero momentum in order to avoid in $\sigma(p^2)$ an IR
singularity proportional to $- \mathcal{D}(0) \,\ln(p^2)$. On the contrary, for $d=3$ and 4, an IR-finite
ghost form-factor $\sigma(p^2)$ is obtained also when $\mathcal{D}(0) > 0$. These results were proven, in
Section \ref{sec:ghost-prop}, using perturbation theory at one loop and, in Section \ref{sec:DSE}, by
considering the DSE for the ghost propagator. Let us stress again that in DSE studies of correlation
functions in minimal Landau gauge, besides using the no-pole condition, a specific boundary condition
is usually imposed on the Gribov ghost form-factor at zero momentum. Here, instead, we have tried to
prove general properties of the Gribov ghost form-factor $\sigma(p^2)$ when the restriction to
the first Gribov horizon is considered.

At the same time, in Section \ref{sec:sigma234}, we have presented closed analytic expressions for the
Gribov form-factor $\sigma(p^2)$ at one loop, considering for the gluon propagator linear combinations
of Yukawa-like propagators (with real and/or complex-conjugate poles). These functional forms, briefly
described in Eqs.\ (\ref{eq:f2dgluon})--(\ref{eq:f4dgluon}), were recently used to fit lattice data of
the gluon propagator in the SU(2) case \cite{Cucchieri:2011ig,Cucchieri:2012gb}. The expressions obtained for
$\sigma(p^2)$ confirm the results presented in Section \ref{sec:ghost-prop}. These expressions also show that,
for the ghost propagator $\mathcal{G}(p^2)$, there is a one-parameter family of behaviors \cite{Boucaud:2008ji,
arXiv:1103.0904} labelled by the coupling constant $g^2$, when it is considered as a free parameter. The
no-pole condition $\sigma(0) \leq 1$ then implies $g^2 \leq g^2_c$, where $g^2_c$ is a ``critical'' value.
For $g^2$ smaller than $g^2_c$ one has $\sigma(0) < 1$ and the ghost propagator is a massive one.
On the contrary, at the ``critical'' value $g^2_c$, i.e.\ for $\sigma(0) = 1$, one finds an IR-enhanced
ghost propagator. As stressed in the Introduction, the physical value of the coupling is expected to
select the actual value of $\sigma(0)$. Present results \cite{Boucaud:2008ji,arXiv:1103.0904} give
$\sigma(0) < 1$ in the four-dimensional SU(3) case.

Our findings imply that a massive gluon propagator cannot be obtained in the two-dimensional case,
in disagreement with some of the results presented in Ref.\ \cite{Huber:2007kc} (see their Table
2).\footnote{A massive solution in the $2d$ case was also obtained in Ref.\ \cite{Zwanziger:2001kw}.
On the other hand, the author of \cite{Zwanziger:2001kw} stressed that a full understanding of the
$2d$ case would require a more detailed investigation.} A possible massive behavior for the gluon
propagator in the $2d$ case was also explicitly conjectured in Ref.\ \cite{Maas:2008ri} as a
Gribov-copy effect. However, since our $2d$ result is valid for any Gribov copy inside the first
Gribov region, we have shown that, at least for the $2d$ gluon propagator $\mathcal{D}(p^2)$ in
the minimal Landau gauge, Gribov-copy effects do not alter our conclusion for the value of the
gluon propagator at zero momentum, i.e.\ $\mathcal{D}(0)$ must vanish. This
observation also represents an explicit counterexample to the identification of the one-parameter
family of solutions for the gluon and ghost DSEs \cite{Boucaud:2008ky,Fischer:2008uz} with different
Gribov copies, as suggested in \cite{Maas:2008ri,Maas:2009se,Maas:2009ph}.

%%%%%%%%%%%%%%%%%%%%%%%%%%%%%%%%%%%%%%%%%%%%%%%%%%%%%%%%%%%%%%%%%%%%%%%%%%%%%%%%%%%%%%%%%%%%%%%%%%%%%%%%%%%%%%%

\section*{Acknowledgments}
D.~Dudal and N.~Vandersickel are supported by the Research-Foundation Flanders (FWO).
A.~Cucchieri acknowledges financial support from the Special Research Fund of Ghent
University (BOF UGent) and thanks CNPq and FAPESP for partial support. We thank T.~Mendes,
J.~Rodriguez-Quintero and S.~P.~Sorella for fruitful comments.

%%%%%%%%%%%%%%%%%%%%%%%%%%%%%%%%%%%%%%%%%%%%%%%%%%%%%%%%%%%%%%%%%%%%%%%%%%%%%%%%%%%%%%%%%%%%%%%%%%%%%%%%%%%%%%%

\appendix

%%%%%%%%%%%%%%%%%%%%%%%%%%%%%%%%%%%%%%%%%%%%%%%%%%%%%%%%%%%%%%%%%%%%%%%%%%%%%%%%%%%%%%%%%%%%%%%%%%%%%%%%%%%%%%%

\section{Angular Integration in the $d$-Dimensional Case}
\label{app:hsc}

Following Appendix B in Ref.\ \cite{Lerche:2002ep} one can easily perform, using hyperspherical coordinates,
the angular integrations necessary for our calculations. To this end let us recall that, in $d$ dimensions,
one has the following relations between Cartesian coordinates $x_i$ (with $i=1,2,\ldots,d$) and hyperspherical
coordinates $r, \phi_j$ (with $j=1,2,\ldots,d-1$):
\begin{eqnarray}
x_1 & = & r \, \cos(\phi_1) \; , \nonumber \\
x_2 & = & r \, \sin(\phi_1) \, \cos(\phi_2) \; , \nonumber \\
x_3 & = & r \, \sin(\phi_1) \, \sin(\phi_2) \, \cos(\phi_3) \; , \nonumber \\
    &   & \quad \ldots \nonumber \\
x_{d-1} & = & r \, \sin(\phi_1) \, \sin(\phi_2) \, \ldots \, \sin(\phi_{d-2}) \, \cos(\phi_{d-1}) \; , \nonumber \\
x_{d} & = & r \, \sin(\phi_1) \, \sin(\phi_2) \, \ldots \, \sin(\phi_{d-2}) \, \sin(\phi_{d-1}) \; .
\end{eqnarray}
The hyperspherical coordinates take values, respectively,
$r \in [0,\infty)$, $\phi_i \in [0,\pi]$ for $i = 1,2,\ldots,d-2$ and $\phi_{d-1} \in [0, 2\pi)$.
At the same time, the volume measure is given by
\begin{equation}
\d V \, = \, r^{d-1} \, \d r\, \d\Omega_d
\end{equation}
with
\begin{equation}
\d\Omega_d \, = \, \sin^{d-2}(\phi_1) \, \sin^{d-3}(\phi_2) \, \ldots \, \sin(\phi_{d-2})
                       \, \d\phi_1 \, \d\phi_2 \, \ldots \d\phi_{d-2} \,\d\phi_{d-1} \; .
\end{equation}
Note that, in the usual three-dimensional case, this notation correspond to $x_1=z$, $x_2=x$ and $x_3=y$.

Here we want to evaluate the integral
\begin{equation}
\int \, \d\Omega_d \, f\left( \phi_1 \right) \; .
\end{equation}
If we indicate with $\Omega_d$ the well-known result
\begin{equation}
\Omega_d \, = \, \int \, \d\Omega_d \, = \, \frac{2 \, \pi^{d/2}}{\Gamma\left(\frac{d}{2}\right)} \; ,
\label{eq:omegaD}
\end{equation}
where $\Gamma\left(x\right)$ is the Gamma function, then we have
\begin{equation}
\int \, f\left( \phi_1 \right) \, \d\Omega_d \, = \, \frac{\Omega_d}{\int_0^{\pi} \, \sin^{d-2}(\phi_1)
   \, \d\phi_1} \, \int_0^{\pi} \, \sin^{d-2}(\phi_1) \, f\left( \phi_1 \right) \, \d\phi_1 \; ,
\end{equation}
where all other angular integrations have already been evaluated.
The integral in the denominator can be written as
\begin{equation}
\int_0^{\pi} \, \sin^{d-2}(\phi_1) \, \d\phi_1 \, = \, \int_{-1}^1 \, \left( 1 - z^2 \right)^{(d-3)/2}
  \, \d z \, = \, \int_0^1 \, t^{-1/2} \, \left( 1 - t \right)^{(d-3)/2} \, \d t
          \, = \, B\left(\frac{d-1}{2}, \frac{1}{2}\right) \; ,
\end{equation}
where $B\left(a,b\right) = \Gamma(a) \, \Gamma(b) / \Gamma(a+b)$ is the Beta function.

In our calculations we consider two different functions $f\left( \phi_1 \right)$, i.e.\
\begin{equation}
f\left( \phi_1 \right) \, = \, 1 - \cos^2(\phi_1)
\end{equation}
and
\begin{equation}
f\left( \phi_1 \right) \, = \, \frac{1 - \cos^2(\phi_1)}{\left[ \,
                          p^2 \,+\, q^2 \, - \, 2 \, p \, q \, \cos(\phi_1) \, \right]^{\nu}}  \; .
\end{equation}
In the first case the integration gives
\begin{equation}
\int \, \left[ 1 - \cos^2(\phi_1) \right] \, \d\Omega_d \, = \,
   \frac{\Omega_d}{B\left(\frac{d-1}{2}, \frac{1}{2}\right)} \,
   \int_0^{\pi} \, \sin^{d-2}(\phi_1) \, \left[ 1 - \cos^2(\phi_1) \right] \, \d\phi_1
\end{equation}
and the integral in the numerator is
\begin{equation}
\int_{-1}^1 \, \left( 1 - z^2 \right)^{(d-1)/2} \, \d z \, = \,
\int_0^1 \, t^{-1/2} \, \left( 1 - t \right)^{(d-1)/2} \, \d t \, = \,
         B\left(\frac{d+1}{2}, \frac{1}{2}\right) \; .
\end{equation}
Collecting these results we find
\begin{equation}
\int \, \left[ 1 - \cos^2(\phi_1) \right] \, \d\Omega_d \, = \, \Omega_d \,
   \frac{B\left(\frac{d+1}{2}, \frac{1}{2}\right)}{B\left(\frac{d-1}{2}, \frac{1}{2}\right)}
          \, = \, \Omega_d \, \frac{d-1}{d} \; ,
\label{eq:firstint}
\end{equation}
where we used $x \, \Gamma(x) = \Gamma(x+1)$. In the second case, i.e.\ when considering the integral
\begin{equation}
\int \, \frac{1 - \cos^2(\phi_1)}{\left[ \,
               p^2 \,+\, q^2 \, - \, 2 \, p \, q \, \cos(\phi_1) \, \right]^{\nu}} \, \d\Omega_d \; ,
\end{equation}
we have
\begin{equation}
\frac{\Omega_d}{B\left(\frac{d-1}{2}, \frac{1}{2}\right)} \,
  \int_0^{\pi} \, \frac{\sin^d(\phi_1)}{\left[ \,
            p^2 \,+\, q^2 \, - \, 2 \, p \, q \, \cos(\phi_1) \, \right]^{\nu}} \, \d\phi_1 \; .
\end{equation}
We can now use the result (see for example formula $3.665.2$ in \cite{GR})
\begin{equation}
  \int_0^{\pi} \, \frac{\sin^{2\mu -1}(\theta)}{\left[ 1 \,+\, a^2 \, \pm \, 2 \, a \, \cos(\theta)
        \right]^{\nu}} \, \d\theta \, = \,
  B(\mu, 1/2) \; _2F_1\left(\nu, \nu -\mu+1/2; \mu+1/2; a^2\right) \; ,
\end{equation}
which is valid for $|a|<1$ and $\text{Re}(\mu) > 0$. Here $_2F_1(a,b;c;z)$ is the Gauss hypergeometric
function (see Appendix \ref{sec:hyper}). Then, we find
\begin{eqnarray}
\int \, \frac{1 - \cos^2(\phi_1)}{\left[ \,
         p^2 \,+\, q^2 \, - \, 2 \, p \, q \, \cos(\phi_1) \, \right]^{\nu}} \, \d\Omega_d & = &
\frac{\Omega_d}{B\left(\frac{d-1}{2}, \frac{1}{2}\right)} \, \frac{1}{q^2} \, B\left(\frac{d+1}{2},
    \frac{1}{2}\right) \; _2F_1\left(\nu, \nu-d/2; 1+d/2; p^2/q^2\right) \\[2mm]
  & = & \frac{\Omega_d}{q^2} \, \frac{d-1}{d} \; _2F_1\left(\nu, \nu-d/2; 1+d/2; p^2/q^2\right)
\label{eq:2F1-uno}
\end{eqnarray}
if $p^2 < q^2$ and
\begin{equation}
\int \, \frac{1 - \cos^2(\phi_1)}{\left[ \,
         p^2 \,+\, q^2 \, - \, 2 \, p \, q \, \cos(\phi_1) \, \right]^{\nu}} \, \d\Omega_d \, = \,
\frac{\Omega_d}{p^2} \, \frac{d-1}{d} \; _2F_1\left(\nu, \nu-d/2; 1+d/2; q^2/p^2\right)
\label{eq:2F1-due}
\end{equation}
if $q^2 < p^2$.

%%%%%%%%%%%%%%%%%%%%%%%%%%%%%%%%%%%%%%%%%%%%%%%%%%%%%%%%%%%%%%%%%%%%%%%%%%%%%%%%%%%%%%%%%%%%%%%%%%%%%%%%%%%%%%%

\section{Properties of the Gauss Hypergeometric Function}
\label{sec:hyper}

Let us recall that the Gauss hypergeometric function $_2F_1(a,b;c;z)$ is defined \cite{GR} for $|z|<1$ by
the series
\begin{equation}
_2F_1(a,b;c;z) \, = \, \sum_{n=0}^{\infty} \, \frac{(a)_n \, (b)_n}{(c)_n} \, \frac{z^n}{n!}
               \, = \, 1 \, + \, \frac{a \, b}{c} \, z \, + \, \frac{a (a+1) b (b+1)}{c (c+1)}
               \, \frac{z^2}{2} \, \ldots \; ,
\label{eq:serie}
\end{equation}
where
\begin{equation}
(a)_n = \frac{\Gamma(a+n)}{\Gamma(a)}
\label{eq:a-n}
\end{equation}
is a so-called Pochhammer symbol. This series is converging for $c \neq 0, -1, -2, \ldots$. It is also
converging for $|z|=1$ if $\Re(c-a-b) > 0$, where $\Re$ indicates the real part. When this condition is
satisfied one can use, for $z=1$, the result (see formula 9.122.1 in Ref.\ \cite{GR})
\begin{equation}
_2F_1\left(a,b;c;1\right) \, = \, \frac{\Gamma(c) \, \Gamma(c-a-b)}{
                                            \Gamma(c-b) \, \Gamma(c-a)} \; .
\label{eq:gausstheorem}
\end{equation}

\vskip 3mm
In Eqs.\ (\ref{eq:2F1-uno}) and (\ref{eq:2F1-due}) the hypergeometric function appears with $c=1+d/2$.
Therefore, the corresponding series is converging inside the unit circle for any dimension $d>0$. At
the same time we have $c-a-b = d+1-2\nu$ and $_2F_1(\nu, \nu-d/2; 1+d/2; z)$ is finite on
the unit circle for $(d+1)/2 > \nu$. Also, using the relation (see equation 9.131.1 in Ref.\ \cite{GR})
\begin{equation}
_2F_1(a,b;c;z) \, = \, (1-z)^{c-a-b} \, _2F_1(c-a,c-b;c;z)
\label{eq:2F1sing}
\end{equation}
we can write
\begin{equation}
_2F_1(\nu, \nu-d/2; 1+d/2; z) \, = \, (1-z)^{d+1-2\nu} \, _2F_1(1+d/2-\nu, 1+d-\nu; 1+d/2; z)
\end{equation}
and for $(d+1)/2 - \nu > 0$ we have that $1+d/2-\nu, 1+d-\nu > 0$, i.e.\ the hypergeometric function
$_2F_1(\nu, \nu-d/2; 1+d/2; z)$ is clearly positive for $z \in [0,1]$.

In the case $\nu=1$ the above results simplify. In particular, we have convergence of the series on
the unit circle for any dimension $d > 1$. Then Eq.\ (\ref{eq:gausstheorem}) yields
\begin{equation}
_2F_1\left(1,1-d/2;1+d/2;1\right) \, = \, \frac{\Gamma(1+d/2) \, \Gamma(d-1)}{
                                            \Gamma(d) \, \Gamma(d/2)}
                                    \, = \, \frac{d/2}{d-1} \; ,
\label{eq:gauss}
\end{equation}
which is positive. Note that for $d=2$ the above result gives $ _2F_1\left(1, 0; 2; 1\right) = 1$.
One can actually check that, when $d=2$, the function $_2F_1\left(1,1-d/2;1+d/2;z\right) \, = \,
_2F_1\left(1, 0; 2; z\right)$ is simply equal to 1 for any value $|z|\leq1$. Indeed, from the series
representation (\ref{eq:serie}) it is obvious that $_2F_1(a,b;c;z)=1$ for $b=0$ (and/or for $a=0$).
The same result can be obtained by considering Eq.\ (\ref{eq:2F1sing}) and the relation (see equation
9.121.1 in Ref.\ \cite{GR})
\begin{equation}
_2F_1(n,c;c;z) \, = \, (1-z)^{-n} \; ,
\label{eq:2F1power}
\end{equation}
yielding
\begin{equation}
_2F_1(a,0;c;z) \, = \, (1-z)^{c-a} \, _2F_1(c-a,c;c;z) \, = \, (1-z)^{c-a} \, (1-z)^{a-c} \, = \, 1 \; .
\label{eq:2F1-D2}
\end{equation}
The hypergeometric function $_2F_1\left(1,1-d/2;1+d/2;z\right)$ is actually very simple for any even
dimension $d=2n$, with $n\geq1$. Indeed, in this case we are considering the function
$_2F_1\left(1,1-n;1+n;z\right)$ and the coefficient $b$ is either zero or negative. This implies that
the series (\ref{eq:serie}) is actually a polynomial in $z$. For example, for $d=4$ we have the very
simple expression
\begin{equation}
_2F_1(1,-1;3;z) \, = \, 1 \, - \, \frac{z}{3} \; .
\label{eq:2F1-D4}
\end{equation}
Note that it is also possible to find a closed form for the hypergeometric function
$_2F_1\left(1,1-d/2;1+d/2;z\right)$ in the $3d$ case. Indeed, by considering the relations (see equations
9.137.8 and 9.137.1 in Ref.\ \cite{GR})
\begin{eqnarray}
0 \, & = & \, _2F_1(a,b;c;z) \, c \,  + \, _2F_1(a+1,b;c+1;z) \, \left(b - c\right)
             \, - \, _2F_1(a+1,b+1;c+1;z) \, b \, \left(1 - z\right) \\[2mm]
0 \, & = & \, _2F_1(a,b;c;z) \, c \, \left[ \, c - 1 - \left(2 c - a - b - 1\right) \, z \, \right]
  \, + \, _2F_1(a,b;c+1;z) \, \left(c - a\right) \, \left(c - b\right) \,  z  \nonumber \\[2mm]
  & & \qquad \qquad \qquad \qquad \qquad \qquad \qquad
         + \, _2F_1(a,b;c-1;z) \, c \, \left(c - 1\right) \, \left(z - 1\right)
\end{eqnarray}
and (see equation 9.131.1 in \cite{GR})
\begin{equation}
_2F_1(a,b;c;z) \, = \, \left(1 - z \right)^{-b} \, _2F_1\left(b,c-a;c;\frac{z}{z-1}\right)
\end{equation}
we can write
\begin{eqnarray}
_2F_1\left(1,-1/2;5/2;z\right) & = & \frac{3}{4} \, _2F_1\left(0,-1/2;3/2;z\right) \, + \,
           \frac{1 - z}{4} \, _2F_1\left(1,1/2;5/2;z\right) \\[2mm]
    & = & \frac{3}{4} \, + \, \frac{1 - z}{4} \, \frac{3 \, \left(1 - z\right)}{2 z} \,
        \left[ \, _2F_1\left(1,1/2;1/2;z\right) \, - \, _2F_1\left(1,1/2;3/2;z\right) \, \right] \\[2mm]
    & = & \frac{3}{4} \, + \, \frac{3 \, \left(1 - z\right)}{8 z} \,
        \left[ \, 1 \, - \, \left(1 - z\right) \, _2F_1\left(1,1/2;3/2;z\right) \, \right] \\[2mm]
    & = & \frac{3}{4} \, + \, \frac{3 \, \left(1 - z\right)}{8 z} \,
        \left[ \, 1 \, - \, \sqrt{1 - z} \; _2F_1\left(1/2,1/2;3/2;\frac{z}{z-1}\right) \, \right] \\[2mm]
    & = & \frac{3}{4} \, + \, \frac{3 \, \left(1 - z\right)}{8 z} \,
        \left[ \, 1 \,-\, \frac{1 - z}{\sqrt{z}} \, \text{arcsinh}\left(\sqrt{\frac{z}{1 - z}} \, \right) \,
                        \right] \; ,
\label{eq:2F1-D3}
\end{eqnarray}
where we have also used Eq.\ (\ref{eq:2F1power}), the relations $_2F_1(0,b;c;z) = 0$ and
(see 9.121.27 in \cite{GR})
\begin{equation}
_2F_1(1/2,1/2;3/2;-z^2) \, = \, \frac{\text{arcsinh} z}{z} \; .
\end{equation}
From the expression (\ref{eq:2F1-D3}) it is easy to check that $_2F_1(1,-1/2;5/2;1)=3/4$ and
that $_2F_1(1,-1/2;5/2;0)=1$, as expected.

\vskip 3mm
Using the above series (\ref{eq:serie}) one can verify that the derivative of $_2F_1(a,b;c;z)$, with
respect to the variable $z$, is given by
\begin{equation}
\frac{\partial}{\partial z} \, _2F_1(a,b;c;z) \, = \, \frac{a \, b}{c} \, _2F_1(a+1,b+1;c+1;z) \; .
\end{equation}
Thus, in the case of interest for us, we have
\begin{equation}
\frac{\partial}{\partial z} \, _2F_1(\nu, \nu-d/2; 1+d/2;z)
              \, = \, \frac{\nu \, \left(\nu-d/2\right)}{1+d/2} \,
                               _2F_1(\nu+1,\nu+1-d/2;2+d/2;z) \; .
\label{eq:2F1dernu}
\end{equation}
These results can be written as
\begin{equation}
\frac{\partial}{\partial z} \, _2F_1(\nu, \nu-d/2; 1+d/2;z)
              \, = \, \frac{\nu \, \left(\nu-d/2\right)}{1+d/2} \, \left(1 - z\right)^{d-2\nu} \,
                               _2F_1(1+d/2-\nu,1+d-\nu;2+d/2;z) \; ,
\label{eq:2F1dernubis}
\end{equation}
where we made use of Eq.\ (\ref{eq:2F1sing}). When the hypergeometric function $_2F_1(\nu, \nu-d/2; 1+d/2;z)$
is finite in the unit circle, i.e.\ for $(d+1)/2 > \nu$, it is clear that this derivative is positive for
$\nu > d/2$ and $_2F_1(\nu, \nu-d/2; 1+d/2;z)$ attains its maximum value at $z=1$, equal to
\begin{equation}
_2F_1(\nu, \nu-d/2; 1+d/2;1) \, = \, \frac{\Gamma(1+d/2) \, \Gamma(1+d-2\nu)}{
                                            \Gamma(1+d-\nu) \, \Gamma(1+d/2-\nu)} \; ,
\label{eq:2F1max}
\end{equation}
and its minimum value, equal to 1, at $z=0$.  On the contrary, the same derivative is negative when $\nu < d/2$.
In this case $_2F_1(\nu, \nu-d/2; 1+d/2;z)$ is largest at $z=0$, with $_2F_1(\nu, \nu-d/2; 1+d/2;0)=1$, and
smallest at $z=1$ with a value given by Eq.\ (\ref{eq:2F1max}). Finally, for $\nu = d/2$ the derivative is null
and the hypergeometric function $_2F_1(\nu, \nu-d/2; 1+d/2;z)$ is equal to 1 for all values of $z \in [0,1]$.
Thus, for $(d+1)/2 > \nu$ we can always write $M' < {}_2F_1(\nu, \nu-d/2; 1+d/2;z) < M'' $, for some positive
constants $M'$ and $M'' $ and with $z$ taking values in the interval $[0,1]$.

For $\nu = 1$ these results again simplify, yielding
\begin{equation}
\frac{\partial}{\partial z} \, _2F_1(1,1-d/2;1+d/2;z) \, = \, \frac{2-d}{2+d} \,
                               _2F_1(2,2-d/2;2+d/2;z) \; .
\label{eq:derivativeF}
\end{equation}
As expected, for $d=2$ this derivative is zero since $_2F_1\left(1, 0; 2; z\right)=1$. The result above
also simplifies for $d=4$, for which we find on the right-hand side of Eq.\ (\ref{eq:derivativeF}) the
value $ - \, _2F_1(2,0;4;z) / 3 = -1/3$, as already known from Eq.\ (\ref{eq:2F1-D4}). At the same time,
for $d=3$ we can write
\begin{equation}
\frac{\partial}{\partial z} \, _2F_1(1,-1/2;5/2;z) \, = \, - \frac{1}{5} \; _2F_1(2,1/2;7/2;z) \, = \,
             - \frac{1}{5} \, \left( \, 1 \, + \, \frac{2 \, z}{7} \, + \,
                                        \frac{z^2}{14} \, \ldots \, \right)
\label{eq:d3derivative2F1}
\end{equation}
and the derivative is clearly negative for any value $z \geq 0$. The same result can actually be proven for
any dimension $d$ larger than 2. Indeed, the hypergeometric function $_2F_1(2,2-d/2;2+d/2;z)$ is finite
in the unit circle for $d > 2$ and using Eq.\ (\ref{eq:2F1sing}) we can easily verify that it is also
positive for $z \in [0,1]$. Thus, the derivative in Eq.\ (\ref{eq:derivativeF}) is negative for $d > 2$ (and
$z \in [0,1]$). As a consequence, under the same hypotheses, we have that the hypergeometric function
$_2F_1(1,1-d/2;1+d/2;z)$ has its maximum value, equal to 1, at $z=0$, and its minimum value, equal to
$d/(2(d-1))$, at $z=1$ [see Eq.\ (\ref{eq:gauss})].

Using the definition (\ref{eq:a-n}) of the Pochhammer symbol $(a)_n$ it is also easy to verify that
\begin{equation}
(a)_n \, = \, a \, (a+1) \, (a+2) \, \ldots \, (a+n-1)
\end{equation}
which implies
\begin{equation}
n \, (a)_n \, = \, a \, \left[ \, (a+1)_n \, - \, (a)_n \, \right] \; .
\end{equation}
Then, using Eq.\ (\ref{eq:serie}), one can prove the relation \cite{site2}
\begin{eqnarray}
z \, \frac{\partial}{\partial z} \, _2F_1(a,b;c;z)
    & = & \sum_{n=0}^{\infty} \, \frac{(a)_n \, (b)_n}{(c)_n} \, n \, \frac{z^n}{n!} \\[2mm]
    & = & a \, \left[ \, \sum_{n=0}^{\infty} \, \frac{(a+1)_n \, (b)_n}{(c)_n} \, \frac{z^n}{n!}
                 \, - \, \sum_{n=0}^{\infty} \, \frac{(a)_n \, (b)_n}{(c)_n} \, \frac{z^n}{n!} \, \right]  \\[2mm]
   & = & a \, \left[ \, _2F_1(a+1,b;c;z) \, - \, _2F_1(a,b;c;z)\, \right] \; .
\end{eqnarray}
In particular, we can write
\begin{equation}
z \, \frac{\partial}{\partial z} \, _2F_1(1,1-d/2;1+d/2;z) \, + \, _2F_1(1,1-d/2;1+d/2;z) \, = \,
                 _2F_1(2,1-d/2;1+d/2;z) \; .
\label{eq:derivativeF2}
\end{equation}
Note that the r.h.s.\ in the above relation is finite in the unit circle for $d > 2$ and, using again
Eq.\ (\ref{eq:2F1sing}), we can verify that it is also positive for $z \in [0,1]$.

\vskip 3mm
The above results, together with the Eqs.\ (\ref{eq:2F1-uno}) and (\ref{eq:2F1-due}), allow to write a
lower and an upper bound for the integral
\begin{equation}
I(p^2,\nu,d,\ell) \, = \, \int_0^{\ell} \, \d q \, \frac{q^{d-1}}{(2\pi)^d} \, \mathcal{D}(q^2)
                               \, \int \, \d\Omega_d \, \frac{1 - \cos^2(\phi_1)}{\left[
                               p^2 + q^2 - 2\,p\,q\,\cos(\phi_1)\right]^{\nu}} \; .
\label{eq:Igeneral-dell}
\end{equation}
Indeed, after considering the angular integration, we have (for $\ell > p$)
\begin{eqnarray}
I(p^2,\nu,d,\ell) & = & \frac{\Omega_d}{(2\pi)^d} \, \frac{d-1}{d} \, \left[ \,
                    \int_0^{p} \, \d q \, q^{d-1} \, \frac{\mathcal{D}(q^2)}{p^2} \,
                               _2F_1(\nu, \nu-d/2; 1+d/2; q^2/p^2) \right. \nonumber \\[2mm]
                       & & \qquad \qquad \qquad \; \left. + \,
                    \int_p^{\ell} \, \d q \, q^{d-3} \, \mathcal{D}(q^2) \,
                                  _2F_1(\nu, \nu-d/2; 1+d/2; p^2/q^2)
                                  \, \right] \; .
\label{eq:Igeneral-dell2}
\end{eqnarray}
Then, for $1+d-2\nu > 0$ we obtain
\begin{equation}
M' \, I_d(p^2,\ell) \, \leq \, I(p^2,\nu,d,\ell) \, \leq \, M'' \, I_d(p^2,\ell)
\label{eq:Igeneral-ell-upper}
\end{equation}
with\footnote{Note that, for a gluon propagator $\mathcal{D}(q^2)$ with a behavior $1/q^2$ at large
momenta, the second integral in Eq.\ (\ref{eq:Id-ell}) is UV divergent if $d \geq 4$ and
$\ell = \infty$.}
\begin{equation}
I_d(p^2,\ell) \, = \, \frac{\Omega_d}{(2\pi)^d} \, \frac{d-1}{d} \, \left[ \,
                    \int_0^{p} \, \d q \, q^{d-1} \, \frac{\mathcal{D}(q^2)}{p^2} \, + \,
                    \int_p^{\ell} \, \d q \, q^{d-3} \, \mathcal{D}(q^2) \, \right] \; .
\label{eq:Id-ell}
\end{equation}
Also, in the limit $p \to 0$, we find (for $d > 1$)
\begin{equation}
\lim_{p \to 0} I_d(p^2,\ell) \, = \, \frac{\Omega_d}{(2\pi)^d} \, \frac{d-1}{d} \, \left[ \,
           \lim_{p \to 0} \frac{p^{d-2} \, \mathcal{D}(p^2)}{2} \, + \,
                \int_0^{\ell} \, \d q \, q^{d-3} \, \mathcal{D}(q^2) \, \right] \; ,
\label{eq:limitI}
\end{equation}
where we used the trapezoidal rule. Clearly, for $\mathcal{D}(0) > 0$, the first term is IR-finite if
$d \geq 2$ while the second term is finite for $d > 2$. Finally, note that for $\nu=1$ the condition
$1+d-2\nu > 0$ simplifies to $d>1$. At the same time, for $d \geq 2$ the inequalities
(\ref{eq:Igeneral-ell-upper}) become
\begin{equation}
\frac{d}{2 \, (d - 1)} \, I_d(p^2,\ell) \, \leq \, I(p^2,\nu,d,\ell) \, \leq \, I_d(p^2,\ell)
\label{eq:I-upper-nu1}
\end{equation}
and for $d=2$ we have $I(p^2,\nu,2,\ell)=I_2(p^2,\ell)$.

%%%%%%%%%%%%%%%%%%%%%%%%%%%%%%%%%%%%%%%%%%%%%%%%%%%%%%%%%%%%%%%%%%%%%%%%%%%%%%%%%%%%%%%%%%%%%%%%%%%%%%%%%%%%%%%

\section{Hypotheses on the $2d$ Gluon Propagator}
\label{app:relax}

In Section \ref{sec:d=2sing} we have proven, using two different approaches, that in the $2d$ case one
needs to set $\mathcal{D}(0) = 0$ in order to have $\sigma(p^2) < + \infty$. The assumptions made for
the gluon propagator were rather general. Indeed, for the first proof one needs, for small momenta $p^2$,
an expansion of the gluon propagator of the type $\mathcal{D}(p^2) \approx \mathcal{D}(0) + B \,
p^{2 \eta} + C p^{2 \xi}$, with $\xi > \eta > 0$ and $\mathcal{D}(0), B$ and $C$ finite. At the
same time, for large momenta $p^2$, we required
\begin{equation}
\lim_{p^2 \to \infty} \, \mathcal{D}(p^2) =
\lim_{p^2 \to \infty} \, \frac{\hat{D}(p^2)}{p^2} \, = \, 0 \; .
\label{eq:Dlim1}
\end{equation}
Let us recall that we are indicating with $\hat{D}(p^2)$ a primitive of $\mathcal{D}(p^2)$ and that
$\mathcal{D}'(p^2)$ is the first derivative with respect to the variable $p^2$. In Section \ref{sec:d=2sing}
above we considered for $\mathcal{D}(p^2)$ a large $p^2$ behavior of the type $1/p^2$. However, it is clear
that a weaker condition can also be used. Indeed, the behavior $\mathcal{D}(p^2) \sim 1/p^{2 \epsilon}$ with
$1 > \epsilon > 0$ also allows to satisfy the above conditions. In order to check this, one should recall that
$\mathcal{D}(p^2) \sim 1/p^{2 \epsilon}$ implies\footnote{Let us recall that, while this (Abelian theorem)
is a correct statement, the converse (also called Tauberian theorem), i.e.\ $\hat{D}(p^2) \sim p^{2-2\epsilon}$
implies $\mathcal{D}(p^2) \sim 1/p^{2 \epsilon}$, is not always true (see for example Ref.\ \cite{site}
and Section 7.3 in Ref.\ \cite{debruijn}). This is why the so-called de l'H\^opital's rule does not always
apply (see also footnote \ref{foot:tauberian}).} $\hat{D}(p^2) \sim p^{2-2 \epsilon}+\text{constant}$ and these
two asymptotic behaviors yield the limits in Eq.\ (\ref{eq:Dlim1}). Under the same hypothesis we can also verify
that the integral [see Eq.\ (\ref{eq:sigmafin2})]
\begin{equation} \label{c2}
\int_{0}^{\infty} \, \d x \, \ln\left( x^{\eta} + M \right) \,
                              \frac{\eta M \left[ \mathcal{D}(x) \, - \, \mathcal{D}(0) \right] \, - \,
                              x \, \left( x^{\eta} + M \right) \, \mathcal{D}'(x)}{x^{1+\eta}}
\end{equation}
is finite. To this end, let us first consider the integral
\begin{equation}
\int_{0}^{p^2} \, \d x \, \ln\left( x^{\eta} + M \right) \,
                              \frac{\eta M \left[ \mathcal{D}(x) \, - \, \mathcal{D}(0) \right] \, - \,
                              x \, \left( x^{\eta} + M \right) \, \mathcal{D}'(x)}{x^{1+\eta}}
\end{equation}
with $0<p$ and $p$ small. In the limit $x \to 0$ the integrand behaves as
\begin{equation}
\ln(M) \, \left[ \, M C \, \left(\eta - \xi\right) \, x^{\xi - \eta - 1}
    \, - \, B \, \eta \, x^{\eta - 1} \, \right]
\end{equation}
and no singularity arises in the integration from $x=0$ to $x=p^2$ (if $\xi > \eta > 0$).  At the same time, for
large $x$ the same integrand behaves as
\begin{equation}
\ln(x) \, \left[ \, \frac{\eta M \mathcal{D}(0)}{x^{1+\eta}} \, - \, \mathcal{D}'(x) \, \right] \; .
\end{equation}
Thus, for sufficiently large $\ell^2$, the integral
\begin{equation}
\int_{\ell^2}^{\infty} \, \d x \, \ln\left( x^{\eta} + M \right) \,
                              \frac{\eta M \left[ \mathcal{D}(x) \, - \, \mathcal{D}(0) \right] \, - \,
                              x \, \left( x^{\eta} + M \right) \, \mathcal{D}'(x)}{x^{1+\eta}}
\end{equation}
can be approximated by
\begin{equation}
I_a(\ell^2) \, = \,\int_{\ell^2}^{\infty} \, \d x \, \ln(x)
  \, \left[ \, \frac{\eta M \mathcal{D}(0)}{x^{1+\eta}} \, - \, \mathcal{D}'(x) \, \right] \; .
\label{eq:Ia}
\end{equation}
After integrating by parts\footnote{Of course, one could also make hypotheses on the first derivative
$\mathcal{D}'(x)$ and avoid the partial integration.} we then find
\begin{equation}
I_a(\ell^2) \, = \, - \ln(x) \, \left.\left[ \, \frac{M \mathcal{D}(0)}{x^{\eta}} \,+\, \mathcal{D}(x) \, \right] \,
    \right|_{\ell^2}^{\infty}  \, + \, \int_{\ell^2}^{\infty} \, \d x \, \left[ \,
         \frac{M \mathcal{D}(0)}{x^{1+\eta}} \,+\, \frac{\mathcal{D}(x)}{x} \, \right] \; ,
\label{eq:Iaparts}
\end{equation}
which is clearly finite under the assumptions made for the gluon propagator $\mathcal{D}(p^2)$.
Finally, the remaining term
\begin{equation}
\int_{p^2}^{\ell^2} \, \d x \, \ln\left( x^{\eta} + M \right) \,
                              \frac{\eta M \left[ \mathcal{D}(x) \, - \, \mathcal{D}(0) \right] \, - \,
                              x \, \left( x^{\eta} + M \right) \, \mathcal{D}'(x)}{x^{1+\eta}} \; ,
\end{equation}
with $p \ll \ell$, can be easily bounded by making the assumption that neither $\mathcal{D}(x)$ nor $\mathcal{D}'(x)$
display a singularity for $x \in [p^2, \ell^2]$. Thus, we can conclude that the integral \eqref{c2} is indeed
finite. We further notice that $I_a(\ell^2)$ is null in the limit $\ell^2 \to \infty$. Going back to the first
proof in Section \ref{sec:d=2sing}, one can verify that the above conditions allow to show that the
Gribov form-factor $\sigma(p^2)$ goes to zero for $p^2$ going to infinity and that the term
$- \mathcal{D}(0) \,\ln(p^2)$ is the only singularity of $\sigma(p^2)$ in the IR limit
$p^2 \to 0$.

The situation is, of course, very similar in the second proof. In this case we considered a finite value for
$\mathcal{D}(0)$ and the limit
\begin{equation}
\lim_{p^2 \to 0} \, \left[ \, p^2 \, \ln(p^2) \, - \, p^2 \, \right] \, \mathcal{D}'(p^2) \, = \, 0 \; .
\label{eq:Dlim3}
\end{equation}
We also imposed, for large momenta $p^2$, the limits
\begin{equation}
\lim_{p^2 \to \infty} \, \left[ \, p^2 \, \ln(p^2) \, - \, p^2 \, \right] \, \mathcal{D}'(p^2) =
\lim_{p^2 \to \infty} \, \ln(p^2) \, \mathcal{D}(p^2) =
\lim_{p^2 \to \infty} \, \frac{\hat{D}(p^2)}{p^2} \, = \, 0 \; .
\label{eq:Dlim4}
\end{equation}
Clearly, any gluon propagator with an IR behavior of the type $\mathcal{D}(p^2) \approx \mathcal{D}(0) + B \,
p^{2 \eta}$, with $\eta > 0$ and with $\mathcal{D}(0)$ and $B$ finite satisfies the limit (\ref{eq:Dlim3}).
Also, if we make the hypothesis $\mathcal{D}'(p^2) \sim 1/p^{2+2 \epsilon}$, with $1 > \epsilon > 0$ for
large values of $p^2$ we have, in the same limit,\footnote{\label{foot:tauberian} As before, one needs to be
careful and make hypotheses on $\mathcal{D}'(p^2)$ and not on $\mathcal{D}(p^2)$. For example, after making
assumptions on the UV behavior of $\mathcal{D}(p^2)$ one could employ de l'H\^opital's rule in order to obtain
\begin{equation}\label{6}
 0 \, = \, \lim_{p^2 \to \infty} \mathcal{D}(p^2) \ln(p^2) \, = \,
  \lim_{p^2 \to \infty} \frac{\mathcal{D}(p^2)}{\frac{1}{\ln(p^2)}} \, = \,
  \lim_{p^2 \to \infty} \frac{\mathcal{D}'(p^2)}{-\frac{1}{p^2 \, \ln^2(p^2)}} \, = \, -
   \lim_{p^2 \to \infty} p^2 \, \ln^2(p^2) \, \mathcal{D}'(p^2)
\end{equation}
and conclude that $\lim_{p^2 \to \infty} \, \left[ \, p^2 \, \ln(p^2) \, - \, p^2 \, \right] \,
\mathcal{D}'(p^2) = 0$ follows from the condition $\lim_{p^2 \to \infty} \, \ln(p^2) \, \mathcal{D}(p^2) = 0$.
However, it is easy to find counterexamples to this result. A classical one is $\mathcal{D}(p^2) = \sin(p^2)/p^2$
for which $\lim_{p^2 \to \infty} \, \left[ \, p^2 \, \ln(p^2) \, - \, p^2 \, \right] \, \mathcal{D}'(p^2)$ is
undetermined even though $\lim_{p^2 \to \infty} \, \ln(p^2) \, \mathcal{D}(p^2) = 0$ clearly holds. Thus, in
order to apply the result in Eq.\ (\ref{6}), additional auxiliary (also called Tauberian) conditions have to be
imposed to the function $\mathcal{D}(p^2)$, for example on $\mathcal{D}'(x)$ or on $\mathcal{D}''(x)$ (see
again Ref.\ \cite{debruijn}). In particular, imposing $\mathcal{D}(p^2) > 0$ is not sufficient since
$\mathcal{D}(p^2) =  \left[2 + \sin(p^2)\right]/p^2$ is also a counterexample. Here we find it simpler
to make hypotheses directly on the asymptotic behavior of $\mathcal{D}'(p^2)$. However, one should also notice
that the large-$x$ behavior $\mathcal{D}'(p^2) \sim 1/p^{2+2 \epsilon}$, with $1 > \epsilon > 0$, could also
imply $\mathcal{D}(p^2) \sim 1/p^{2 \epsilon} + \text{constant}$ (see for example Ref.\ \cite{site}). Thus,
we also have to impose explicitly the condition $\mathcal{D}(p^2) \to 0$ for $p^2 \to \infty$.} $\mathcal{D}(p^2)
\sim 1/p^{2 \epsilon}$ and $\hat{D}(p^2) \sim p^{2-2 \epsilon}$ and one can easily prove the results in Eq.\
(\ref{eq:Dlim4}). As a consequence, we can also verify that the integral [see Eq.\ (\ref{eq:sigma0due})]
\begin{equation}
I_b(p^2) \, = \, \int_{p^2}^{\infty} \, \d x \,  \left[ \, x \, \ln(x) \, - \, x \, \right]
                  \, \mathcal{D}''(x)
\label{eq:Dlim5}
\end{equation}
is finite for any $p^2 \geq 0$. Indeed, we can integrate by parts\footnote{Again, instead of integrating by
parts, we could also make hypotheses on the second derivative $\mathcal{D}''(x)$.} obtaining
\begin{eqnarray}
I_b(p^2) & = & \left[ \, x \, \ln(x) \, - \, x \, \right] \, \mathcal{D}'(x) \Bigr|_{p^2}^{\infty} \, - \,
                        \int_{p^2}^{\infty} \, \d x \, \ln(x) \,\mathcal{D}'(x) \\[2mm]
         & = & \left[ \, p^2 \, \ln(p^2) \, - \, p^2 \, \right] \, \mathcal{D}'(p^2) \, - \,
                        \int_{p^2}^{\infty} \, \d x \, \ln(x) \,\mathcal{D}'(x) \; .
\label{eq:Ib}
\end{eqnarray}
Note that the integral on the r.h.s.\ of Eq.\ (\ref{eq:Ib}) also appears in the second term of Eq.\
(\ref{eq:Ia}). Thus, using Eqs.\ (\ref{eq:Iaparts}) and (\ref{eq:Dlim4}) we have that, for large $p^2$, the
integral $I_b(p^2)$ is finite and $\lim_{p^2 \to \infty} I_b(p^2) = 0$. At the same time, for $p^2$ going
to zero we have $\mathcal{D}''(p^2) \sim p^{2 \eta - 4}$ and the integrand in Eq.\ (\ref{eq:Dlim5}) behaves
as $\left[ \, \ln(x) \, - \, 1 \, \right] x^{\eta -1}$. Thus, with $\eta > 0$, no singularity arises\footnote{For
the first term this can be easily shown by integrating by parts.} from the integration at $x=0$. This result
completes the conditions necessary (in the second proof) to show that $\sigma(p^2)$ goes to zero at large momenta
and that the IR singularity $- \mathcal{D}(0) \,\ln(p^2)$ appears in the limit $p^2 \to 0$.

Finally, due to the well-known results
\begin{equation}
\lim_{x \to 0} \, x^{\epsilon} \, \ln^a(x) \, = \,
\lim_{x \to \infty} \, \frac{\ln^a(x)}{x^{\epsilon}} \, = \, 0
\end{equation}
for $\epsilon > 0$, it is clear that the above proofs can also be generalized to asymptotic behaviors that
include logarithmic functions. At large momenta these logarithmic corrections could be present, for example,
if one uses, as an input in the evaluation of $\sigma(p^2)$, a gluon propagator obtained in perturbation
theory beyond the tree-level term.

%%%%%%%%%%%%%%%%%%%%%%%%%%%%%%%%%%%%%%%%%%%%%%%%%%%%%%%%%%%%%%%%%%%%%%%%%%%%%%%%%%%%%%%%%%%%%%%%%%%%%%%%%%%%%%%

\section{The $d=4$ Case Using a MOM Scheme}
\label{sec:mom}

Let us start from Eq.\ (\ref{eq:Igeneral-dell2}), with $\nu=1$ and $\ell=\infty$, and subtract $\sigma(\mu^2)$
from $\sigma(p^2)$, where $\mu$ is a fixed momentum. Then we can write
\begin{eqnarray}
\frac{\sigma(p^2) \, - \, \sigma(\mu^2)}{g^2 N_c} & = &
   \frac{\Omega_d}{(2\pi)^d} \, \frac{d-1}{d} \,  \Biggl\{ \,
          \int_0^{p} \, \d q \, q^{d-1} \, \mathcal{D}(q^2)
                 \, \frac{ _2F_1\left(1, 1-d/2; 1+d/2; q^2/p^2\right) }{p^2} \nonumber \\[2mm]
   & & \qquad \qquad \quad \, - \,
          \int_0^{\mu} \, \d q \, q^{d-1} \, \mathcal{D}(q^2)
                 \, \frac{ _2F_1\left(1, 1-d/2; 1+d/2; q^2/\mu^2\right) }{\mu^2} \nonumber \\[2mm]
   & & \qquad \qquad \quad \, + \,
          \int_{p}^{\mu} \, \d q \, q^{d-3} \, \mathcal{D}(q^2)
                 \, _2F_1\left(1, 1-d/2; 1+d/2; p^2/q^2\right) \nonumber \\[2mm]
   & & \qquad \qquad \quad \, + \,
          \int_{\mu}^{\infty} \, \d q \, q^{d-3} \, \mathcal{D}(q^2) \, \left[ \,
                 _2F_1\left(1, 1-d/2; 1+d/2; p^2/q^2\right) \right. \nonumber \\[2mm]
   & & \qquad \qquad \qquad \qquad \qquad \, - \, \left.
                 _2F_1\left(1, 1-d/2; 1+d/2; \mu^2/q^2\right) \, \right] \, \Biggr\} \; .
\label{eq:4dmom}
\end{eqnarray}
In the case $d=4$ we can use the result (\ref{eq:2F1-D4}) in Appendix \ref{sec:hyper}. Thus, the last integral
in the above expression becomes
\begin{equation}
\frac{ \mu^2 \, - \, p^2 }{3} \, \int_{\mu}^{\infty} \, \d q \, \frac{\mathcal{D}(q^2)}{q} \; ,
\label{eq:UV4d}
\end{equation}
which is UV-finite for any gluon propagator that goes to zero at large momenta. The apparent linear
divergence for large $p^2$ in the above integral is of course canceled by the third integral in Eq.\
(\ref{eq:4dmom}) above, i.e.\ by
\begin{equation}
\int_{p}^{\mu} \, \d q \, q \, \mathcal{D}(q^2) \, \left( 1 \,-\,\frac{p^2}{3 q^2} \, \right) \, = \,
\int_{\mu}^{p} \, \d q \, \frac{\mathcal{D}(q^2)}{3 q} \, \left( p^2 \, - \, q^2 \right) \; .
\end{equation}
Then, for $p^2 \to \infty$ and for $\mathcal{D}(q^2) \sim 1/q^2$ at large momenta one only gets a
logarithmic contribution $-\ln(p)$, as expected.

Using the above formula (\ref{eq:4dmom}), the proof that $\sigma(p^2)$ is IR-finite for $\mathcal{D}(p^2) > 0$
can be obtained as in Section \ref{sec:d-exact}. Indeed, for $p^2$ going to zero we have to consider only the
first and the third integrals\footnote{Clearly, the second integral does not depend on $p^2$ and the last one
is regular at $p^2 = 0$ [see Eq.\ (\ref{eq:UV4d})].} in Eq.\ (\ref{eq:4dmom}). Then, using again the result
(\ref{eq:2F1-D4}) we can write
\begin{equation}
\int_0^{p} \, \d q \, q^3 \, \frac{\mathcal{D}(q^2)}{p^2} \, \left( 1 \,-\,\frac{q^2}{3 p^2} \, \right)
\, + \, \int_{p}^{\mu} \, \d q \, q \, \mathcal{D}(q^2) \, \left( 1 \,-\,\frac{p^2}{3 q^2} \, \right)
\, \leq \, \int_0^{p} \, \d q \, q^3 \, \frac{\mathcal{D}(q^2)}{p^2}
\, + \, \int_{p}^{\mu} \, \d q \, q \, \mathcal{D}(q^2)
\end{equation}
and no singularity arises in the limit $p^2 \to 0$ if $\mathcal{D}(0) > 0$. At the same result one
arrives by setting $\mathcal{D}(q^2) = \mathcal{D}(0)$ and by integrating explicitly the l.h.s.\ in the
above equation.

%%%%%%%%%%%%%%%%%%%%%%%%%%%%%%%%%%%%%%%%%%%%%%%%%%%%%%%%%%%%%%%%%%%%%%%%%%%%%%%%%%%%%%%%%%%%%%%%%%%%%%%%%%%%%%%

\end{document}